%% file: M_dwarfs_FINAL.tex
\documentclass[useAMS,usenatbib]{mnras}
\usepackage{graphicx}
\usepackage{lscape}
\usepackage{longtable}
\usepackage{epstopdf}
\usepackage{txfonts}
\usepackage{color}

\def\aj{AJ}%
\def\araa{ARA\&A}%
\def\apj{ApJ}%
\def\apjl{ApJ}%
\def\apjs{ApJS}%
\def\ao{Appl.~Opt.}%
\def\aap{A\&A}%
\def\mnras{MNRAS}%
\def\pasp{PASP}%
\def\pasj{PASJ}%
\title[Dust settling in Cha I and TWA]{Physical parameters of late M-type members of Chamaleon I and TW Hydrae Association: Dust settling, age dispersion and activity\thanks{Based on the ESO observing program 077.C-0815(A)}}
\author[A. Bayo, et al.]{A. Bayo$^{1,2}$\thanks{E-mail: amelia.bayo@uv.cl}, D. Barrado$^{3}$, F. Allard$^{4}$, T. Henning$^{2}$, F. Comer\'on$^{5}$, M. Morales-Calder\'on$^{3}$, \newauthor A. S. Rajpurohit$^{6}$, K. Pe\~na Ram\'irez$^{7,8}$ and J. C. Beam\'in$^{1,8}$ \\
\\
$^{1}$Instituto de F\'isica y Astronom\'ia, Facultad de Ciencias, Universidad de Valpara\'iso, Av. Gran Breta\~na 1111, 5030 Casilla, Valpara\'iso, Chile, Chile\\
$^{2}$Max Planck Institut f\"ur Astronomie, K\"onigstuhl 17, 69117, Heidelberg, Germany\\
$^{3}$Depto. Astrof\'isica, Centro de Astrobiolog\'ia (INTA-CSIC), ESAC campus, Camino Bajo del Castillo s/n, E-28692 Villanueva de la Ca\~nada, Spain\\
$^{4}$Centre de Recherche Astronomique de Lyon (CRAL), \'Ecole Normale Sup\'erieure de Lyon, 69364, Lyon, France\\
$^{5}$European Southern Observatory, Alonso de C\'ordova 3107, Vitacura, Santiago, Chile\\
$^{6}$Astronomy \& Astrophysics Division, Physical Research Laboratory, Ahmedabad 380009, India\\
$^{7}$Instituto de Astrof\'isica. Pontificia Universidad Cat\'olica de Chile (IA-PUC), E-7820436 Santiago, Chile.\\
$^{8}$Millennium Institute of Astrophysics, Santiago, Chile
}
\begin{document}

\date{Accepted xxx. Received xxx}

\pagerange{\pageref{firstpage}--\pageref{lastpage}} \pubyear{2014}

\maketitle

\label{firstpage}

\begin{abstract}
Although mid-to-late type M dwarfs are the most common stars in our stellar neighborhood, our knowledge of these objects is still limited. Open questions include the evolution of their angular momentum, internal structures, dust settling in their atmospheres, age dispersion within populations. In addition, at young ages, late-type Ms have masses below the hydrogen burning limit and therefore are key objects in the debate on the brown dwarf mechanism of formation.
In this work we determine and study in detail the physical parameters of two samples of young, late M-type sources belonging to either the Chamaeleon I Dark Cloud or the TW Hydrae Association and compare them with the results obtained in the literature for other young clusters and also for older, field, dwarfs.
We used multi-wavelength photometry to construct and analyze SEDs to determine general properties of the photosphere and disk presence. We also used low resolution optical and near-infrared spectroscopy to study activity, accretion, gravity and effective temperature sensitive indicators.
We propose a VO-based spectral index that is both temperature and age sensitive. We derived physical parameters using independent techniques confirming the already common feature/problem of the age/luminosity spread. In particular, we highlight two brown dwarfs showing very similar temperatures but clearly different surface gravity (explained invoking extreme early accretion). We also show how, despite large improvement in the dust treatment in theoretical models, there is still room for further progress in the simultaneous reproduction of the optical and near-infrared features of these cold young objects.
\end{abstract}


\section{Introduction}

M dwarfs are the most common stars in our stellar neighborhood \citep{Salpeter55,Chabrier03,Henry06}, but they are also among the least well understood. While the number of known, faint M-type stars and brown dwarfs has increased dramatically and they have even become common targets to the search for habitable earth-like planets \citep{Berta13,Dressing15, carmenes15}, our understanding of their fundamental properties has not progressed at the same speed. Some of the main open questions can be grouped into the following key interrelated topics: interiors, the mass-luminosity relation, complex atmospheres for low temperatures, and the earliest stages of formation.

In comparison with solar-type stars, there is a dramatic difference in internal structure affecting spectral types later than $\sim$M3. These cool objects are fully convective bodies, and therefore the classical $\alpha \Omega$ dynamo does not operate anymore. This difference in the internal structure most likely has implications in the angular momentum evolution, the rotation-activity connection, and maybe also in the joint evolution of the pre-main low-mass stars with their circumstellar disks. 

As an example, while evidence grows towards disk-locking \citep{Shu94, Bouvier97} being the key ingredient to compensate the spin-up of pre-main sequence (PMS) stars  \citep{Affer13}, the bi-modality in rotational periods found between accreting (CTTs) and not accreting (WTTs) stars has not been confirmed for the lower mass domain yet. 

In addition, although M dwarfs are extremely common, their luminosity function suggests that not all subclasses are equally populated. \cite{Dobbie02} compiled such functions for several young clusters and reported a common feature to all of them, a significant lack of M7-M8 sources. Further observational evidence can be found in \cite{Stauffer99, Barrado02b} for the {$\alpha$} Persei Open Cluster, in \cite{Barrado01b} for IC2391, \cite{Luhman08} for Chamaeleon I or \cite{Bayo11} for Collinder 69.  These clusters cover an age range of several tens of Myr and their M members are still in the PMS phase.
The explanation proposed by \cite{Dobbie02} was that this feature is the consequence of a drop in the mass-luminosity (M/L) relation that could be caused by the formation of dust in the atmospheres at these temperatures.

One of the implications of such a drop in the M/L relationship would be that the masses of objects cooler than M8 could have been systematically underestimate, which would affect the shape of the lower end of the Initial Mass Function (IMF). However, a different interpretation is given in \cite{Thies08}, where this dip in the luminosity function is explained as the outcome of a change in binarity fraction and properties resulting from a different mechanism of formation for some very-low mass stars and brown dwarfs (see below).

Vast theoretical effort has been invested in the understanding of the dust settling problem and the production of synthetic spectra that reproduce the features of late M dwarfs both in the optical and near-infrared. The next natural step is to confront theory with observations and few examples are already available in the literature (see for example \citealt{Rajpurohit13} for old field M-dwarfs and \citealt{Bonnefoy14} for younger M and L-type objects) but since most of these works can only be performed on small samples, further studies are still mandatory to asses the goodness of the newly availably grids of synthetic spectra. In this work we will focus on the progress achieved in this respect with the ``BT-Settl" grid by \cite{Allard12} to characterize not only the effective temperature of M dwarfs and young objects, but also their surface gravity (directly related, in principle, with their age). 

Besides the unique interior and atmospheric characteristics of these objects, the M spectral class (at young ages) is a mix of low-mass stars and brown dwarfs, and therefore, as a class, it is affected by the still open debate on which is the dominant mechanism of formation of substellar objects.

Molecular cloud fragmentation is accepted as the initial step on the formation of low-mass stars, but since the typical Jean mass of molecular clouds is $\sim$1M$_{\odot}$, objects below the Hydrogen-burning limit ($\sim$0.072M$_\odot$) cannot form as scaled-down version of the former. Several scenarios are proposed in the literature to overcome this caveat either invoking new mechanisms like dynamical interactions \citep{Reipurth01}, massive-disks fragmentation \citep{Goodwin07, Stamatellos07}, or photoevaporation \citep{Whitworth04}, or modifying the initial conditions introducing turbulence so that the Jeans mass decreases \citep{Padoan02, Hennebelle08}.

Finally, M-type sources belonging to young clusters have been reported to show similar or even larger luminosity/age dispersion than earlier spectral type members (see for example \citealt{Barrado01a,Zapatero02,daSilva09,Bayo11}). This dispersion  (if real and not a consequence of observational uncertainties) can have two interpretations; either the members of these clusters are not coeval and therefore the star formation process is much slower than expected from the timescales of shock-dominated turbulence. Or some process during the formation of the individual members of the clusters makes  objects with the same age and effective temperature to exhibit remarkably different luminosities. The later scenario is preferred by \cite{Baraffe10,Vorobyov15} where non--steady accretion at the very early stages of star formation can account for the luminosity spread. 

To try to shed light on at least some of these questions, we have compiled and studied in detail a sample of young very low mass stars and brown dwarfs belonging to the Chamaeleon I dark cloud (Cha I) and the TW-Hydra association (TWA). We only selected spectroscopically confirmed members to the two associations with late M spectral types. We gathered the available information in the literature regarding photometry, rotational velocities, ages, distances and low resolution optical spectroscopy (see Table~\ref{sources}). We combined those data with new spectroscopic near-infrared (NIR) low resolution observations and archival Spitzer IRAC and MIPS data. We also obtained NIR low resolution spectroscopy for a set of field objects to use as templates of ``old M dwarfs".

In short, Cha I is the most active star forming cloud from the Chamaeleon complex \citep{Reipurth91}. It is nearby (160 pc; \citealt{Wichmann98,Knude98}), the extinction is low as compared to other star forming regions \citep{Cambresy97,Cambresy99}, and it lies at relatively high galactic latitude, implying a moderate density of background objects. The estimated age is $\sim$2 Myr with a few Myr difference in isochronal ages reported by \cite{Luhman07b} between the two knots where its members are preferentially located (see \citealt{Luhman08} for the most up-to-date census of members).

We have selected a sample of 11 very low--mass stars and brown dwarfs from \citet{Comeron00}, lying in the central region of the cloud. Their spectral types are in the range M6-M8; some show prominent H$\alpha$ emission; the projected rotational velocities have been estimated by  \cite{Joergens01} for eight of them; two of the sources have been reported to actually be binary systems with extremely different separations: Cha H$\alpha$ 8 ($\lesssim$1 AU, \citealt{Joergens10}) and Cha H$\alpha$ 2 ($\sim$35 AU \citealt{Vogt12}); and finally,
the accretion properties for three of the selected sources were recently addressed in \citet{Manara16}. 

TWA, is a nearby ($\sim$50 pc) moving group which includes a few dozen stars and brown dwarfs. The age has been estimated to be between $\sim$8 and 20 Myr \citep{Kastner97, Stauffer95, Soderblom98, Weintraub00, Makarov01, Reza06, Barrado06, Ducourant14, Bell15}. We have selected three brown dwarfs from this association, namely 2MASS J1139511-315921, SSSPMJ1102-3431, and 2MASS J1207334-393254 (hereafter 2M1139, PMJ1102 and 2M1207), with spectral types M8, M8.5, and M8, respectively. One of them, 2M1207, is of particular interest due to its complexity: it is a double system (brown dwarf + planetary mass companion, \citealt{Chauvin04}) where the brown dwarf is still undergoing active accretion \citep{Mohanty03b} and harbours a circum-substellar disk \citep{Sterzik04}. 

\input{sources.tex}

This work is organized as follows: In Section~\ref{data} we describe the compiled data, including the optical spectroscopy and give details on the new NIR low resolution spectroscopy. In Sections~\ref{spt} and ~\ref{temperatures}  we estimate near-infrared spectral types, effective temperatures, and interstellar extinction. The latter two based on comparisons of the SED and optical+infrared spectra with models. 
 In Section~\ref{age} we study different age indicators for the sample. In Section~\ref{HR} we construct a distance independent Hertsprung-Russel diagram and discuss possible causes for age discrepancies. In Section~\ref{aar} we study the accretion and activity indicators of the sample and relate them to their rotational velocities (for Cha I members). Finally in Section~\ref{conclussions} we summarize our work before concluding.

\section[]{Literature data, new observations and data reduction}

\label{data}

As mentioned in the Introduction, besides obtaining new NIR spectroscopy, we have searched in the literature for public data useful for our analysis. For the sample of Cha I sources, \cite{Comeron00} provided reddening-independent spectral types and optical photometry (both shown in Table~\ref{sources}). Besides, \cite{Joergens01} obtained projected rotational velocities for eight of them. Regarding the TWA sources, \cite{Gizis02} derived spectral types for 2M1207 and 2M1139 and \cite{Scholz05a} estimated the spectral type of SSSPMJ1102. Finally, the SIMBAD database provides spectral types for all the field M-dwarfs considered in this work.

We have had access to previously published low-resolution optical spectroscopy for all sources in the Cha I and TWA samples but one (SSSPMJ1102) and we provide a summary of the characteristics of these data in the following subsection.

\subsection{Mid-infrared photometry}
\label{MIRphot}

We have used Spitzer IRAC 3.6, 4.5, 5.8 and 8.0~$\mu$m and MIPS 24~$\mu$m observations from the archive to build more populated SEDs for the confirmed members from \cite{Comeron00}.
Observations were obtained as part of a GTO program in October, 2004 and February, 2006, for IRAC and MIPS, respectively. All the mosaics cover an area that contains the objects of interest. The eleven Cha~I targets were detected in the four IRAC bands (although Cha H$\alpha$ 11 photometry has not been extracted because the region was saturated by a nearby object) but only four of them had detectable fluxes at MIPS 24~$\mu$m. Aperture photometry was performed on the mosaics using the task PHOT under the IRAF\footnote{IRAF is distributed by the National Optical Astronomy Observatory, which is operated by the Association of Universities for Research in Astronomy, Inc. under contract to the National Science Foundation.} environment. For the IRAC images we used an aperture radius of 3 pixels and the sky was computed using a circular annulus 4 pixels wide starting at a radius 3 pixels away from the center. Zero point fluxes of 280.9, 179.7, 115.0 and, 64.13~Jy and aperture corrections of 1.124, 1.127, 1.143, and 1.234 (provided by the SSC) were used for IRAC channels 1 through 4 in order to compute the magnitudes. For the MIPS 24~$\mu$m image we used an aperture of 5.31 pixels and a sky annulus from 8.16 to 13.06 pixels. The zero point flux and aperture correction applied were 7.14~Jy and 1.167, respectively.  The magnitudes and errors obtained are shown in Table~\ref{sources:IRAC} (Note that the additional calibration uncertainty of $\sim$2\% has already been included in the quoted errors). For most of the sources, \cite{Luhman08a} provided values for the [5.8] and [8.0] bands, and the agreement with our photometry is within the error bars.

\input{IRAC_Phot.tex}

\subsection{Optical spectroscopy}
\label{OP_spec}

\cite{Comeron00} provided low resolution long-slit optical spectroscopy for all the Cha I sources in our sample. The observations were performed with EMMI at the NTT telescope  on 17 and 18 April 1999, with a spectral coverage from 6000 to 10000\AA~and a resolving power of $\sim$270. 

Regarding the TWA sample, \cite{Barrado06} provided higher resolution optical spectroscopy (still moderate, R$\sim$2600) obtained with the Boller \& Chivens (B\&C) spectrograph at the Magellan/Baade telescope on March 11, 2003 with a spectral coverage of 6200-7800\AA.

\subsection{Near infrared spectroscopy}
\label{NIR_spec}

The observations of the three samples (Cha I, TWA and field M dwarfs) were part of the ESO program 077.C-0815(A) carried out on April 25--27 2006 with the ESO NTT near-IR spectrograph/imaging camera SOFI (Son Of ISAAC). We observed our targets with the two low-resolution grisms (red and blue) to roughly cover the $JHK$ bands. The blue grism covers the spectral region between 0.95--1.63 $\mu$m and the red grism the region between 1.53--2.52 $\mu$m. The corresponding spectral resolutions are 930 and 980 respectively for a 0.6$''$ slit.

The telescope was nodded 30$''$ along the slit between consecutive positions following the usual ABBA pattern. 
In addition to the program sources we observed several atmospheric standards with airmasses similar to those of the science objects. These spectra were used in combination with a high-resolution model of Vega to remove the hydrogen absorption features in the stellar spectra, as described by \citet{Vacca03}, and to estimate the instrumental response.
A xenon lamp provided the wavelength calibration for our data (consistent with the OH airglow calibration) with an accuracy of 1.2 \AA~for the blue grism, and 2 \AA~for the red one (corresponding to 1/5 of the size of the pixel).

To reduce the data we used IRAF and performed the standard steps: flat-field the individual frames, subtract pairs of nodded observations, align individual exposures corresponding to each grism and target, combine the corresponding frames into one blue and one red image per object, extract an co-add the one dimensional spectra, and correct for instrumental response.

We finally combined the blue and the red grism spectra for each object and eliminated regions of deep atmospheric absorptions from our analysis as not satisfactory corrections were obtained in these regions.
The useful spectral ranges are therefore: 9800--11000 \AA, 11600--13300 \AA, 14950--17500 \AA, and 20600--23500 \AA.

\subsection{Flux calibration}
\label{calibration}

For each epoch and instrumental setup we carried out the flux calibration via fit of the synthetic photometry (calculated as described in \citealt{Bayo08}) to the known fluxes for standards taken during the same nights.

Unfortunately flux standards were not available for the optical spectroscopy for the TWA members.  This will limit our analysis in Secion~\ref{specfit} but the optical spectra will still be useful to study activity and accretion in Section~\ref{aar}.

For the Cha I sources, we found a good agreement in the stitching of the optical and NIR spectra and to check our flux calibration we compared the synthetic photometry of the science targets with the $R_C$ and $I_C$ magnitudes from \cite{Comeron00} and the $JHKs$ magnitudes from \cite{2MASS}.

\begin{figure*}
\resizebox{\hsize}{!}{\includegraphics{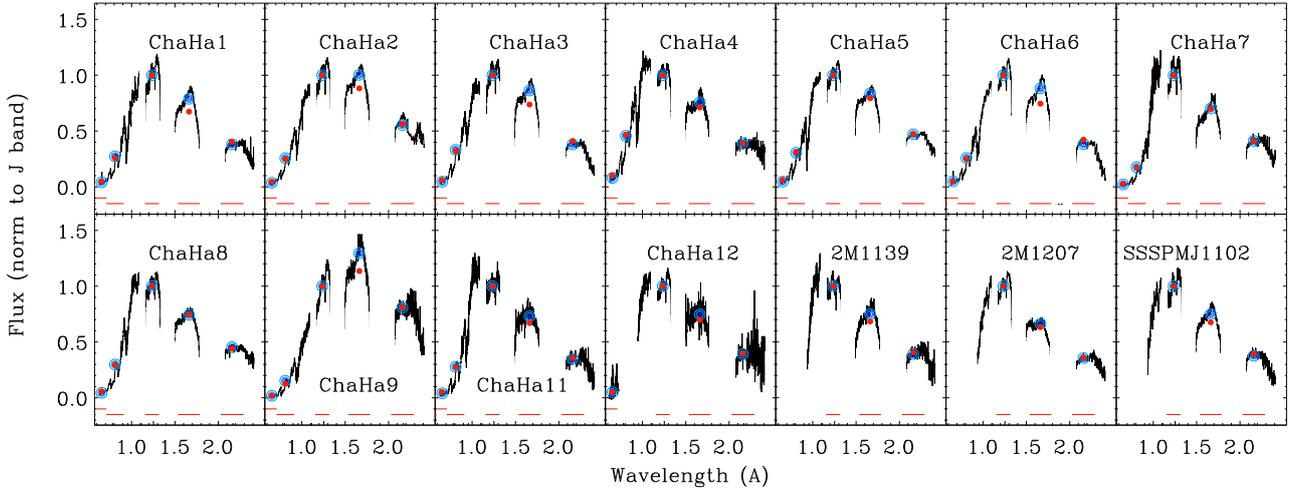}}
\caption{
Flux calibrated spectra (normalized to the J-band flux) of the Cha I and TWA members (black solid line). The observed photometry (also normalized to the J-band) is displayed as red circles and the synthetic photometry as shaded blue circles. The size of the circles showing the synthetic photometry have been scaled so that they represent uncertainties of $\sim$5\% in the H band. The horizontal red lines illustrate the effective bandwidth of the filters involved in the flux calibration.}
\label{fig:ChaIop}
\end{figure*}

In Fig.~\ref{fig:ChaIop} we display our calibrated spectra (normalized to the J band flux) with black solid lines and on top of them with red circles the 2MASS and \cite{Comeron00} photometry (the latter only for the Cha I objects) and with blue shaded circles the synthetic photometry calculated for the same filter sets. 

The comparison returns differences between the observed and synthetic photometry compatible with the 2MASS errors ($\sim3\%$ mean differences in flux) for the field dwarfs. On the other hand, these differences are larger for the young targets reaching a maximum of $\sim$15\% for the Cha I sample and $\sim$10\% for the TWA members. We attribute these larger differences to the intrinsic variable nature of young sources, what is supported by the fact that the objects showing the most intense H$\alpha$ emission are also those where the synthetic photometry deviates the most from the observed one.

To be conservative, since we are stitching data from different epochs we will assume in the comparison with models that our flux calibration has an accuracy of 15\% for Cha I members and 10\% for TWA ones.

\section[]{Spectral type determination}
\label{spt}

\begin{figure*}
\resizebox{\hsize}{!}{\includegraphics{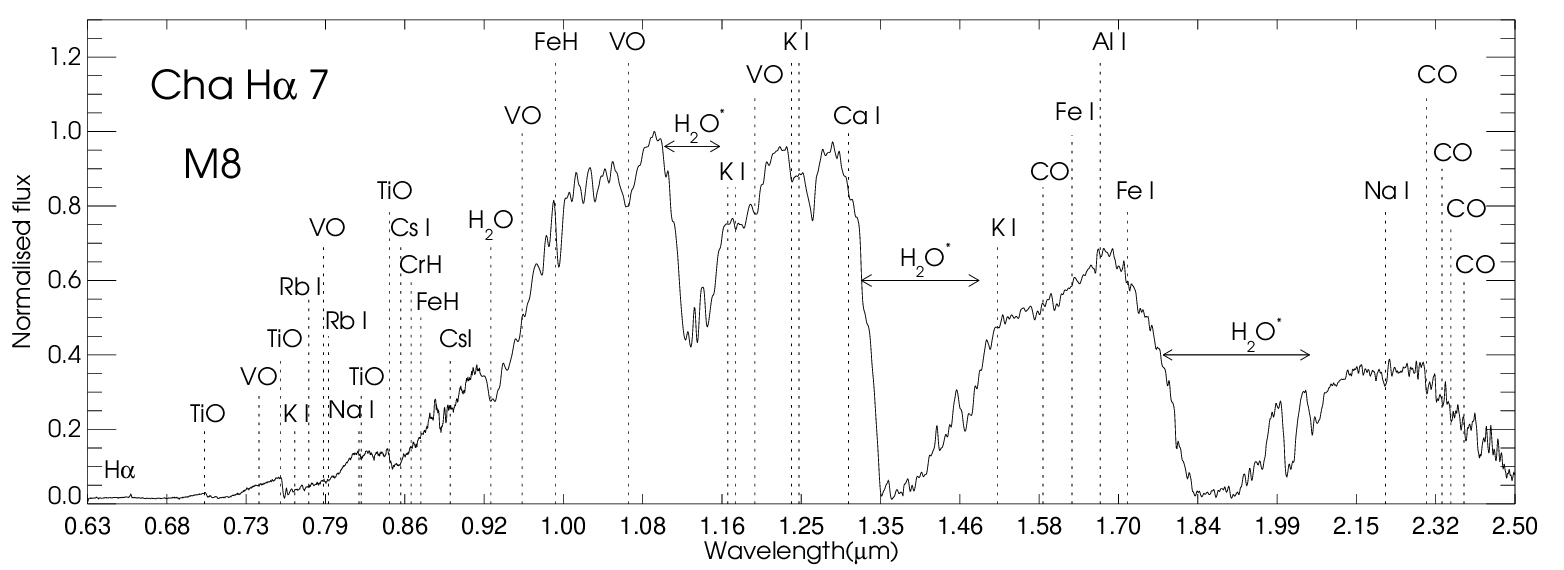}}
\caption{Spectral identifications for the M8 Cha I member Cha H$\alpha$ 7. We used mainly \protect\cite{Jones94,Kirkpatrick93, Geballe96} and \protect\cite{Allard97} for the spectral feature recognition.}
\label{specFeat}
\end{figure*}

M-type objects display very rich-feature spectra. In Fig.~\ref{specFeat}, we show an example of this abundance by identifying a collection of lines and molecular bands on the optical + NIR spectra of the  M8 Cha I member Cha $\alpha$ 7.

In particular, the optical spectra of M dwarfs are distinctively characterized by strong molecular absorption bands. The most notorious correspond to the titanium and vanadium oxides (the latter for spectral types M7 and later) and their intensity, as well as the slope of the pseudo-continuum, have set the bases of spectral indexes that provide quantitative means to estimate spectral subtypes (see for example \citealt{Reid95,Martin96,Martin99,Cruz02,Riddick07}).

A caveat for these indexes is that optical spectra of red faint objects can become very hard to obtain in regions affected by large amounts of interstellar extinction. Fortunately, the NIR spectrum (less sensitive to extinction) of M-type sources is also rich in molecular features that scale with temperature such as H$_2$O, CO, FeH and VO, see \cite{Jones94}. 

\subsection{The NIR water bands}

The most prominent NIR temperature sensitive features are the water bands at 1.4, 1.85 and 2.5$\micron$, with the drawback that the effect of the earth's atmosphere at these wavelengths is maximum and the transmission at the center of these bands at La Silla during our run was practically null. We must note however that the absorption bands coming from the stelar atmosphere are significantly broader than the terrestrial ones (given the much higher temperatures involved) and therefore the wings of the bands (their slope) can still be used for spectral classification.

The dependence of the reddest of these water bands with spectral type was characterized via the reddening independent $Q$ index by \cite{Wilking99}. Later on, \cite{Comeron00} presented averaged H+K spectra of the Cha I members studied in this paper and defined the $I_{H_{2}O}$ reddening independent index focused on the 1.8\micron~band. The conclusion regarding the utility of this index was that the narrow spectral type range analyzed prevented them from achieving any strong verdict. An independent work by   \cite{Gomez02} compare the results obtained with the $I_{H_{2}O}$ and $Q$ indexes and found a good agreement between the two estimations. Besides, \cite{Bayo11} revisited the slope of the  $I_{H_{2}O}$ vs spectral type relation with a larger sample of field dwarfs achieving a classification with 1.5 subtype accuracy.

However, these water indexes still can suffer from two main issues: the choice of extinction law assumed to achieve reddening independent relations; and the possible dependence of the water-bands morphology with age (see \citealt{Bayo11} and references therein). Concerning the extinction law, there are significant differences between correcting the flux ratios taken at several spectral bands using the ``classical" \cite{Rieke85} relation or the one revisited by \cite{Fitzpatrick99} for the near and mid-infrared. Besides, the assumption of a particular $R_v$ value can add a further source of uncertainty. To avoid these issues we have slightly modified the ($f_1,f_2,f_3,f_4$) bands selected in \cite{Comeron00} to have centers at 1.685, 1.75, 2.085 and 2.15\micron~(so that the ``continuum-band" pairs have negligible differences in extinction), kept the same 0.05\micron~widths for the spectral bands and redefine the $I_{H_{2}O}$ simply as $(f_1/f_2)\times(f_3/f_4)$.

\begin{figure}
\resizebox{\hsize}{!}{\includegraphics{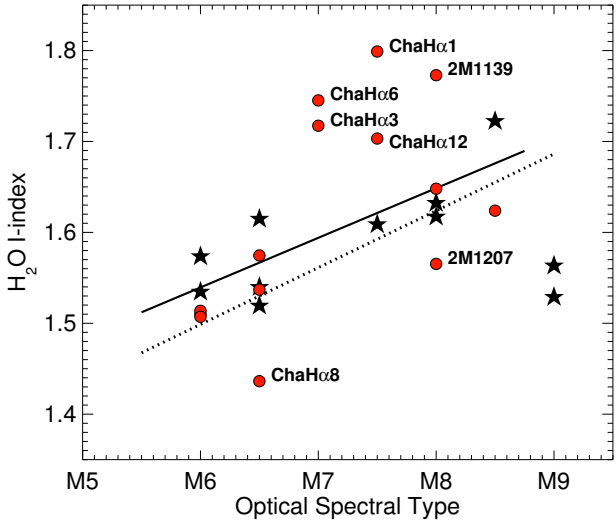}}
\caption{Optical spectral type vs revisited $I_{H_{2}O}$ index diagram. Field dwarfs are displayed as filled black five point star symbols and Cha I and TWA members as filled red circles. The original relation proposed by Comer\'on et al. (2000) is shown with a dotted line and the solid line corresponds to a linear fit to the field dwarf sample with spectral types earlier than M9. The young sources not following the trend are highlighted with the source name as a label.}
\label{Iindex}
\end{figure}

The revised water index values vs. the spectral type of all the sources in our sample is shown in Fig.~\ref{Iindex}: while the field dwarfs with spectral types earlier than M9 follow a linear trend within the 1.5 subspectral class uncertainty, this is not the case for the young objects (Cha I and TWA members). The sources showing the largest differences in the water index with respect to the corresponding field dwarfs values are highlighted with their IDs as labels in the same figure. Most of these sources show higher values of the water index than the field dwarfs with the exception of Cha H$\alpha$ 8 and 2M1207 that fall below this trend. Visual comparison of other temperature sensitivity features in the  Cha H$\alpha$ 8 spectrum with sources sharing the same spectral type in our sample does not reveal any significant differences. However that is not the case of 2M1207. 

\begin{figure*}
\includegraphics[width=13.cm]{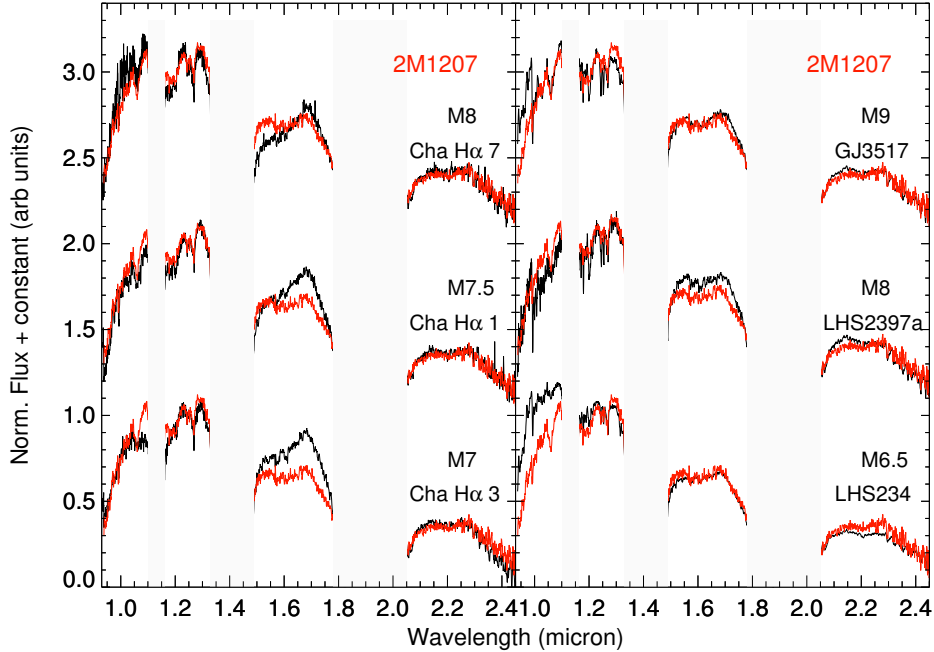}
\caption{Comparison of the NIR spectrum of 2M1207 (in red, M8 spectral type) with objects with similar spectral type from the Cha I (black solid lines on left panel) and field (right panel) samples. Note how the H-band is a better match to the field sources unlike the remaining features that are more similar to those present in the Cha I sources.}
\label{2M1207}
\end{figure*}

In Fig.~\ref{2M1207} we show the comparison of the NIR spectrum of 2M1207 with other sources with similar spectral types from the Cha I (we selected sources showing the lowest level of extinction, see the following sections) and field samples:  while most of the features are a good match to those in the spectrum of Cha H$\alpha$7 (like for example the typical gravity sensitive z and J band shapes of 1-12 Myr sources), it is obvious that the H-band features agree much better with those of the older field dwarfs. This could imply an older age for 2M1207 than previously estimated, but the alkali line analysis of the same spectrum (see Section~\ref{age}) suggests that 2M1207 cannot be much older than 10 Myr (in agreement with the estimation based on Li I from \citealt{Barrado06}). An alternative possibility to an older age could be related to the effect of the surface gravity on the clouds in the atmosphere of 2M1207 that would translate in a different IR spectral shape. 
Finally, the fact that 2M1207 is undergoing active accretion while Cha H$\alpha$ 7 is not, could also be the origin of this different H-band shape.

\subsection{The NIR VO bands}

As mentioned before, the water bands, although being the most prominent ones, are not the only temperature sensitive features present in NIR spectra of M dwarfs. Other examples include the CO bands at the red end of the K-band and also FeH at 0.99\micron \citep{Jones94}. Our observations are not suitable to study any of these molecules because while the former comprises  the reddest edge of the spectrum where the data quality decreases significantly, the latter resides at the wavelength range where we perform the merging with the optical spectra and therefore uncertainties arise from the individual spectra themselves (edges of the detectors, etc) as well as from the merging process.

Moreover, VO is responsible for two absorption bands at 1.05 and 1.2\micron, respectively ($VO1$ and $VO2$ from now on), which is a very suitable wavelength range to be analyzed on our spectra for practical reasons, but also for reasons related to the nature of the young sources: the $\sim$1\micron~spectral region is much less affected by extinction than the optical regime, but at the same time less sensitive to possible infrared excess due to the presence of a circumstellar disk than, for example, the K-band spectral region. 

\begin{figure}
\resizebox{\hsize}{!}{\includegraphics{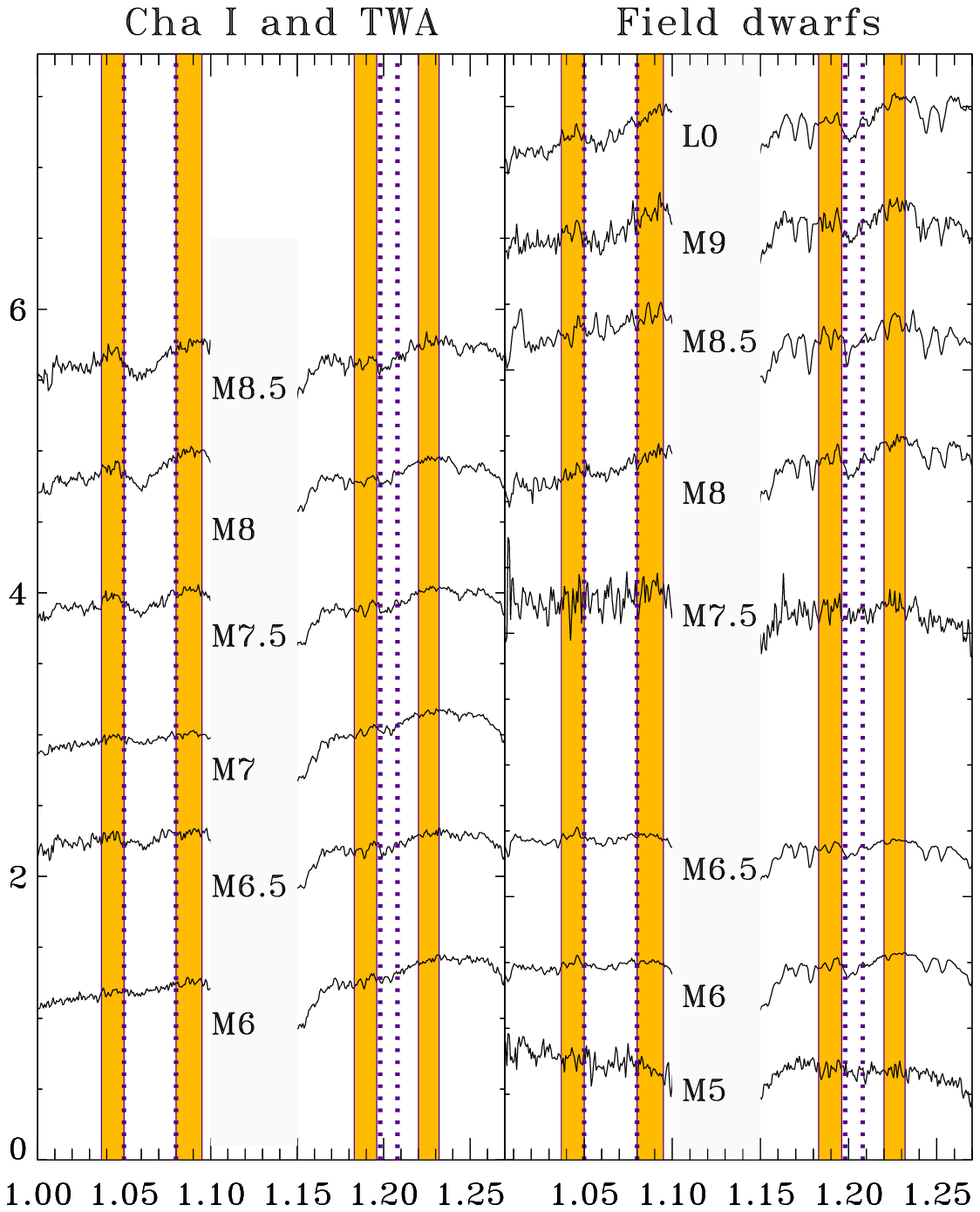}}
\caption{Close-in view of the VO absorption bands at 1.05 and 1.2\micron for representative spectral type of sources in the two samples (Cha I and TWA members in the left panel and field dwarfs in the right panel). We have shaded in yellow the areas used to define the pseudocontinuum set as reference to measure the depth of the bands.}
\label{VOdefinition}
\end{figure}

Figure~\ref{VOdefinition} shows a cut of several spectra in the young and old (left and right panel, respectively) samples where these absorption bands are highlighted with vertical dashed lines confining them. Qualitatively, this figure shows that while the depth of $VO1$ scales with spectral type among the young sources, the same band seems to be insensitive to temperature for the field dwarfs. On the contrary $VO2$ shows the opposite trend being more sensitive to changes in temperature among the older sample. We have characterized the depths of these bands by estimating the ratio of the measured density flux in the absorption band with respect to the expected flux from a linear fit to a pseudo-continuum defined with two adjacent bands. Table~\ref{indexparam} summarizes the wavelength ranges used to define each one of the bands: ``in-band region", and references one and two for the pseudocontinuum determination. 

\begin{table}
\begin{center}
\caption{Spectral ranges used in the definition of the indexes characterizing the $VO1$ and $VO2$ absorption bands.} 
\label{indexparam}
\begin{small}
\begin{tabular}{lccc}
 &  ``In-band" range& Reference 1 range& Reference 2 range  \\
 & (\micron) &(\micron) & (\micron)\\
\hline
\hline
$VO1$ & [1.05, 1.08] & [1.037, 1.05] & [1.08, 1.095] \\
$VO2$ & [1.198, 1.208] & [1.183, 1.196] & [1.22, 1.232] \\
\hline
\hline
\end{tabular}
\end{small}
\end{center}
\end{table}

\begin{figure}
\resizebox{\hsize}{!}{\includegraphics{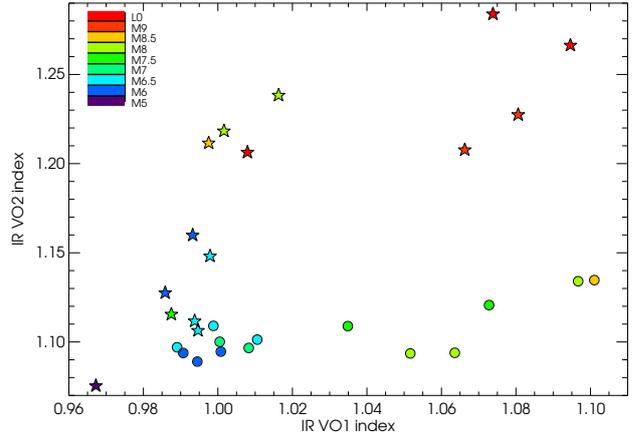}}
\caption{$VO1$ vs $VO2$ indexes for the sample of young (Cha I and TWA members, filled circles) and old (field dwarfs, filled stars) sources. All symbols are color coded according to the spectral type of the object (see legend).}
\label{VOindex}
\end{figure}

Figure~\ref{VOindex} shows that in the ($VO1$, $VO2$) index space, field dwarfs and members of Cha I and TWA, define distinct sequences, and therefore these indexes can be used to discriminate between old and young sources and, in  the case of young sources to estimate spectral types with 1 subspectral type uncertainty.

\section{Dust settling: effective temperature and interstellar extinction via model fitting including different dust treatment}
\label{temperatures}

To determine the effective temperatures of the Cha I and TWA members (and interstellar extinction for the former) we have followed a two step approach: first we have gathered multi-wavelength photometry and performed SED fits with VOSA \citep{Bayo08, Bayo14}, and then we have used the results from VOSA ($\chi^2$ minimization and posterior probability functions for the parameters) to define a finer grid of models to perform the fit directly to the available spectroscopy in each case (optical + NIR for the Cha I members, and NIR alone for the TWA sources).

\subsection[]{SED model fitting}
\label{sec:SEDfit}

With the starting point of the optical photometry from \cite{Comeron00} (for the ChaI sources) and the estimated Spitzer IRAC and MIPS photometry given in Table~\ref{sources:IRAC}, we used VOSA to further populate the SEDs of our sources. The complete list of catalogs accessible through VOSA is provided in \cite{Bayo13}, and in this case, counterparts to all sources were found in the DENIS (\citealt{DENIS}, although some of the measurements exhibit large error-bars caused by clouds passing by during the observation according to the quality flags of the catalog), 2MASS \citep{2MASS} and WISE \cite{WISE} catalogs.

As shown in Fig.~\ref{SEDfit}, two sets of MIR photometry are available for all our sources apart from ChaH$\alpha$11, with MIR photometry (both for Spitzer and WISE) critically affected by a saturated nearby star, and SSSPMJ1102, for which no IRAC photometry is available. Given the similarities\footnote{http://wise2.ipac.caltech.edu/docs/release/prelim/expsup/figures/sec4\_3gf4b.gif} between the first two channels (I1 and I2) of Spitzer/IRAC and those of WISE (W1 and W2), we compared both measurements looking for signs of MIR variability. Since the photometric systems and filters are similar but not directly comparable, we used a simple RayleighÐJeans approximation and the SVO filter service\footnote{http://svo2.cab.inta-csic.es/theory/fps/index.php} to estimate the expected W1 and W2 fluxes from the measured I1 and I2. Obviously this approximation only makes sense for objects that do not show excess at these wavelengths and therefore we only performed the exercise on the photometry of sources for which VOSA detects no excess at the given wavelength ($\sim$3.6 and $\sim$4.5 $\mu$m). In Table~\ref{MIRvar} we show these comparisons and we can conclude that there is no significant variations between the two epochs of observations for any of the sources.

\input{MIR_var.tex}

Besides the initial photometry, we also provided VOSA with the distance to the sources (needed for VOSA to estimate the panchromatic bolometric correction) and a range of A$_{\rm V}$ to be probed in the fitting process. As mentioned in the Introduction, Cha I is located at $160\pm15$ pc and TWA at $50\pm10$pc. We used these values for all our sources but for 2MJ1207 for which the more precise value of $54\pm3$ pc from \cite{Faherty09} was available.

Regarding interstellar extinction, a 2MASS star count extinction map of the Camaeleon I cloud (constructed in similar manner than \citealt{Cambresy97} and presented in \cite{Lopez13b}) suggest maximum values of A$_{\rm V}$ below 11.7 mag for the lines of sights of all the Cha I members under study. Thus the probed range for extinction by VOSA for Cha I members is [0.0, 11.7] in a 0.585 mag step (where the step is automatically determined by VOSA). 

Furthermore, the relative proximity and the surprising (given the youth of the members of the association) absence of significant interstellar or intramolecular cloud extinction associated with the TW Hya Association (see \citealt{Tachihara09} for a detailed search of remnant clouds) allow us to fix the Av to 0 magnitudes in the fit of the TWA brown dwarfs

Once the SEDs were compiled, and given the late-M nature of the sources, from the list of models available in VOSA the best choice is the BT-Settl collection \cite{Allard12} that correspond to an integral treatment of the dust in the atmosphere of cool objects. This integral treatment in principle obsoletes the limiting cases of the COND (total gravitational settling assumed and due to the condensation of species involving Ti, V, Ca and Fe, the molecular opacity sources disappears) and DUSTY (inefficient gravitational settling assumed, meaning that the dust is distributed according to the chemical equilibrium predictions) previous approached \citep{Allard03, Allard12}. In addition, the BT-Settl models do not enforce grains to be in equilibrium with the gas phase (as is the case of the DUSTY and COND models), and so, the gas phase opacities reflect the depletion of elements from the gas phase caused by grain growth. To illustrate this improvement we carried out parallel fits with the three dust treatments, not only in the SED fit but in the spectral fittings and the propagation to the HR diagram.

Concerning the additional (besides interstellar extinction) parameters space explored for each collection of models, we imposed no constrain on the effective temperature, and since we are using broad band photometry for the fit, we allowed the $\log(g)$ to vary between 3.5 and 4.5 (typical range for young cool VLMs and BDs) but we refer to Section~\ref{specfit} for the spectroscopic determination of the surface gravity. Solar metallicity was fixed which should be a good approximation given our grid step in metallicity combined with the values of the sightly sub--solar metallicity members of Cha I studied by \cite{Spina14}.

\subsubsection{Comparison of the fitting approaches}

VOSA follows two approaches for the SED fit \citep{Bayo08, Bayo13}: on the one hand a minimization of the squared differences ($\chi^2$ fit) with synthetic SEDs (calculated from the grids of synthetic spectra for the same instrument/filter configurations as the observations) and on the other hand a Bayesian statistical analysis following \cite{Kauffmann03} that provides as output the projected probability distribution functions (PDFs) for each parameter of the grid of synthetic spectra.

\begin{figure*}
\resizebox{\hsize}{!}{\includegraphics{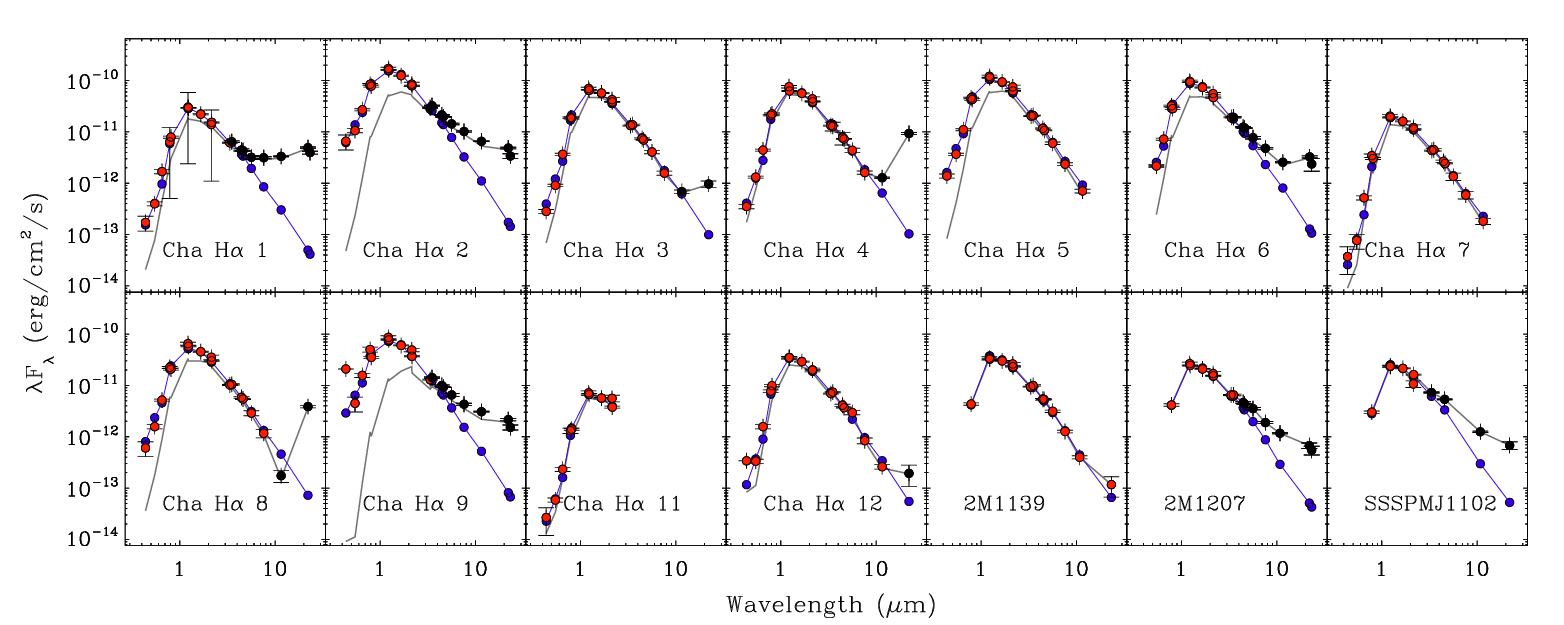}}
\caption{Compiled SEDs for all the sources analyzed in this work. In most cases the wavelength coverage includes observations from the blue optical domain to the mid-infrared (see Table~\ref{sources:IRAC} for the IRAC and MIPS). The light-grey solid line corresponds to the ``raw" (non-extinction corrected) photometry, red and black filled circles show the observed photometry corrected from the best fitting value of the extinction. Finally, the blue filled circles connected by a solid line of the same color displays the best fitting model (in terms of $\chi^2$ minimization) from the BT-Settl collection. The fittings do not consider the regions where some of the sources show infrared excesses. Errors bars are plotted on top of the extinction-corrected photometry.}
\label{SEDfit}
\end{figure*}

\begin{figure*}
\includegraphics[width=8.cm]{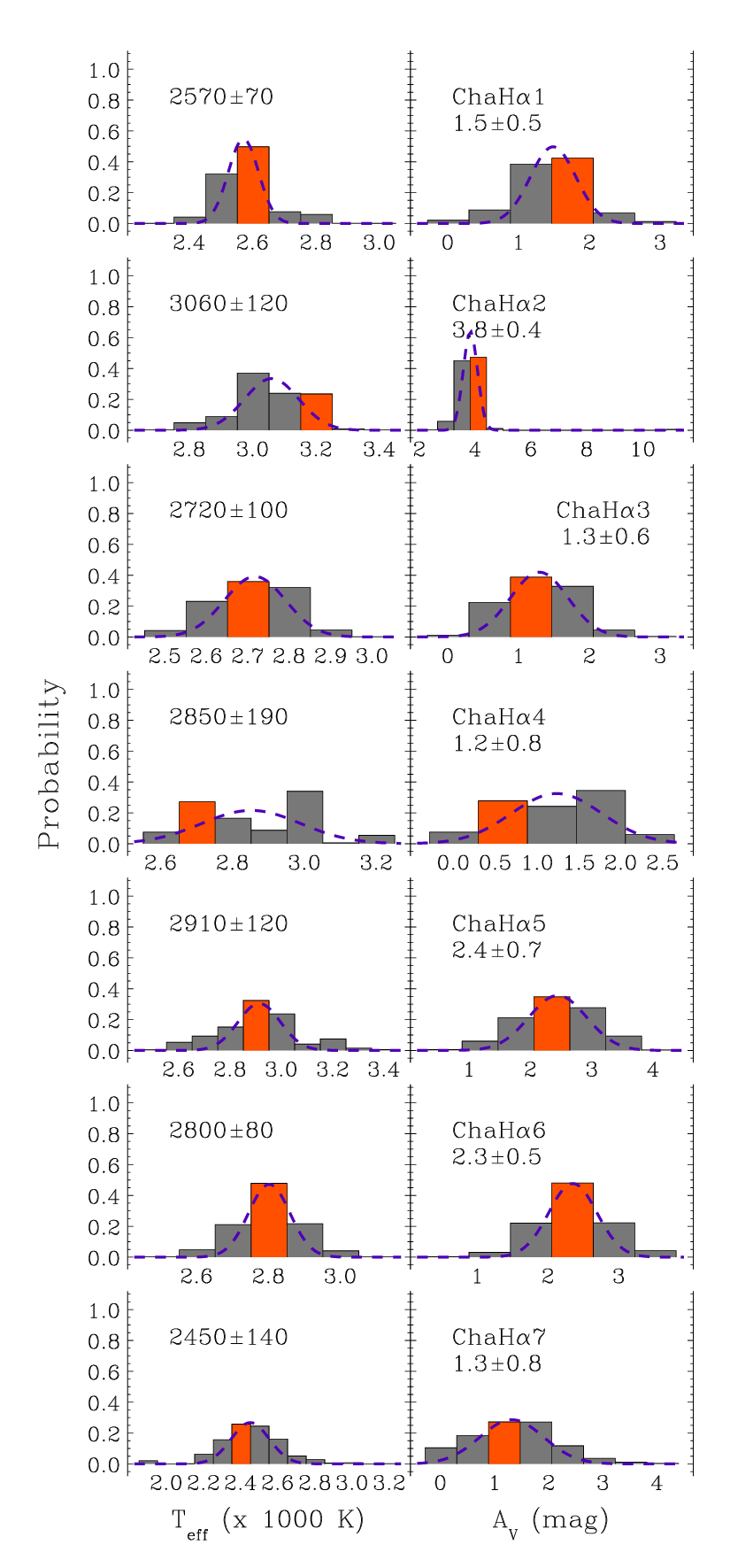}
\includegraphics[width=8.cm]{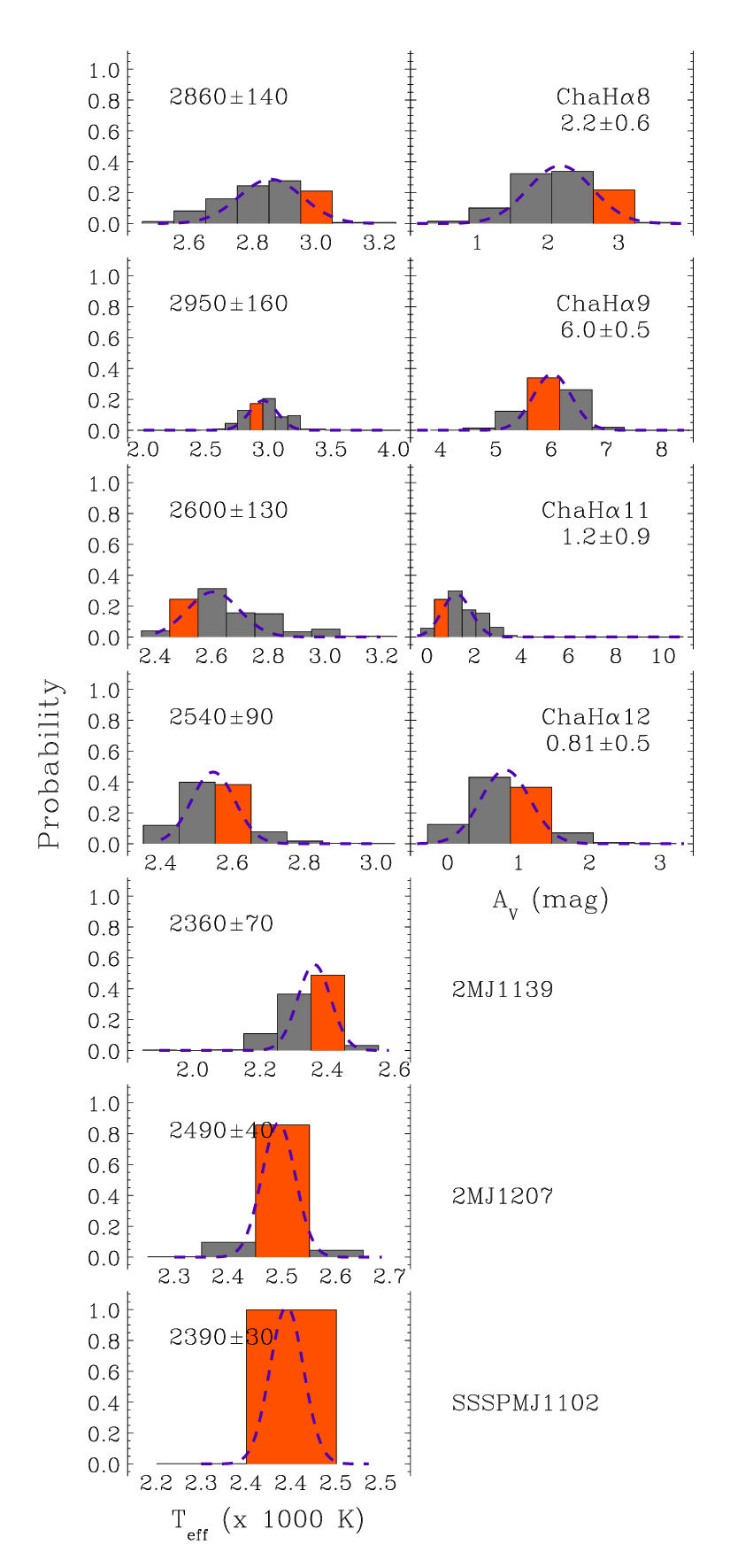}
\caption{Gaussian fits (in purple) to the posterior probability functions (in grey) for T$_{\rm eff}$ and A$_{\rm V}$ provided by VOSA for the BT-Settl collection of models. The bin corresponding to the best fits in terms of $\chi^2$ are highlighted in orange and the basic parameters of the gaussian fit (mean and $\sigma$) are provided as plot-labels.}
\label{Bayesgaussfit}
\end{figure*}

In Table~\ref{chahachi} we provide a summary the results obtained with these two fitting approaches for the three dust treatment scenarios: whenever the Bayes and $\chi^2$ determined parameters do not agree we have included both determinations in the table (separated by ``/"). In addition, in Fig.~\ref{SEDfit} and Fig.~\ref{Bayesgaussfit}, we display examples of the corresponding graphical outputs for the BT-Settl collection of models (providing the best results as we will discuss further on). Respectively, Fig.~\ref{SEDfit}  shows the observed SEDs (``raw" data with a grey line and interstellar-extinction-corrected photometry with red and back filled circles) along with the best fitting model in the sense of $\chi^2$ minimization (blue filled circles joined with a solid blue line), and Fig.~\ref{Bayesgaussfit} illustrates the Bayesian analysis for the most relevant free parameters (T$_{\rm eff}$ and A$_{\rm V}$). In the latter figure we highlight (in orange) the parameter estimations from the $\chi^2$ minimization.

\input{SED_fit_param.tex}

\begin{figure}
\resizebox{\hsize}{!}{\includegraphics{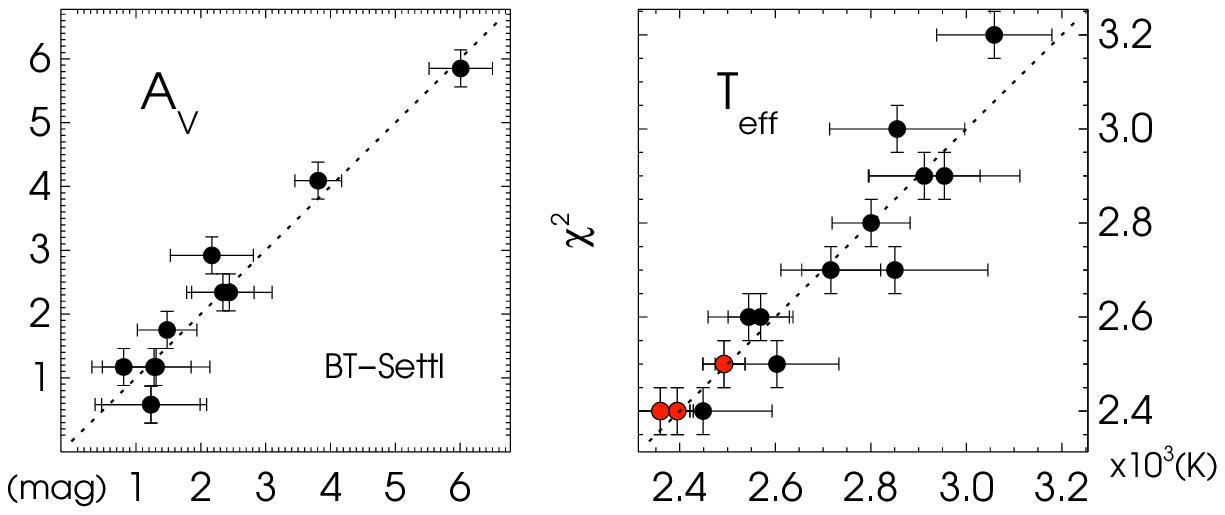}}
\resizebox{\hsize}{!}{\includegraphics{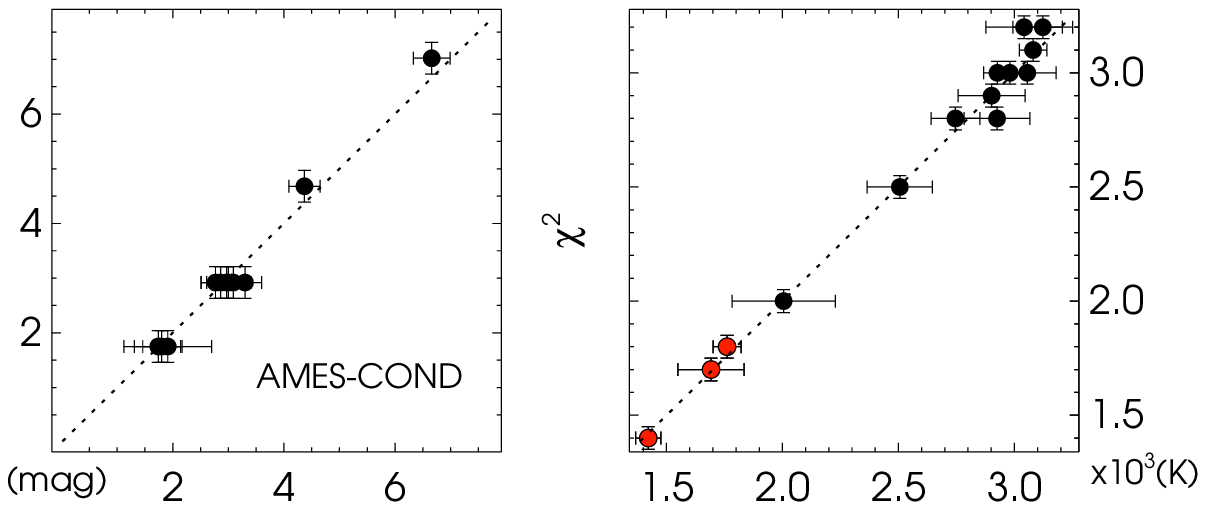}}
\resizebox{\hsize}{!}{\includegraphics{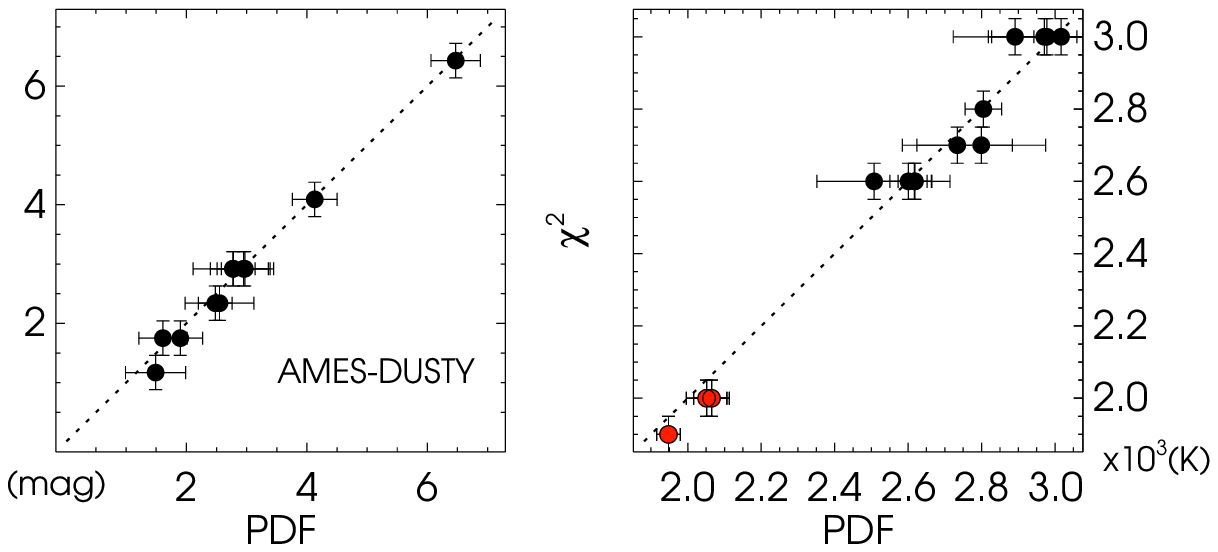}}
\caption{Comparison of the estimated A$_{\rm V}$ and T$_{\rm eff}$ via minimization of the squared differences and gaussian fit to the posterior distribution functions (see for an example Fig.~\ref{Bayesgaussfit}) for the three dust treatments. Both estimations agree within the error-bars calculated as half the parameter step in the grid of models and the $\sigma$ of the gaussian fit to the PDF.}
\label{paramComparBayesChi}
\end{figure}

To further confirm/quantify the suggestion from Table~\ref{chahachi} that, regardless of the chosen collection of models, the two statistical approaches followed by VOSA provide consistent estimations of the reddening and effective temperatures, we show a direct comparison in Fig.~\ref{paramComparBayesChi} of the two estimations. For this comparison, we assume that the uncertainty in the $\chi^2$ determined parameters correspond to half the step of the grid for the given parameter ( i.e. 50 K, and 0.585 mag, for T$_{\rm eff}$ and A$_{\rm V}$, respectively) and for the Bayes approach we do simple gaussian fitting to the PDFs (see Fig.~\ref{Bayesgaussfit} for an example) and assume 1$\sigma$ uncertainties. 

From this comparison we can conclude that both approaches agree within the uncertainties. We do observe however that the Bayes approach is very beneficial in identifying objects for which the range of ``most probable" parameters values are larger than the grid steps (in fact this is one of the motivations of introduccing this new approach in VOSA) or whenever there is degeneration in the fit. A clear example of the latter is ChaH$\alpha$4 where the analysis of the PDFs for A$_{\rm V}$ and T$_{\rm eff}$ clearly suggest a strong degeneracy between both parameters.

\subsubsection{Implications of the dust treatment}

Given this good agreement among the statistical approaches and the comment from the previous section, we will use the Bayes approach results (characterized via the simple gaussian fits) for the comparisons between the different flavors of dust treatment.

\begin{figure*}
\includegraphics[width=12.85cm]{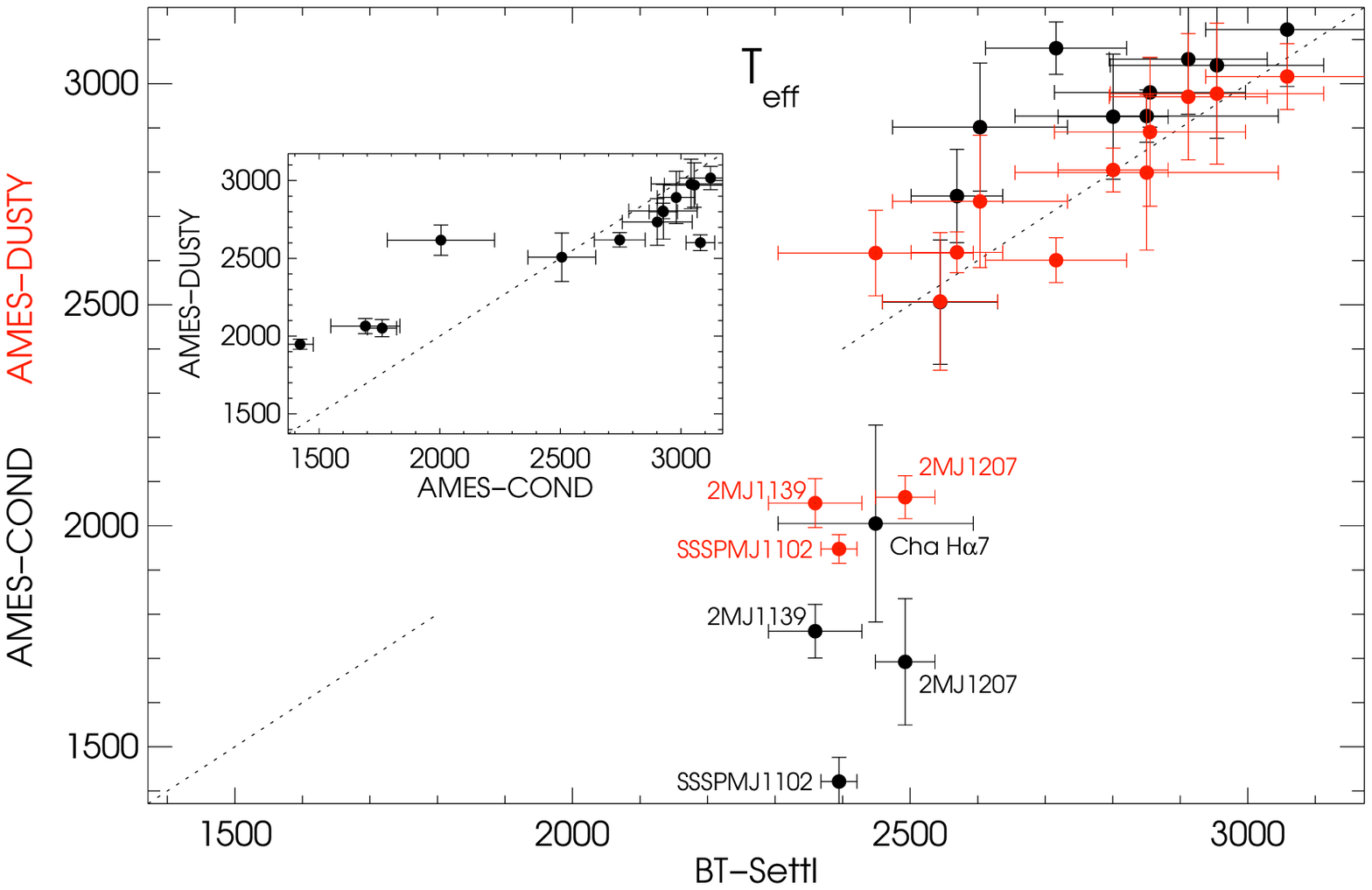}
\includegraphics[width=12.85cm]{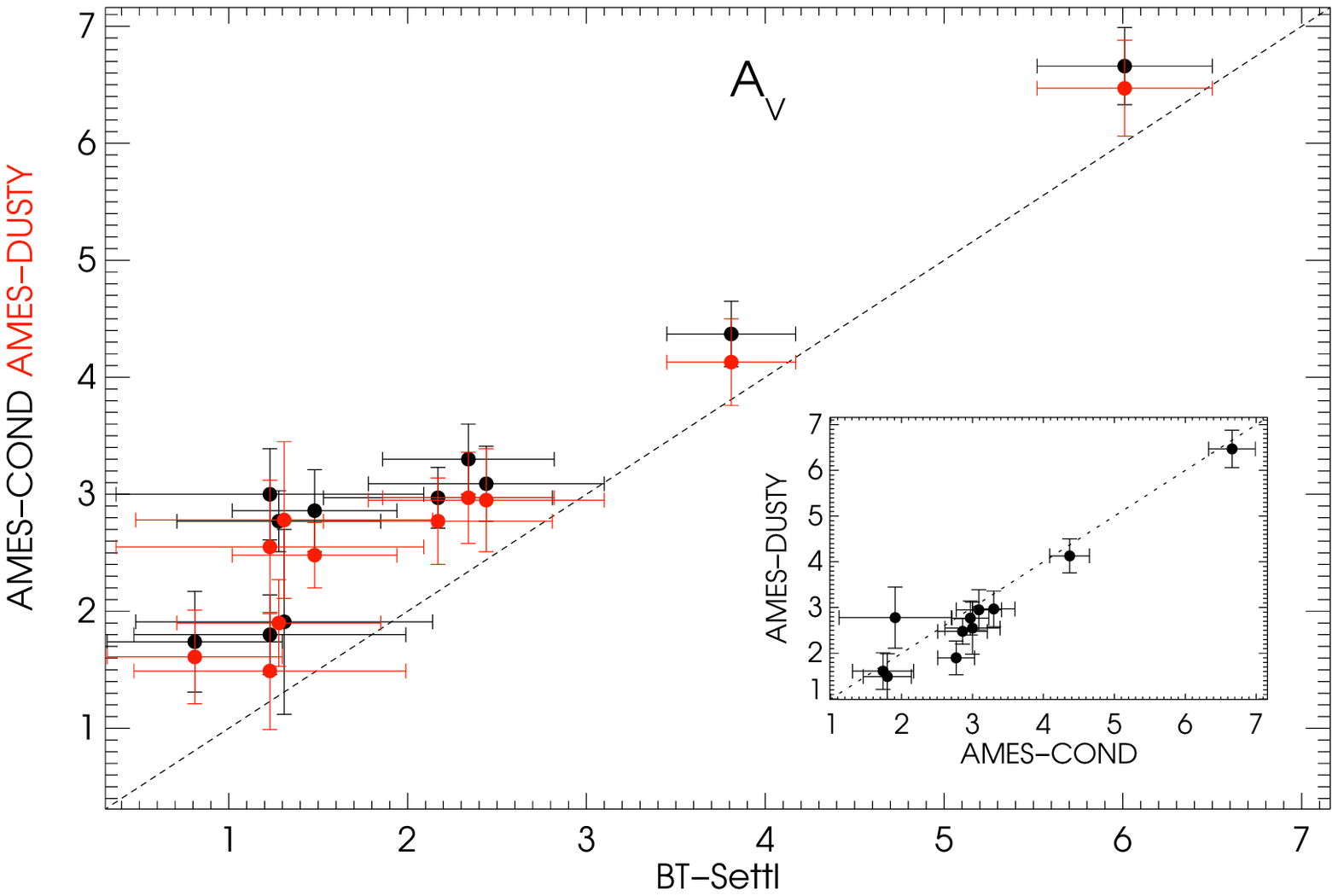}
\caption{Comparison of the estimated A$_{\rm V}$ (bottom panel) and T$_{\rm eff}$ (top panel) via gaussian fit to the posterior distribution functions (see Fig.~\ref{paramComparBayesChi} and text for details) for the three different dust treatments. For both parameters under study In the large panels we show the values obtained with the BT-Settl collection vs AMES-COND (in blak) and AMES-DUSTY (in red) and in the smaller panels we display the comparison AMES-COND vs AMES-DUSTY.}
\label{paramComparDustTreat}
\end{figure*}

In Fig.~\ref{paramComparDustTreat} we show the comparison of the values of T$_{\rm eff}$ and A$_{\rm V}$ obtained for each one of the three dust treatments / model collections. Two main features can be extracted from these comparisons: a systematic $\sim$0.5 step in magnitude between the A$_{\rm V}$ estimations obtained from the AMES-COND and DUSTY collections (higher) from those obtained with he BT-Settl collection and a more dramatic feature consistent in a very significant underestimation of the effective temperature (that translate in underestimated masses) with the COND and DUSTY collections versus the values estimated with the BT-Settl integral dust treatment for temperatures below 2600 K.

We will not address in detail the differences in A$_{\rm V}$ estimations because although the ones obtained with BT-Settl are systematically lower than the other two, still the difference is within the magnitude step considered by VOSA (0.585 magnitudes).

Besides, the dramatic feature observed for estimated temperatures below 2600 K corresponds to the expected behavior given the dust treatment assume for each collection: both, total gravitational settling and inefficient gravitational settling result in SEDs bluer than those obtained with the integral treatment of dust, and therefore the best matching SEDs to the observations in the first two cases correspond to much colder temperatures than in the latter. In Section~\ref{specfit} we will see how these very cold temperatures obtained with AMES-COND and AMES-DUSTY that may reproduce the SEDs to a certain degree, do not reproduce the main spectral features of the sources.

\subsubsection{Comparison of the BT-Settl results with the literature}

Besides the SED fitting process, VOSA uses the best fitting model (in terms of minimuum $\chi^2$) to estimate a panchromatic bolometric correction to provide the bolometric luminosity of the object taking into account the distance (and uncertainty) provided by the user (or obtained from VO services). With the estimated effective temperatures and bolometric luminosities (and their respective uncertainties) VOSA interpolates in the VO compatible isochrones and evolutionary tracks available: those from \cite{Baraffe03} for the COND approximation, \cite{Chabrier00} for the DUSTY one, and (until state of the art isochrones are released), a combination of \cite{Baraffe98,Baraffe03} for the BT-Settl dust treatment, providing individual estimations of ages and masses (see Table~\ref{chahachi}).

Since the AMES-COND and AMES-DUSTY results yield extremely low temperatures but surprisingly high luminosities, for most of those objects, their location in the HR diagram place them well above the 1 Myr isochrone and therefore no estimation of the mass can be provided by VOSA. 

\begin{figure}
\resizebox{\hsize}{!}{\includegraphics{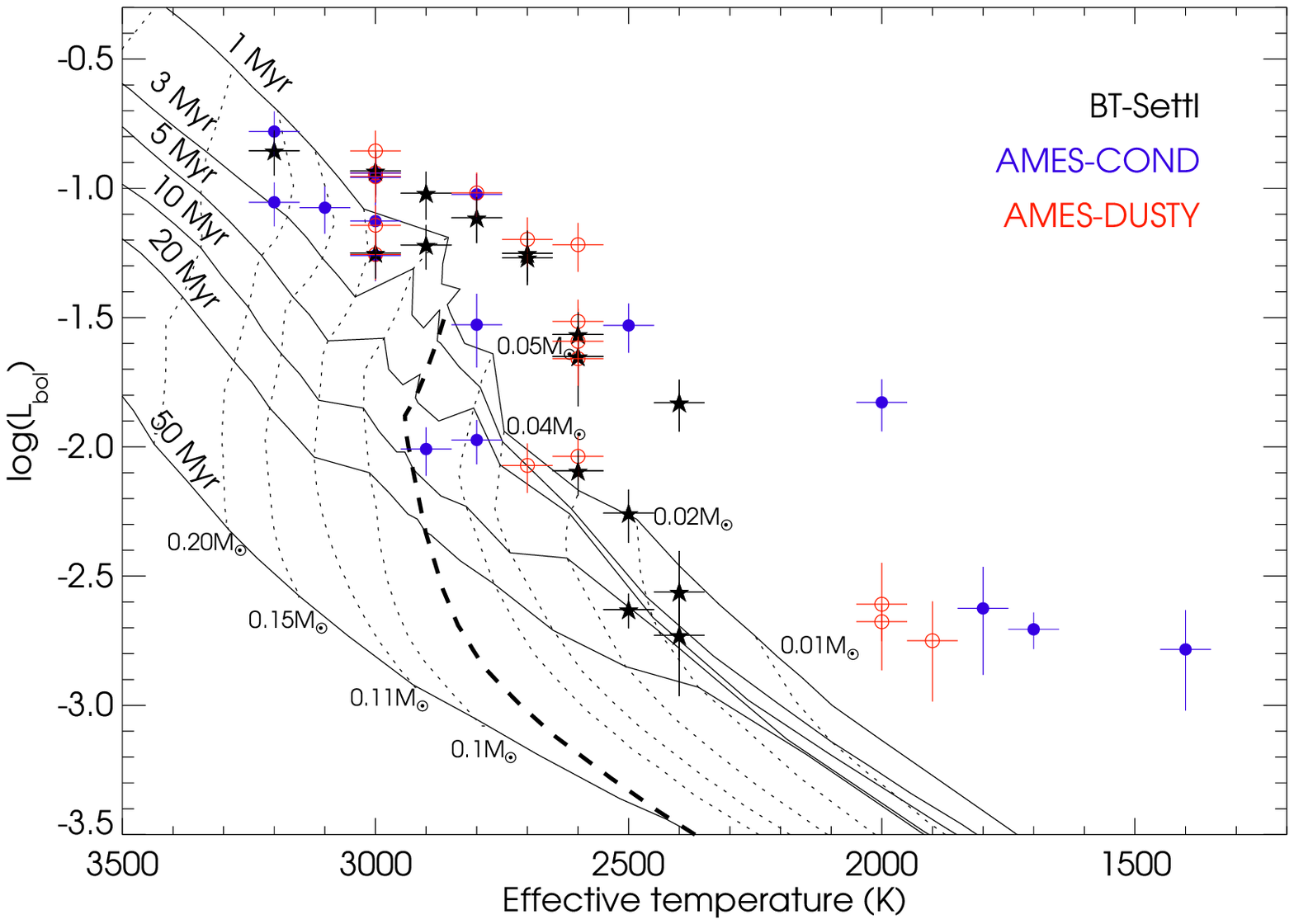}}
\caption{HR diagram with isochrones and evolutionary tracks for the BT-Settl models\protect\footnotemark \protect\citep{Allard12}
 where we display the results obtained for each approach to the dust treatment: black five-point-filled stars for BT-Settl, red open circles for AMS-DUSTY and blue filled circles for AMES-COND. The substellar boundary is highlighted with a thick dashed line.}
\label{HRSEDs}
\end{figure}

\begin{figure}
\resizebox{\hsize}{!}{\includegraphics{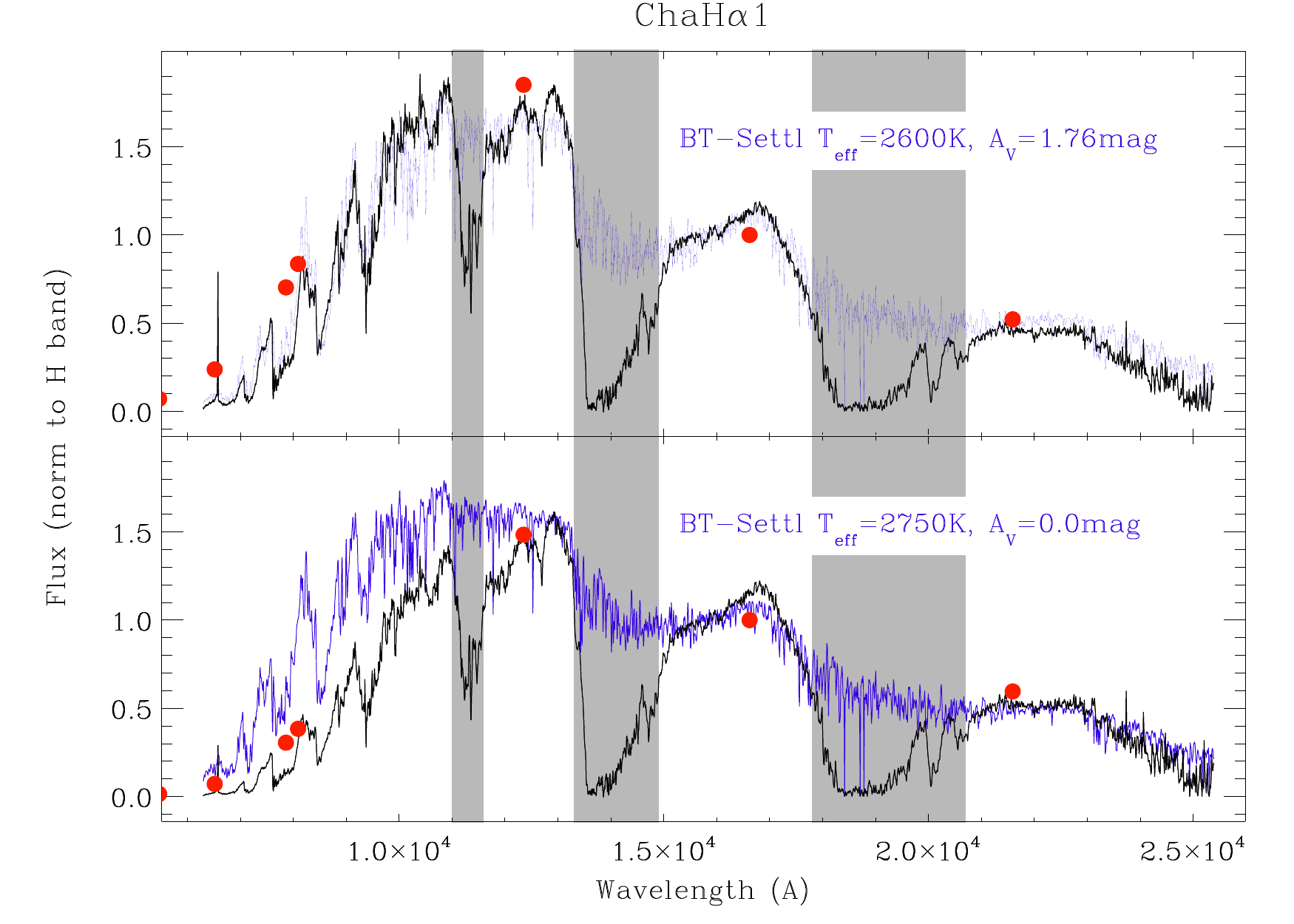}}
\resizebox{\hsize}{!}{\includegraphics{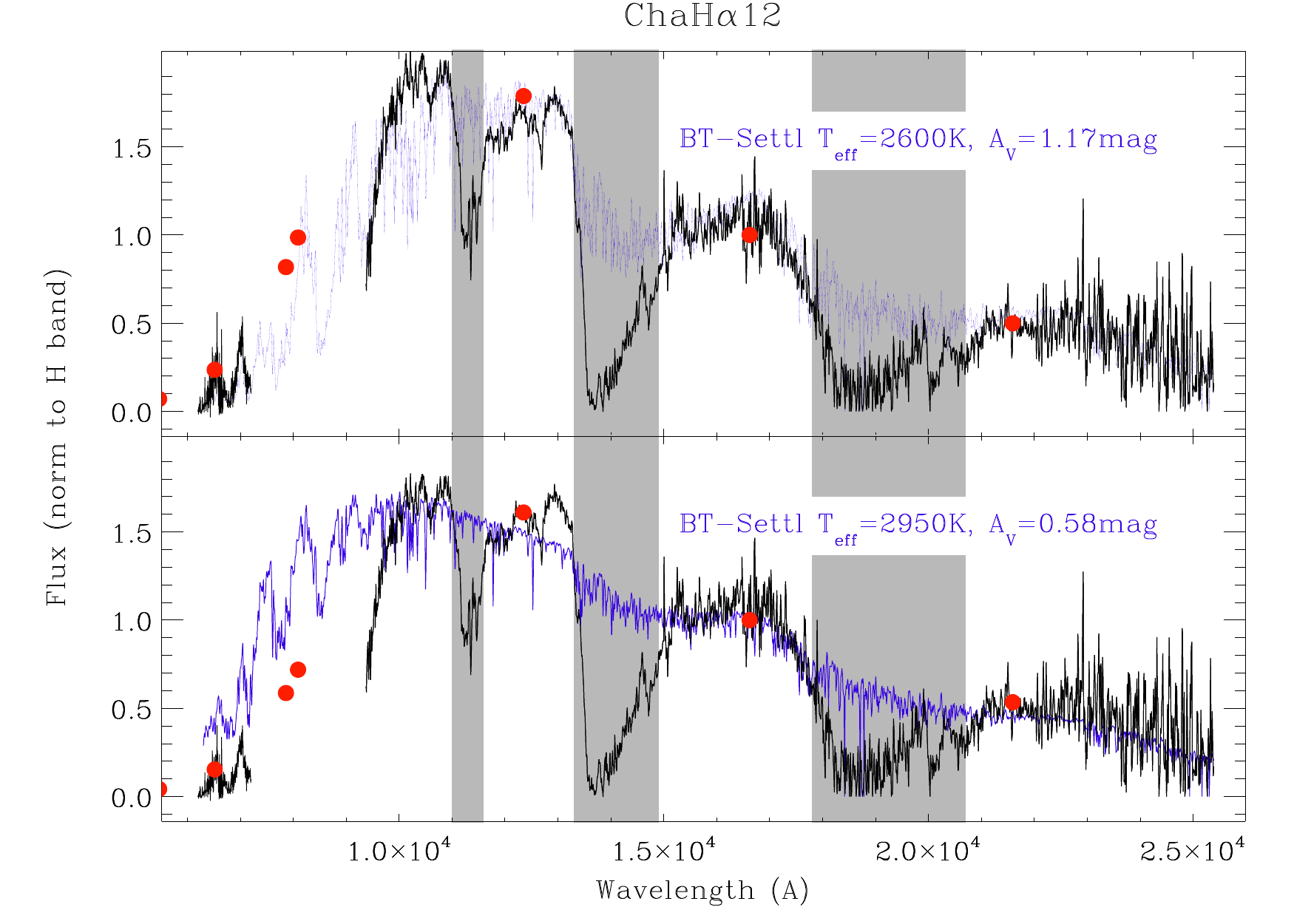}}
\caption{Comparison of the best fitting parameters determined with VOSA (upper panels) and those provided in Luhman et al. (2007) (lower panels) for ChaH$\alpha$1 and ChaH$\alpha$12. In all cases the observed spectra are displayed in black (optical + near-infrared), de-reddened with the corresponding extinction value and normalized to the flux in the H-band. The SED composed and fitted by VOSA (also normalized to the H-band and de-redenned) is displayed with red dots and the synthetic spectra with the corresponding temperatures are overplotted in blue. The wavelength ranges with low atmospheric transmission, where the observed spectra cannot be properly corrected, are shaded in grey.}
\label{comparaLuhman}
\end{figure}

On the other hand, if we focus on the results obtained with the BT-Settl collection, the parameters estimated in this manner for the TWA objects are in good agreement with the literature \citep{Gizis02}, but the effective temperatures estimated by VOSA for the Cha I members are systematically lower (same applies to the masses) than those provided by \cite{Luhman07b}. In the next subsection we use our optical and near-infrared spectra to refine these determinations, but in Fig.~\ref{comparaLuhman} (where no further fitting procedure than the SED fit one had been carried out), we show how at least for some cases, models corresponding to the BT-Settl-VOSA-estimated (T$_{\rm eff}$, A$_{\rm V}$) pairs (upper sub-panels in the two cases) reproduce better the observe spectra (in black) than those using the estimations from \cite{Luhman07b}. 

This fact has already been pointed out in \cite{Rajpurohit13}, where the authors provide evidence that the differences with previous studies arise from the use of the NextGen limiting case collection (atmosphere that do not include dust settling, as, for example, in \citealt{Luhman07b}) in the fits. In fact, from the comparison shown in Fig.~\ref{Teffscale} (where we use the refined effective temperatures obtained from the fit to the optical + near infrared spectra, see next section for details) it is obvious that our estimations show a similar trend than those from \cite{Rajpurohit13}, obtained also for the BT-Settl collection, but via comparison of high-resolution optical spectroscopy of a sample of M dwarfs with the models.

\footnotetext{http://phoenix.ens-lyon.fr/Grids/BT-Settl/}
\subsection[]{Spectra model fitting}
\label{specfit}

From the SED fits we have shown that the BT-Settl models tend to reproduce better the observations especially for temperatures below 2600K, where the dust treatment becomes critical. To better constrain these differences and test the ability of the models to reproduce the spectral features of young M dwarfs in the optical and NIR, we have performed a direct fit to the spectroscopic observations described in Section 2.2 and 2.3.

We have first prepared a grid of synthetic spectra probing the parameter space around 5$\sigma$ of the best fitting values obtained from the VOSA fit for each object (in terms of effective temperature and extinction, and probing $\log(g)$ valus between 3.0 and 5.0 dex). We then degraded the resolution of the original models to match our observations. We did not have to consider the rotational widening of the lines since, given the low resolution of our observations, the instrumental broadening is the dominant effect for our observations. Finally we performed a linear interpolation to decrease the step in effective temperature to 50 K.

For the fitting process to the previously describe grid we considered three cases: fitting simultaneously the whole range of observations (in most cases, the Chamaeleon sources, this means optical and NIR spectra), or fitting independently the optical spectra and the NIR spectra. Whenever relevant, for the comparison with models, we masked out areas of the spectra affected by effects present in the observed data but not in the theoretical models (basically, strong telluric absorptions and the H$\alpha$ emission). The results obtained for the three models and three approaches are displayed in Figures~\ref{bestfitspec} and~\ref{bestfitspec2} and summarized in Table~\ref{specfitparam}. 

\begin{figure*}
\includegraphics[width=5.5cm]{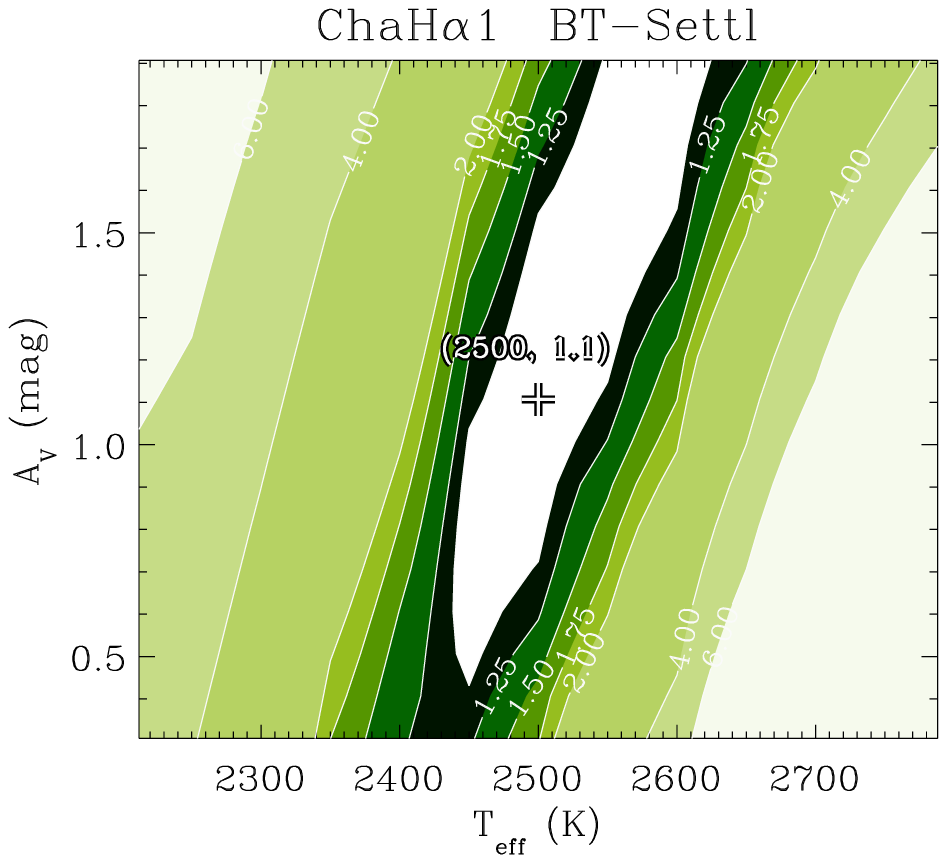}
\includegraphics[width=5.5cm]{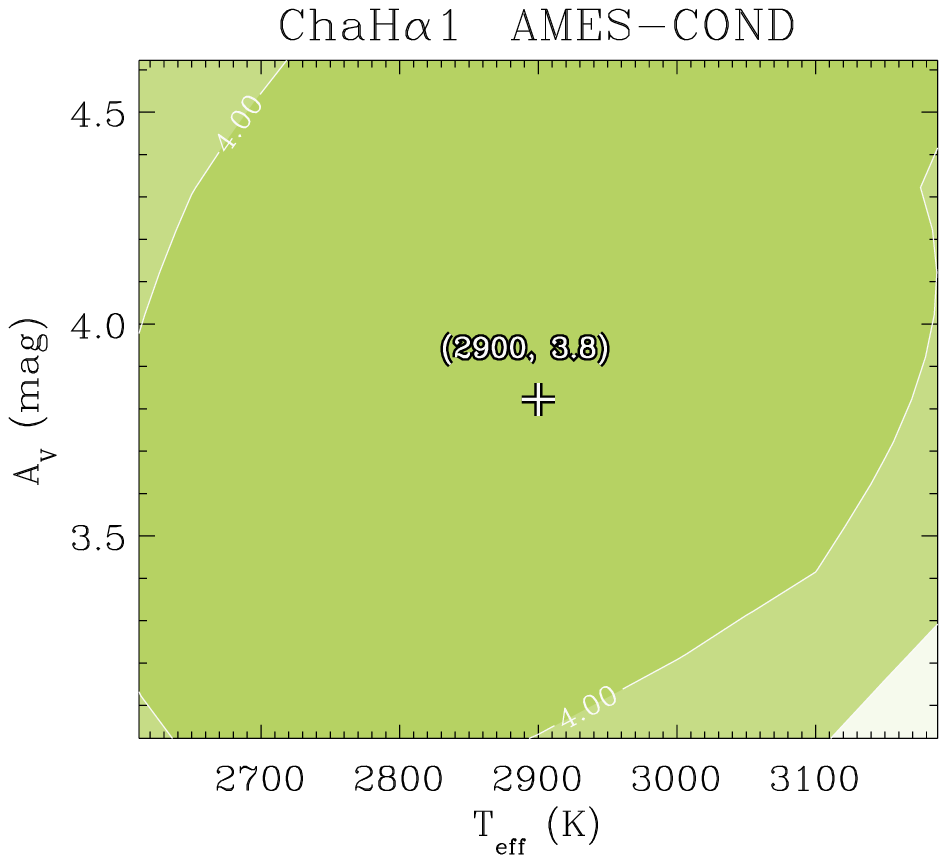}
\includegraphics[width=5.5cm]{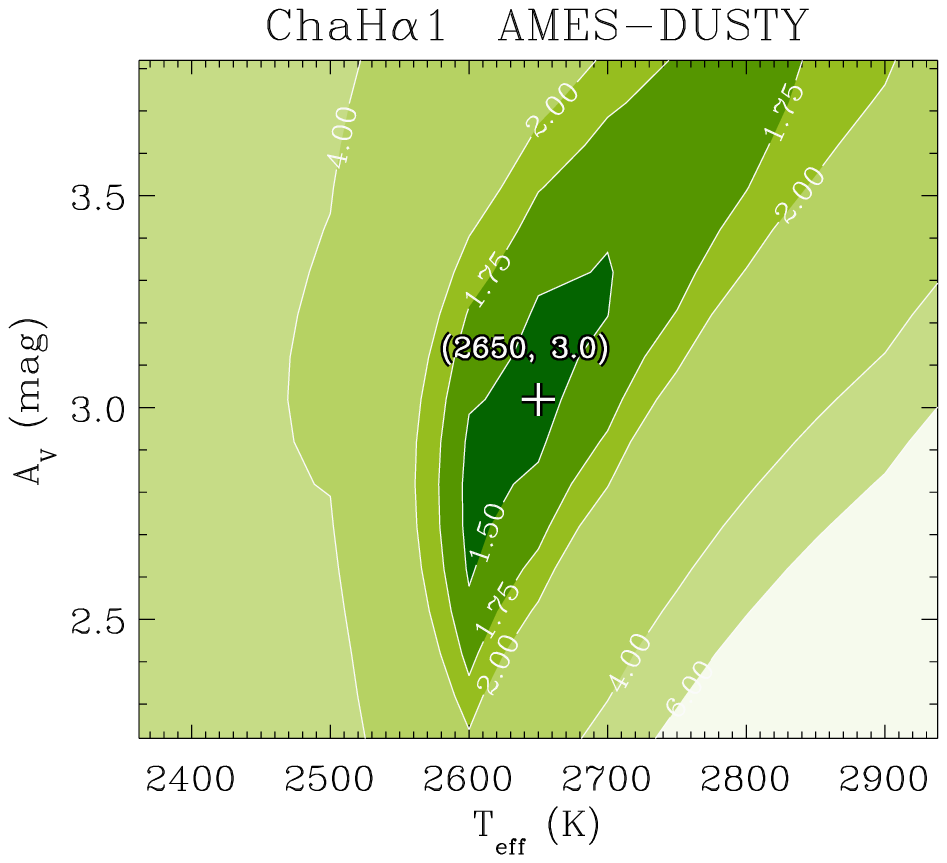}
\includegraphics[width=5.5cm]{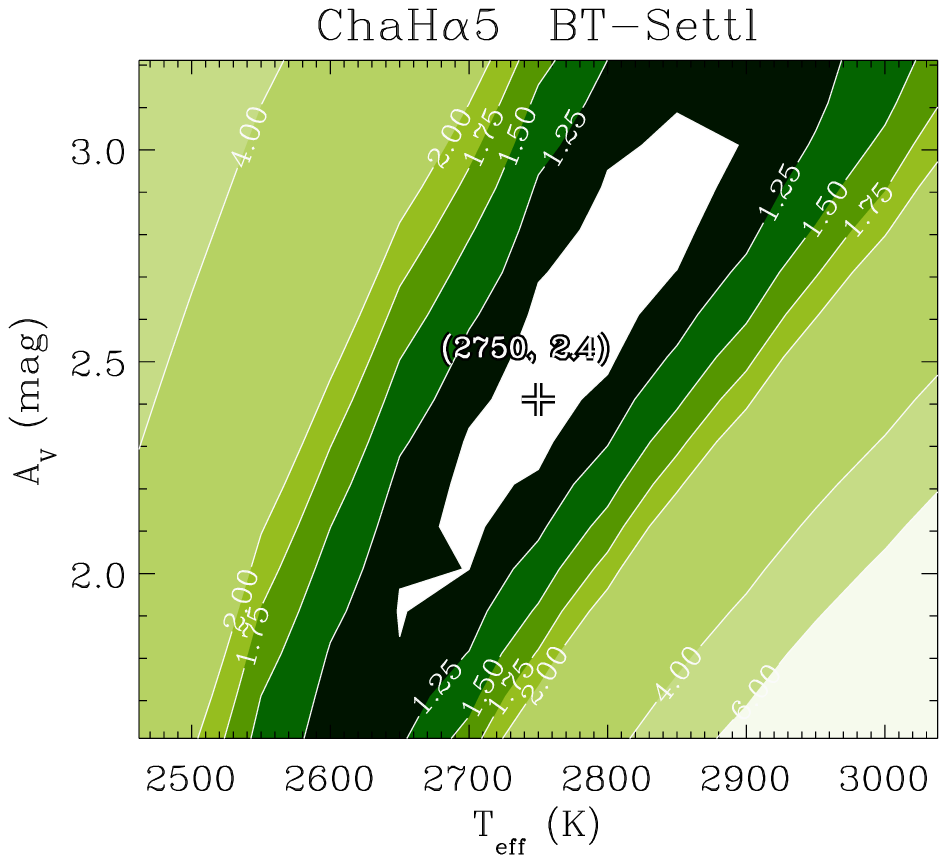}
\includegraphics[width=5.5cm]{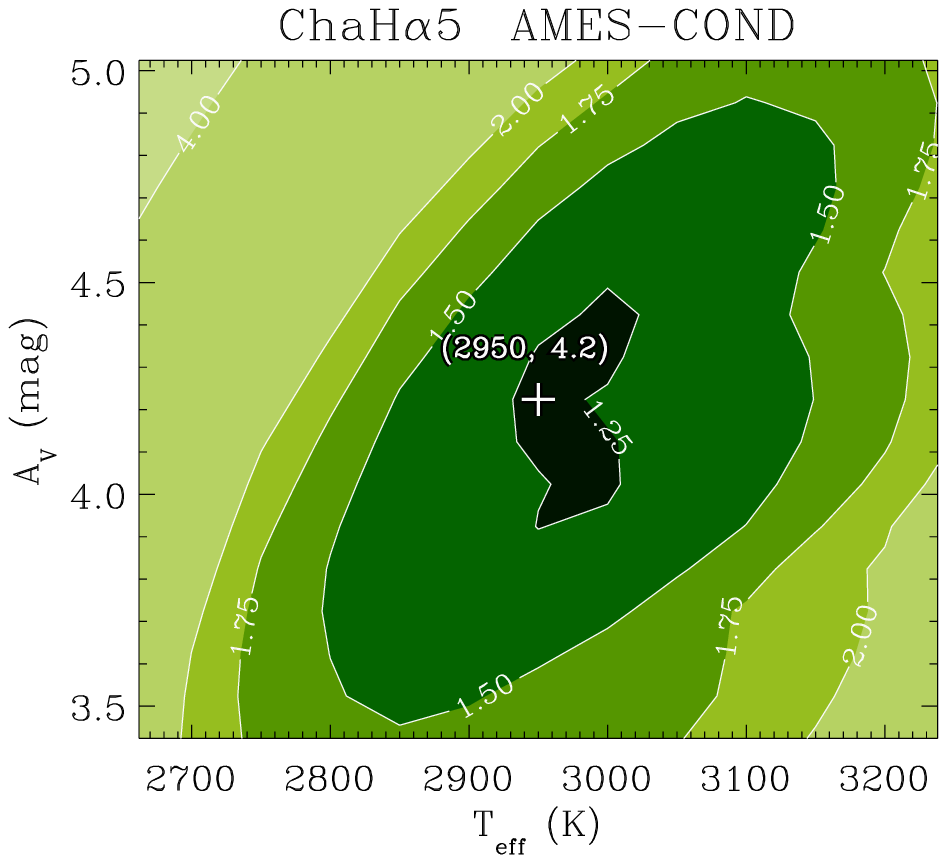}
\includegraphics[width=5.5cm]{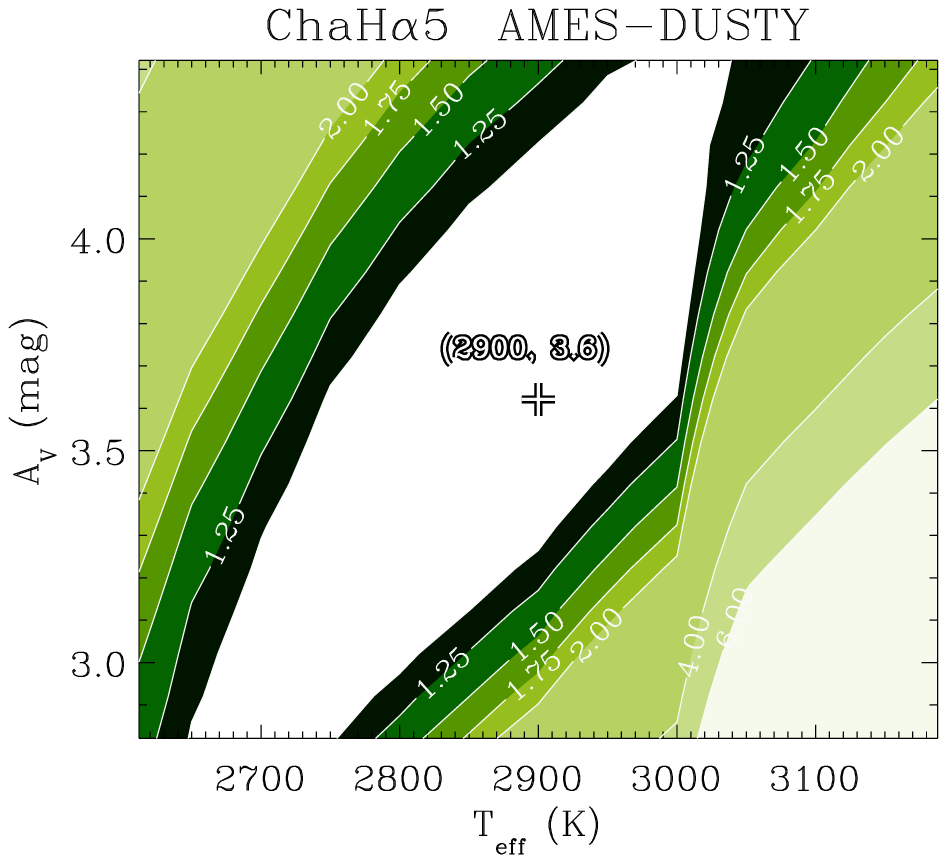}
\caption{Individual slices for constant $\log(g)$ values of 4.0 dex for the $\chi^2$ cubes obtained from the spectral fits. We just show two objects as examples of the variations in the $\chi^2$ maps that we obtain depending on the dust treatment. The white area highlights the parameter space are with $\dot{\chi^2}\le 1$ }
\label{chislices}
\end{figure*}

\begin{figure*}
\resizebox{\hsize}{!}{\includegraphics{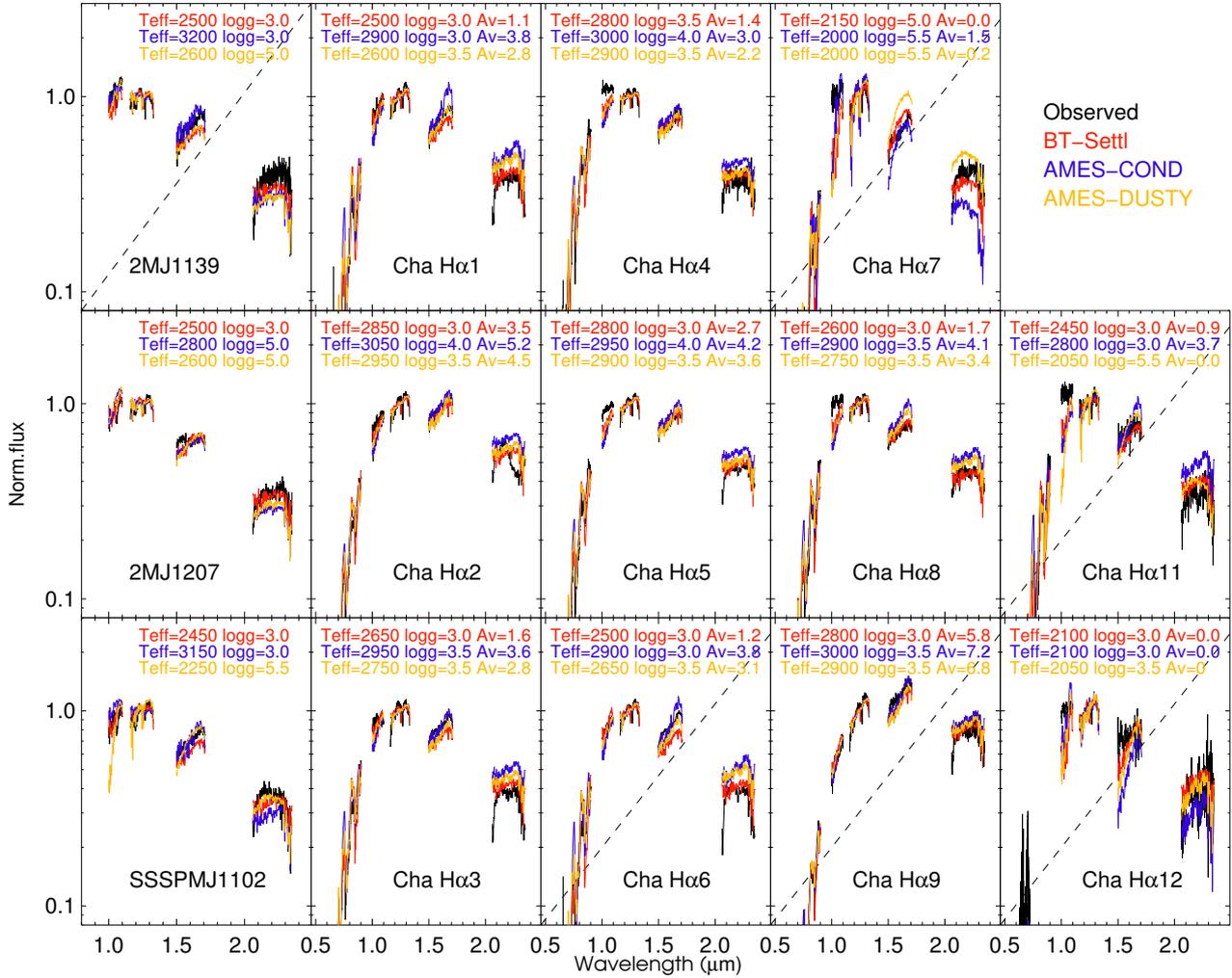}}
\caption{Best fitting synthetic spectra for each dust treatment plotted on top of the observed one (each theoretical spectrum has been reddened to the best fitting A$_{\rm V}$ value). We display the observations in black, the best BT-Settl fitting model in red, the best AMES-COND fitting model in blue and, finally, the best fitting AMES-DUSTY model in dark yellow. Objects for which $\dot{\chi^2} > 1$ (for the BT-Settl collection) are highlighted with a dashed line crossing the corresponding panel.}
\label{bestfitspec}
\end{figure*}

\begin{figure*}
\includegraphics[viewport=0 0 685 675,clip=,height=11.85cm]{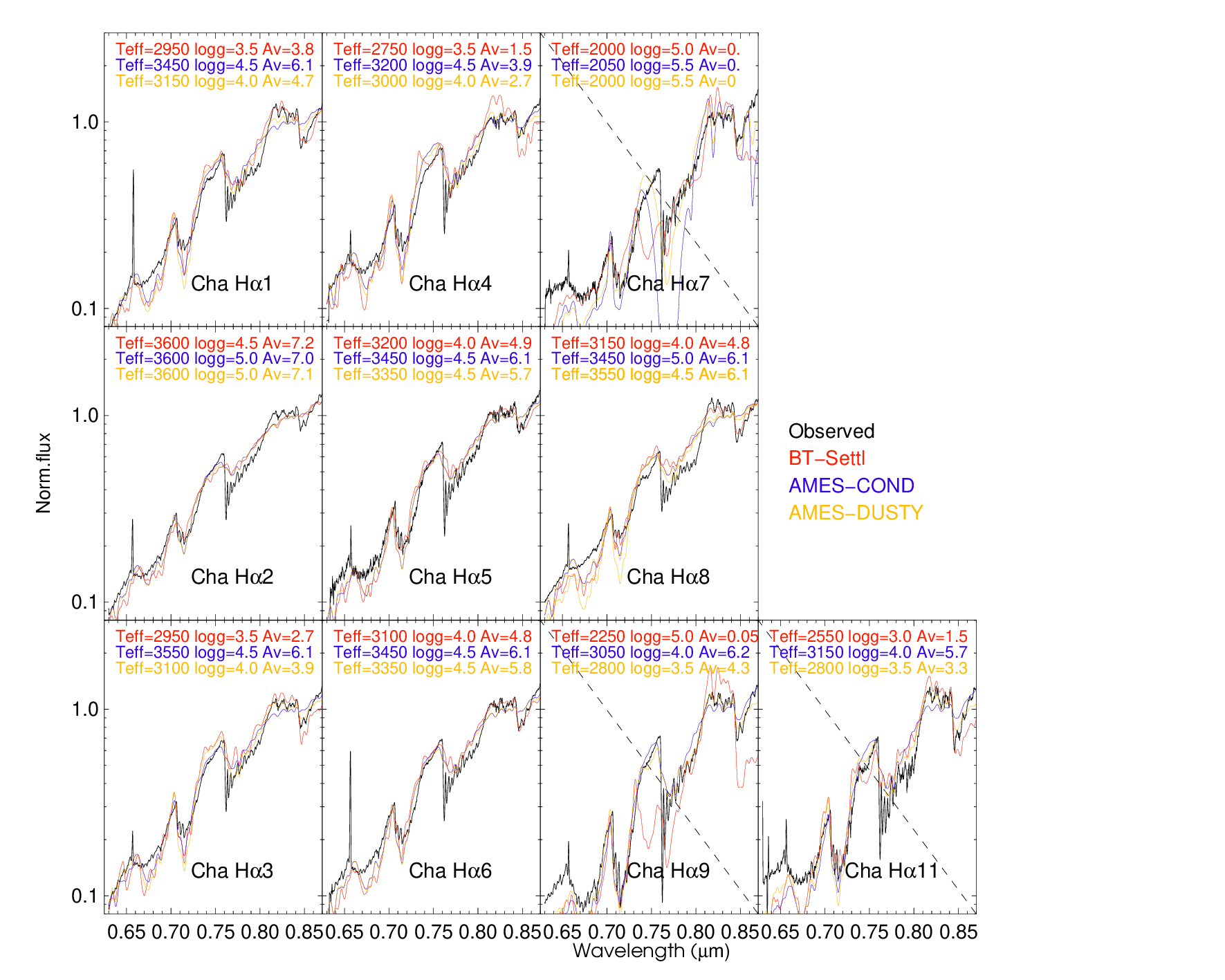}
\includegraphics[viewport=40 0 830 675,clip=,width=13.2cm]{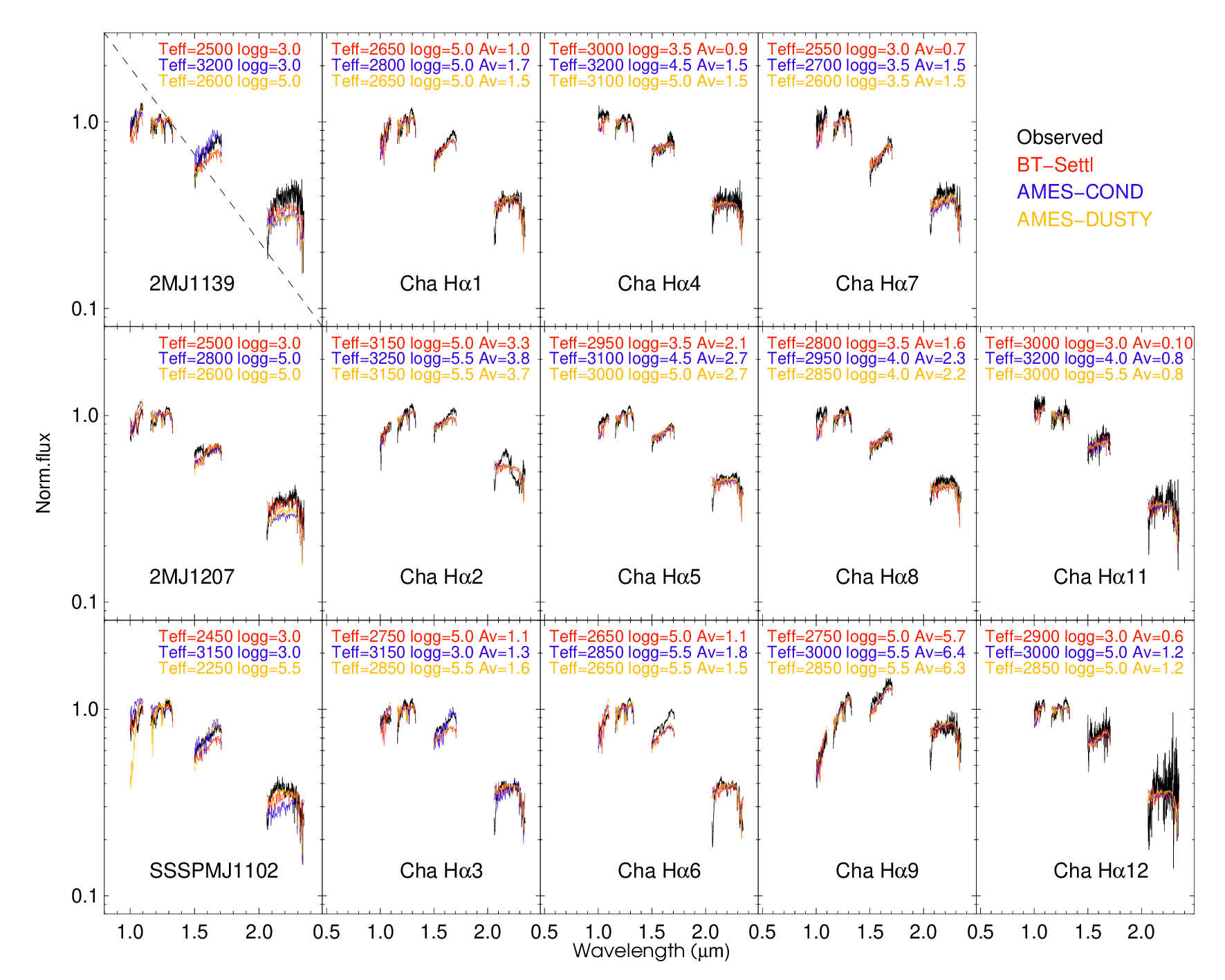}
\caption{{\bf Top: }Best fits for the three collections of models (observations in black, BT-Settl model in red, and AMES-COND and DUSTY in blue and orange, respectively) taking into account only the optical spectroscopy (normalized to the pseudo-continuum at $\sim$8500\AA) in the analysis. {\bf Bottom:} Best fits for the three collections of models (same color code as in the left panels) obtained taking into account only the NIR spectroscopy (spectra normalized to the J-band).}
\label{bestfitspec2}
\end{figure*}

Technically, the spectral fit was carried out with the same approaches than the SED fit; i.e., via minimization of the $\chi^2$ and via computation of the posterior distribution functions of the main fitting parameters: T$_{\rm eff}$ and $A_{\rm V}$ (we refer to the next section for the more precise determination of $\log(g)$). In Fig.~\ref{chislices} we display slices of the $\chi^2$ cubes obtained in the fitting process for Cha H$\alpha$1 and Cha H$\alpha$5 for the three dust treatments and considering the simultaneous fit approach. As can be seen, the result of this brute force approach does not always return a reduced $\chi^2$ value of one. In Figures ~\ref{bestfitspec} and~\ref{bestfitspec2} we highlight this fact (for the case of the BT-Setll models) by overplotting a dashed line crossing the corresponding panel. In Table~\ref{specfitparam} we provide the parameters corresponding to the minimum reduced $\chi^2$ and the uncertainties accepting 10\% variations from the minimum $\chi^2$, which in all cases agree with the most likely parameters from the PDF. This latter statement does not apply to the $\log(g)$ determination because the general spectra fitting process is not sensitive to this parameter (similar to the SED fit, resulting in flat PDFs) and we show better determinations via gravity sensitive lines fit further on.

\input{best_fit_spec_table_comparison.tex}

The comparison of the three dust treatments indicates that while the BT-Settl models, in general, reproduce better the observed data than the extreme DUSTY and COND approaches; there is still room for improvement, and for several objects (namely Cha H$\alpha$ 6, Cha H$\alpha$ 7, Cha H$\alpha$ 9 and Cha H$\alpha$ 11, 2MJ1139) the spectral features over the spectral range analyzed are not well reproduced resulting in fits with large $\dot{\chi^2}$. A different case is Cha H$\alpha$ 12, where the SNR of the optical spectra is the lowest of the sample not allowing a reliable fit to that part of the spectrum.

\begin{figure*}
\includegraphics[width=8.8cm]{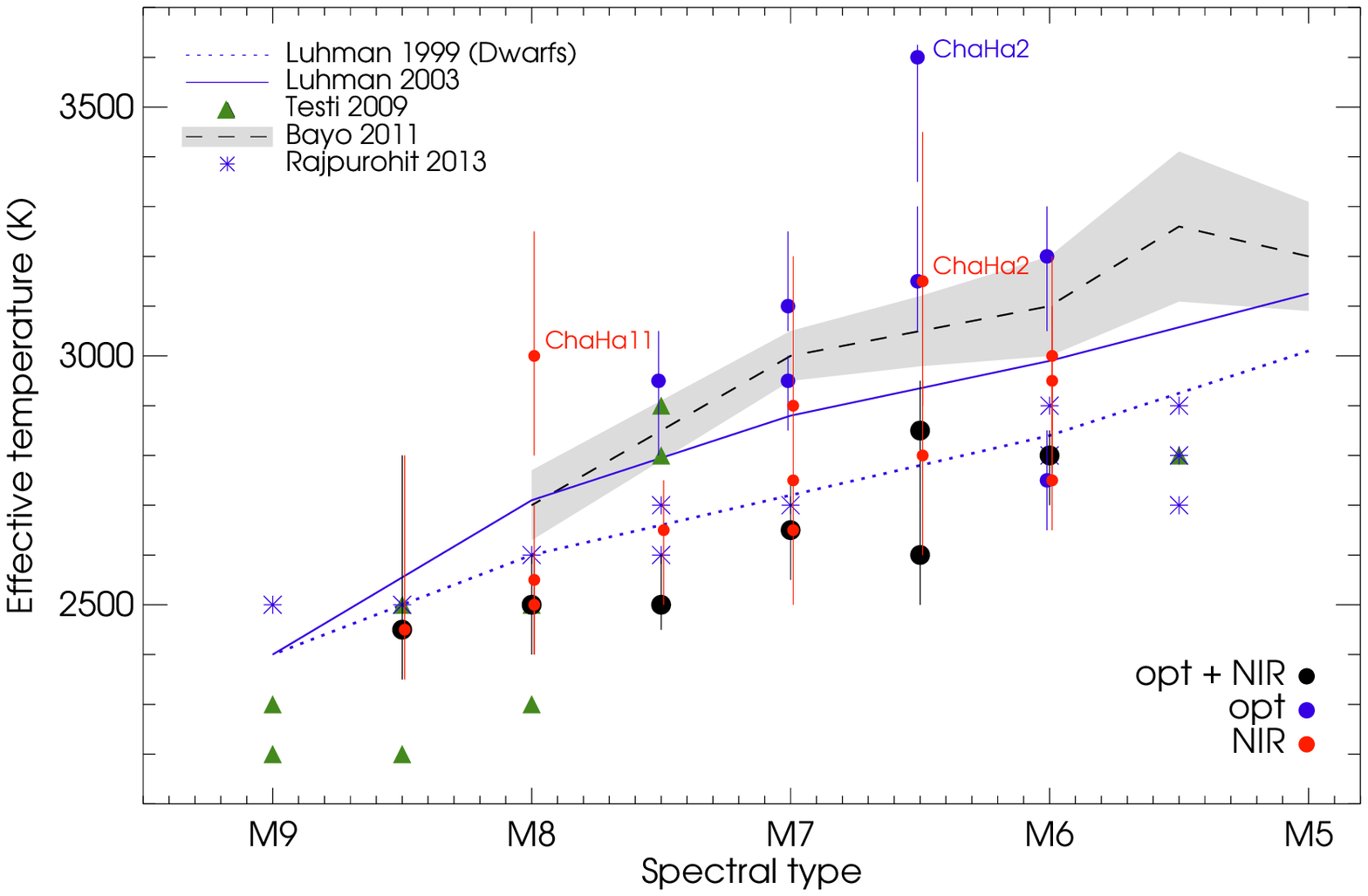}
\includegraphics[width=8.8cm]{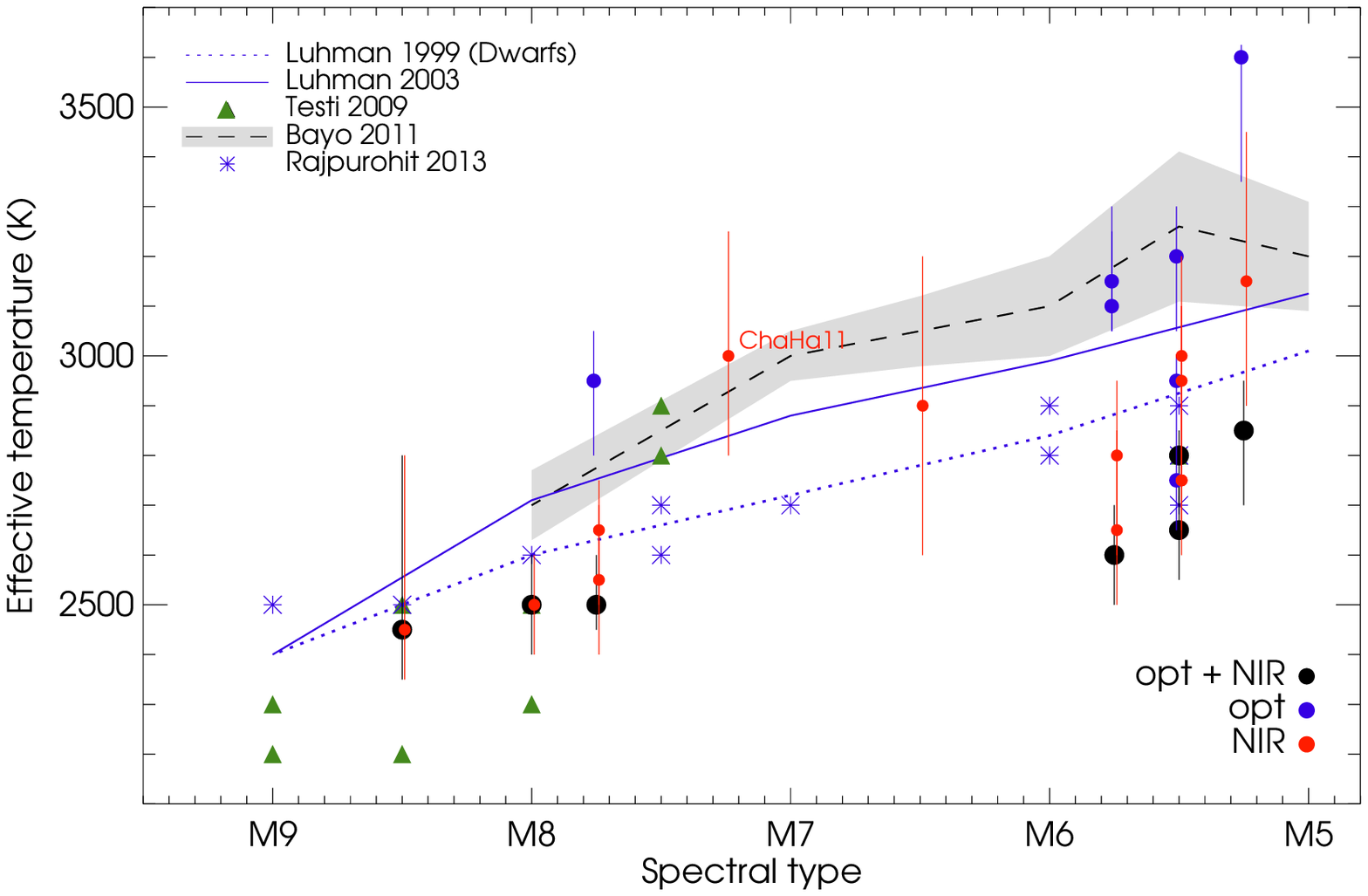}
\caption{Effective temperature scales previously published and the values obtained in this work via spectral fit against the BT-Settl collection of models (solid dots in black, blue and red depending on the spectra used for the fit: optical + NIR, optical alone or NIR alone, respectively). We note that regarding our determinations, we only display objects with best-fit fulfilling $\dot{\chi^2}<1.0$. Regarding literature data: scales by \protect\cite{Luhman03} and \protect\cite{Luhman99} are displayed with blue solid and dotted lines, respectively. Blue asterisks show the estimations by \protect\cite{Rajpurohit13} with the BT-Settling approach for field dwarfs, a dashed black line shows the results from \protect\cite{Bayo11} for the $\sim$5-7 Myr old cluster Collinder 69 (where the effective temperatures were determined via SED fit using atmospheric models with the limiting dust treatments: AMS-DUSTY and AMES-COND), and finally filled green triangles show the estimations by \protect\cite{Testi09}. The two clear outliers (Cha H$\alpha$ 2 and Cha H$\alpha$ 11) are dicussed in the text. In the left panel we have used the spectral types provided in \protect\cite{Comeron00} for the Chamaeleon sources while in the right panel we used the \protect\cite{Luhman07b} determinations. The spectral types have been slightly shifted to better visualize the overlapping determinatinos and uncertainties.
}
\label{Teffscale}
\end{figure*}

Regarding differences obtained when using the optical, NIR, or both sets of data simultaneously, we can conclude that there is a significant contrast depending on the approach used, even for those objects where the quality of the fit is good (i.e., $\dot{\chi^2}\le1.0$) for all cases. In Fig.~\ref{Teffscale} we show the temperature scales obtained with the BT-Settl collection for the ``only optical", ``only NIR" and ``optical + NIR" cases (including exclusively objects with good quality fits) and how those compare with the literature. We have included two panels in the figure because in a number of cases from the Chamaeleon sample there are large discrepancies between the spectral types reported in \cite{Comeron00} and those from \cite{Luhman07b}.

On the one hand, as has been reported already in the literature, there is a trend towards determination of higher temperatures when using only optical data (blue solid dots in the figure) vs. including the NIR data in the fit. These higher temperature determinations agree within the errors and dispersion with the temperature scale for young sources proposed in \cite{Bayo11}, based in optical spectral type determinations and optical+NIR+MIR SED fit. 

On the other hand, the temperatures determined using simultaneously the optical and NIR data are slightly colder than those reported by \cite{Rajpurohit13} which may seem surprising given the older nature of the field-dwarf sample analyzed in the latter work (although in agreement with the results from \citealt{Pecaut13} for earlier M-type sources). This result could just be due to the fact that \cite{Rajpurohit13} use 
only optical data (high-resolution spectroscopy) in their fits. The inclusion of NIR spectra in the fits of these older objects will most likely yield lower temperatures. Such analysis is undergoing and will be presented in \cite{Rajpurohit16}.

Finally, regarding the two sources of spectral types, those reported in \cite{Luhman07b} translate in lower dispersion in effective temperature determined per spectral type than those reported in \cite{Comeron00}. Besides, our spectral fits yield in general lower values for T$_{\rm eff}$ than those from \cite{Luhman03} for the Cha I sample (same trend commented in the previous section for the SED fits), that we attribute to the dust settling treatment in the NextGen models used in \cite{Luhman03}.

Especial mention may deserve Cha H$\alpha$ 2, for which a very large dispersion in effective temperature is found depending on the wavelength range used for the fit with that derived from the optical spectra being higher than expected from its spectral type (in comparison with the rest of the sample and with the temperature scales published in the literature). We attribute this dispersion to the fact that Cha H$\alpha$ 2 is a close periodic binary \citep{Vogt12,Cody14} not resolved in our spectra and different components of the system dominate in different regions of the spectrum.

\section[]{Age indicators}
\label{age}

In order to estimate the age of each object of the sample we analyzed gravity sensitive alkali lines and compared their intensity with models computed for different surface gravity values.  
Taking into account the available data and resolution, we used the K I J-band doublet at 1.244 \& 1.252 $\micron$ \citep{Reid01} for these comparisons. In the previous section we have showed that, despite some room for improvement, the BT-Settl dust treatment reproduces best simultaneously the optical and NIR spectral features of these young sources and therefore that is the collection used for the gravity determination. Regarding the effective temperature used for each object, we performed several tests using those determined from NIR + optical spectral fitting and NIR alone and we did not find significant differences. The results presented in this section are those obtained with the "NIR only" fitting approach to guarantee homogeneity through the Cha I and TWA samples.

For each object, we took for four different values of surface gravity ($\log(g) = 3.5, 4.0, 4.5 ~{\rm and}~ 5.0$), the BT-Settl model corresponding to the best fitting temperature according to the NIR fit (models that were already adapted to our resolution by the convolution of the original synthetic spectra with Gaussian profiles with FWHM $\simeq$ 25 \AA). We reddened them with the the best fitting value of A$_{\rm V}$, because, although this step was not critical since we will do the comparison over the continuum-normalized spectra, we wanted to avoid possible discrepancies in continuum determination depending on the amount of reddening present for the different objects. Finally we determine a local continuum by fitting a low order polynomial to the observed data and used the same pseudo-continuum to normalize both the observed and theoretical data.

In Table~\ref{PysicPar1} we present the surface gravity values estimated in this matter along with masses, radius and distances obtained combining these estimations with isochrones and the dilution factor (M$_{\rm d}$) estimated via full SED fit in Section~\ref{sec:SEDfit} (via the simple Eqs.~\ref{eq:distance} and \ref{eq:radii}, where $M_{\rm d} = F_{\rm Model} / F_{\rm Obs}$). For most sources, the theoretical spectra that reproduces best the observed doublet is the one corresponding to $\log(g)$ of 4.0 dex, but there are some exceptions discussed in the next subsection. 

\begin{equation}
\label{eq:radii}
\log g = 4.44 + \log M(M_{\odot}) - 2\log R(R_{\odot})
\end{equation}
\begin{equation}
\label{eq:distance}
d({\rm pc}) = 2.26 \times 10^{-8} R_{*}(R_{\odot}) \sqrt{\left(\frac{F_{\rm Model}}{F_{\rm Obs}}\right)}
\end{equation}

\input{logg_table.tex}

\subsection[]{The Hertsprung-Russel Diagram and the age dispersion}
\label{HR}

\begin{figure}
\resizebox{\hsize}{!}{\includegraphics{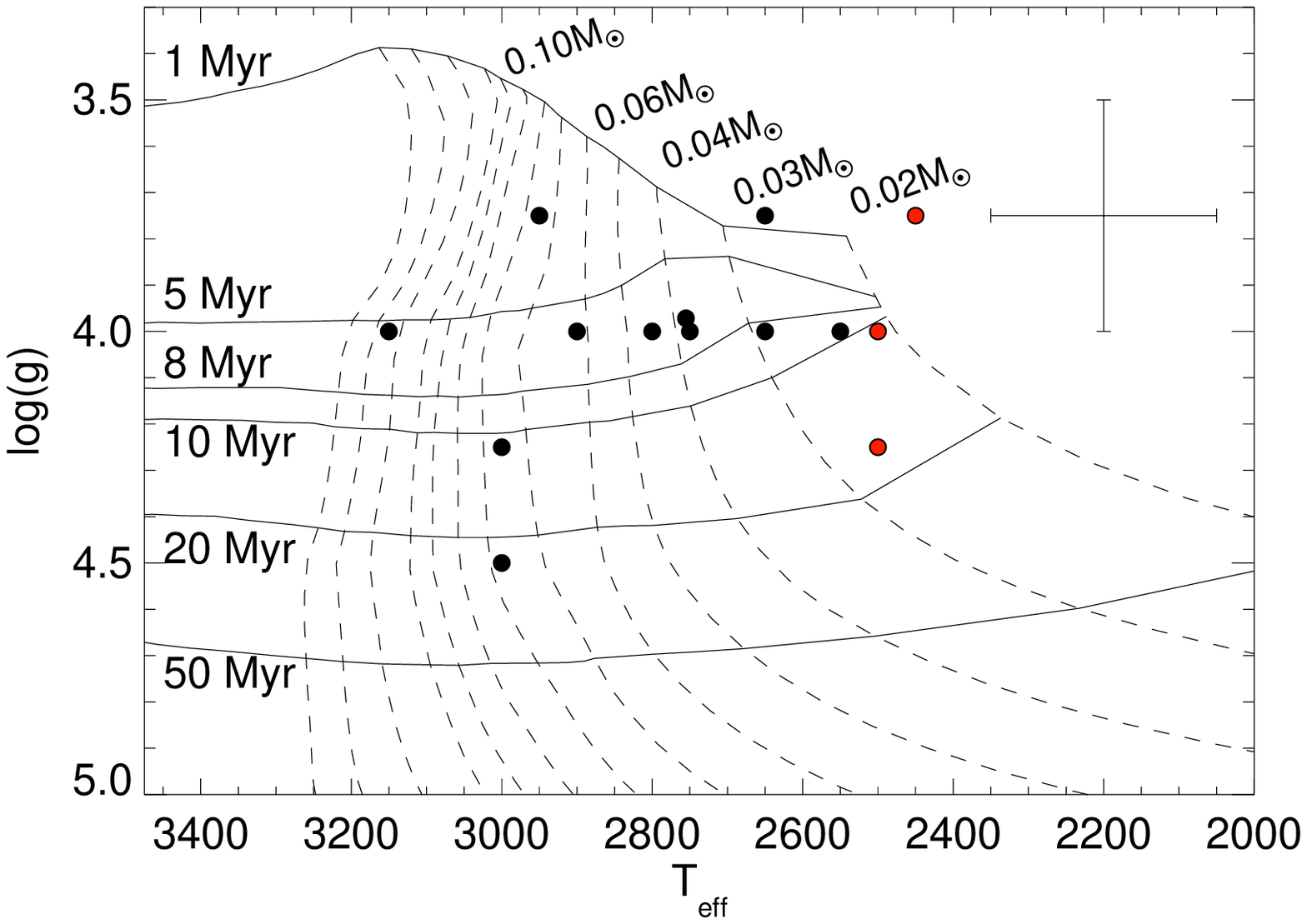}}
\caption{T$_{\rm eff}$ vs $\log(g)$ diagram for the Cha I (black filled circles) and TWA (red filled circles) sources. The effective temperature corresponds to the best fitting BT-Settl model to the NIR spectra (Sec.~\ref{specfit}) and the $\log(g)$ values come from the K I doublet comparisons (Sec.~\ref{age}). We interpolated in the isochrones and evolutionary tracks displayed \citep{Baraffe98, Baraffe02} to provide age and mass estimations that are independent of the distance to the sources. In the upper right side of the figure we display the mean error-bars.}
\label{loggteff}
\end{figure}

As mentioned before, we have combined the $\log(g)$ determinations with the effective temperatures derived in Section~\ref{specfit} (those coming from the best BT-Settl fitting model to the NIR spectrum of each source) to build a pseudo-HR diagram where no assumption on the distance to the targets has to be made. We show this T$_{\rm eff}$ vs. $\log(g)$ diagram in Figure~\ref{loggteff} along with isochrones and evolutionary tracks that are a combination of those from \cite{Baraffe98} and \cite{Baraffe02}. 

We performed linear interpolation among the isochrones and evolutionary tracks to provide the age and mass estimations presented in Table~\ref{PysicPar1}. As will be discussed later, we observed a large age dispersion in both populations of targets (the Cha I and the TWA samples), as was already reported by \cite{Luhman07b} for Cha I, and has been discussed in other young clusters like Collinder 69 \citep{Bayo11}. This dispersion could have its origin in different early accretion history (see \citealt{Baraffe10} for the proposed scenario).

In addition, even taking into account the non-negligible error bars, we can conclude that, according to these evolutionary tracks, nine of our sources are substellar, and one is clearly stellar (Cha H$\alpha$ 2).

To compare the results from Section~\ref{sec:SEDfit} with those of Section~\ref{specfit} in the estimation of fundamental parameters, we calculated the radii of the sources following two different approaches: on the one hand we combined (via Eq.~\ref{eq:distance}) the distances available in the literature (160pc for the Cha I sample, from \citealt{Wichmann98,Knude98}, and 18, 54 and 22pc for 2MJ1139, 2MJ1207 and SSSPMJ1102, respectively, from \citealt{Faherty09}) with the dilution factor estimated by VOSA. On the other hand, we used the isochrone-masses described above and the $\log(g)$ values from the previous subsection (via Eq.~\ref{eq:radii}). Both sets of estimations are summarized in Table~\ref{PysicPar1}.

The first approach yields unrealistic large values of the radii for almost all objects from the Cha I sample, not only when compared with the second approach, but also with the upper-limits estimated in \cite{Joergens01} (that take into account the rotational velocities of the sources). Furthermore, for the TWA sources, we get values in agreement with the literature (see for example \citealt{Bonnefoy14}). 

Also in Table~\ref{PysicPar1} we compute the distances expected from the radii estimated with the second approach and the dilution factor from VOSA. The discrepancies in radii and distances could be explained by uncertainties in the distances (note that in the case of more precise distances, the TWA sample, the discrepancies are much smaller or inexistent as in the case of 2M1207), uncertainties/outdated interior models (especially complicated to calculate for ages younger than $\sim$10 Myrs) and overall uncertainty in the dilution factor induced by variability of the sources. 

\subsubsection[]{The particular case of ChaH$\alpha$8 and ChaH$\alpha$11}

To illustrate the age dispersion found in the Cha I sample, in this section we compare in more detail the spectra from ChaH$\alpha$8 and ChaH$\alpha$11. We have chosen these two objects because they have similar properties: range of effective temperatures estimated in Section~\ref{specfit}, range of spectral types (when considering both, \citealt{Comeron00, Luhman07b}) and moderately intense H$\alpha$ emission \citep{Comeron00}. In spite of these similarities, while or ChaH$\alpha$8 the K I doublet analysis suggests a surface gravity of 4.0 dex (like most of the Cha I sources), the larges value (4.5 dex) is found for ChaH$\alpha$11.

\begin{figure}
\includegraphics[viewport=0 0 257 400,clip=,width=4.05cm]{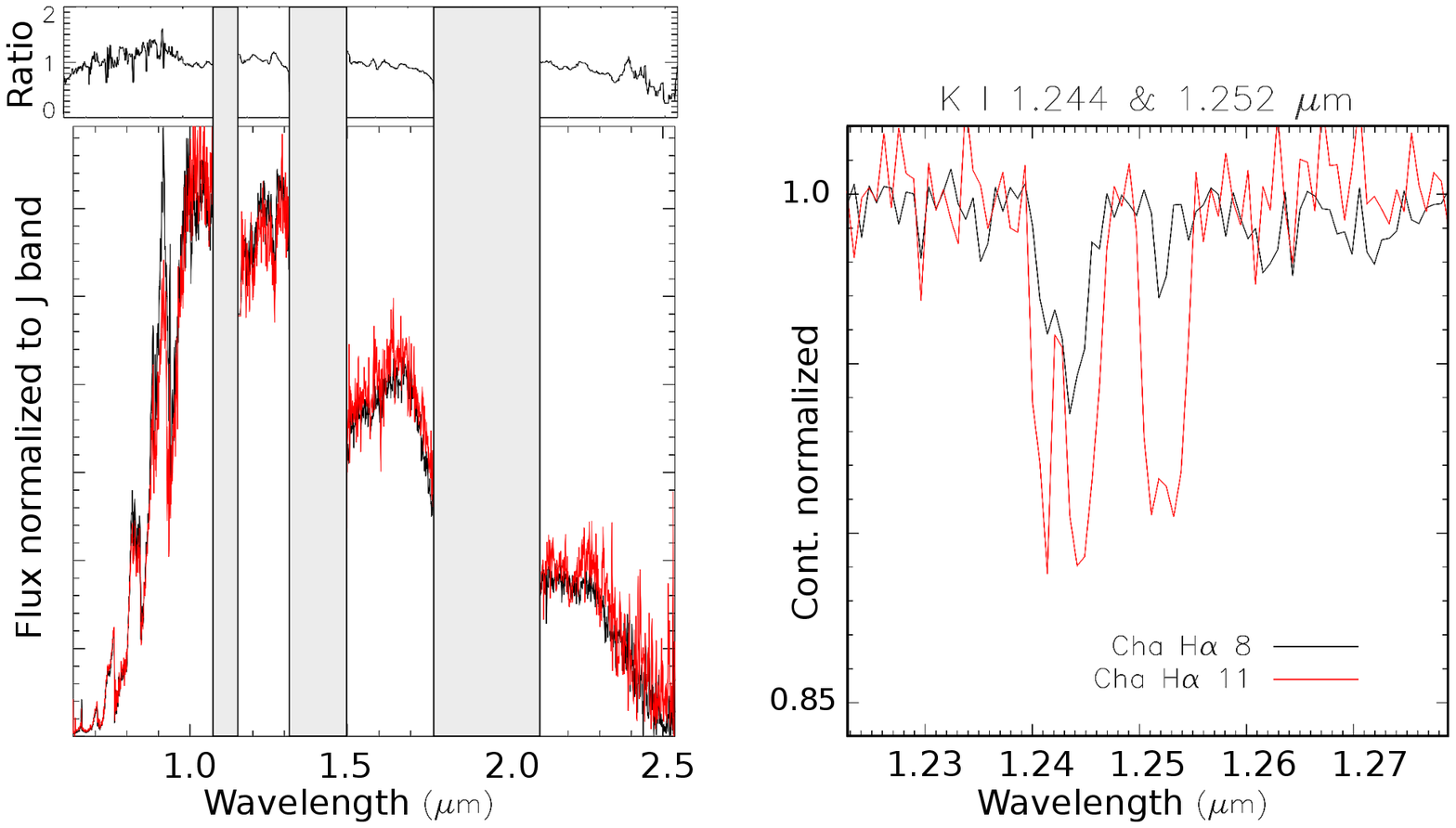}
\includegraphics[viewport=280 0 536 400,clip=,width=4.05cm]{2ages_complete_FINAL.eps}
\caption{{\bf Left:} Direct comparison of the optical + near-infrared spectra of Cha H$\alpha$ 8 (in black) and Cha H$\alpha$ 11 (in red). Both spectra have been dereddened according to the A$_{\rm V}$ values given in Table~\ref{specfitparam} corresponding to the optical + NIR fitting approach. In the upper panel, the ratio between both spectra is provided. In both panels, the areas where we have not been able to correct for the strong telluric bands are covered with white rectangles. {\bf Right:} Detail of the continuum normalized spectra for both sources in the spectral region of the K I doublet analyzed in Section~\ref{age}. The color code is the same as in the left panel.}
\label{chaha8chaha11}
\end{figure}

In the left panel of Fig.~\ref{chaha8chaha11} we show the optical + NIR spectra for both sources (the spectra have been dereddened with the best fitting values of A$_{\rm V}$ from Section~\ref{specfit} with the optical + NIR fitting approach, and then normalized to the J-band flux) for direct comparison and in the top part of that panel a detail of the intensity ratio across the whole wavelength range. With this panel it is clear that both spectra share the main broad features. On the other hand, in the right panel we show a comparison of the same spectra, continuum normalized in the K I doublet spectral region. Although the difference in surface gravity determinations among both sources is only twice the uncertainties given in Table~\ref{PysicPar1}, we must note that the quoted uncertainty is given by the $\log(g)$ step in the BT-Settl grid and with this comparison we show that those uncertainties are over-estimated, at least in this case, where the intensity of the doublet in one spectrum doubles that of the other one, translating in an age difference estimation of 15 Myrs.

\section[]{Accretion, activity and rotation}
\label{aar}

Since the models we have used to characterize our sample of sources are ``simple photospheres", they do not include, by design, all effects occurring in other outer layers of the stars and brown dwarfs as, for example, chromospheric activity. Still, we can try to gain knowledge about these other physical processes (activity, accretion, etc.) analyzing emission lines in our observations, additional parameters, as rotational velocity, or disk presence (as inferred from infrared excess over the expected photospheric flux).

\input{HaEW_Table.tex}

One of the most common emission lines reported in young low-mass stars and brown dwarfs is H$\alpha$.While the presence of this kind of emission in normal low-mass main-sequence stars is a sign of chromospheric activity, the presence of H$\alpha$ emission in T Tauri stars with disks and brown dwarfs seems to be a much more complicated problem due to the fact that both, activity and accretion, can be responsible for the observed emission \citep{Joergens03}. 

In the case of high-resolution spectroscopy observations, there is the possibility to decompose the emission line in different profiles and velocities in order to unravel the physical mechanism behind it. This detailed analysis is not possible with low-resolution observations, but one can still use saturation criteria such the one from \cite{Barrado03} to determine whether activity alone can explain the intensity of the H$\alpha$ emission for a given spectral type or if it is necessary to invoke accretion.

\begin{figure}
\resizebox{\hsize}{!}{\includegraphics{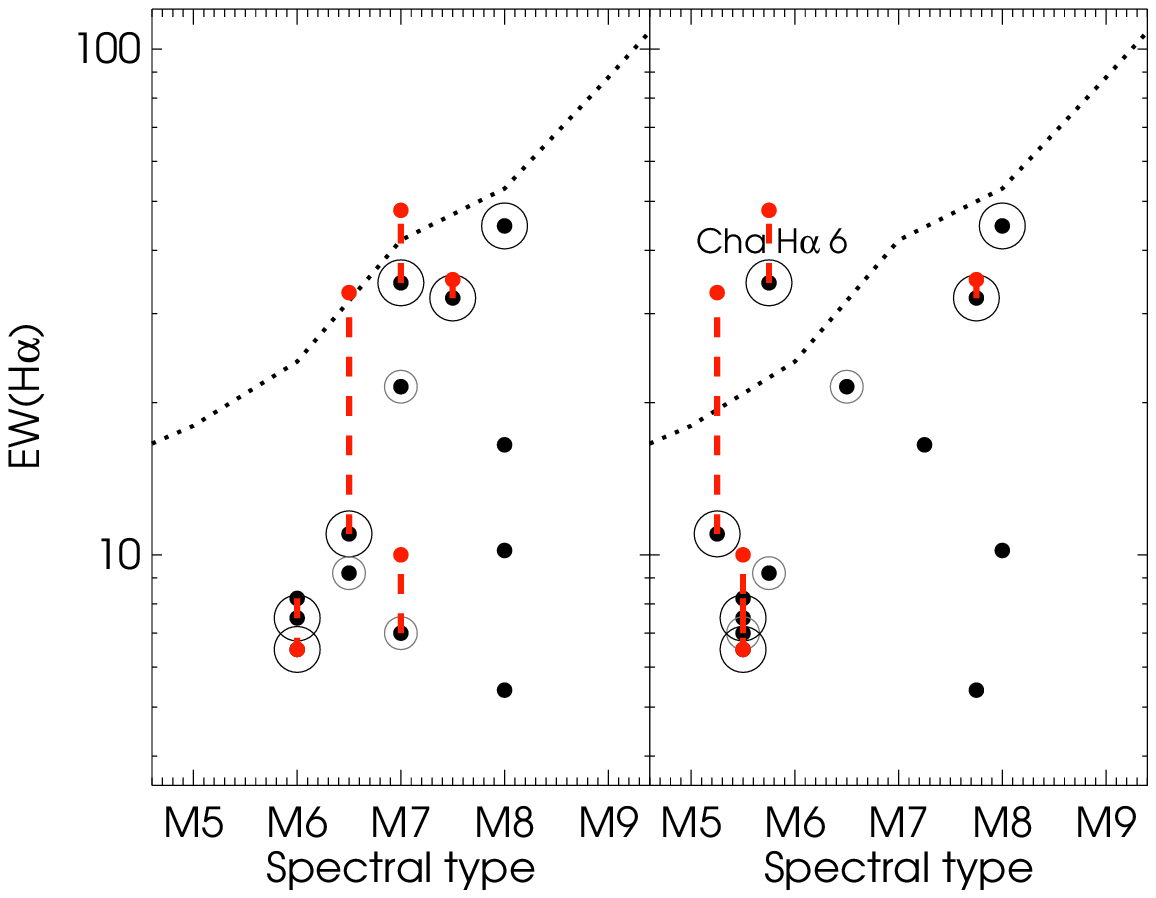}}
\caption{Equivalent width of H$\alpha$ vs. spectral type for the objects for which we have this kind of information. In the case of the Cha I sample we present our revised measurements of equivalent width performed on the spectra presened in \protect\cite{Comeron00}. In the left panel we use the spectral types reported in \protect\cite{Comeron00} for the Cha I sources and in the right panel, those provided by \protect\cite{Luhman07b}. In both cases we overplot (as a dotted line) the saturation criterion proposed by \protect\cite{Barrado03} to disentangle between accretion and activity. Also, for the Cha I sample, in both panels we added the equivalent widths of H$\alpha$ provided in \protect\cite{Natta04} with red dots connected (with red dashed lines) to our revised measurements.}
\label{acc_crit}
\end{figure}

Strong H$\alpha$ emission has been reported for most of the sources in our sample (see \citealt{Comeron00,Natta04,Barrado04}, etc.). However, for the Cha I sample, we noted large differences between the values quoted in \cite{Comeron00} and those provided in \cite{Natta04}. Although at first sight this could be attributed to the difference in the resolution of the observations, in \cite{Bayo11} is shown in detail how even extreme differences in resolution cannot account for this dispersion in the values. On the other hand, the differences could be just reflecting true physical variability of the sources. We repeated the measurements with the automatic procedure developed in \cite{Bayo11} and found values much closer to those provided in \cite{Natta04} that we report in Table~\ref{Halpha}.

The only object for which our revised measurement confirms a value very different from that of \cite{Natta04} is Cha H$\alpha$2. This is the highest mass member of the Cha I sample and the difference in H$\alpha$ emission between the two epochs, results in Cha H$\alpha$2 laying above or below the saturation criterion by \cite{Barrado03} (i.e. classified as accreting or non-accreting source). This young low-mass star (0.14-0.25 M$_{\odot}$, according to the previous sections) shows strong infrared excess (see Table. 2 and Fig. 7), signpost of disk harbouring, and the discrepancies among the H$\alpha$ emission intensities between the two epochs could be explained by episodic accretion.

In addition, Cha H$\alpha$ 6 was classified as accretor by \cite{Natta04}, and it is also classified as accretor by the saturation criterion by\cite{Barrado03} (when using the spectral type derived in \citealt{Luhman07b} and right at the border when using the spectral type from \citealt{Comeron00}). In this case the equivalent withs of H$\alpha$ from the two different epochs of data agree, what could point towards this lower mass object (brown dwarf, $\sim$0.03 M$_{\odot}$, according to previous sections) undergoing more steady accretion. The remaining sources are classified as non-accreting (in the same manner than, for those sources in common, \citealt{Natta04}).

\begin{figure}
\resizebox{\hsize}{!}{\includegraphics{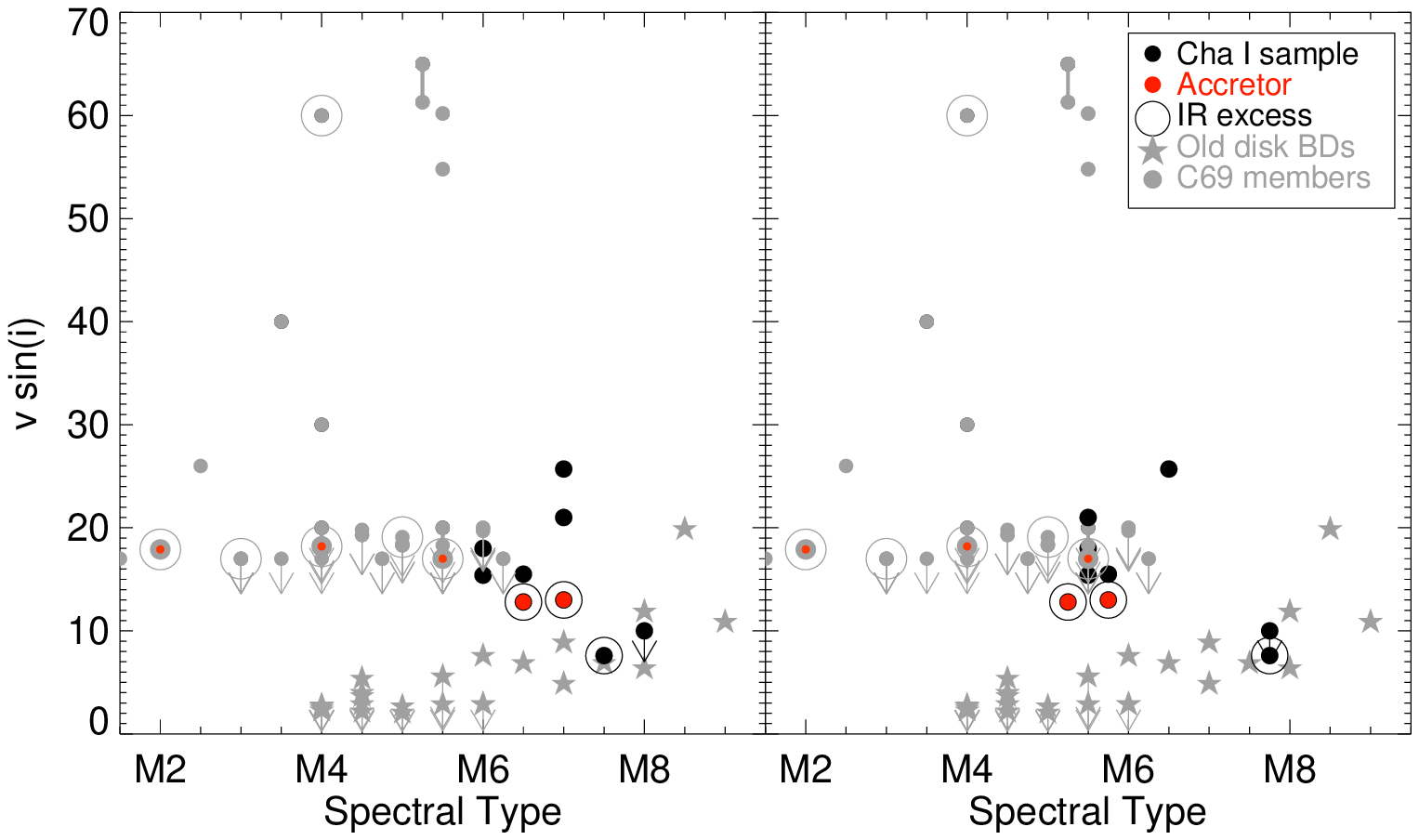}}
\caption{Spectral type vs projected rotational velocity for three different samples of objects: in gray 5-pointed filled stars, old disk population brown dwarfs from \protect\cite{Mohanty03}; in gray filled circles, young members of Collinder 69 from \protect\cite{Bayo11}, and in black filed circles, the objects from the Cha I sample. As in the previous figure, in the left panel we use the spectral types from \protect\cite{Comeron00} and in the right panel those from \protect\cite{Luhman07b} for the Cha I sample (and in both cases we use the $v\sin(i)$ values from \protect\citealt{Joergens01}). Besides, for the two samples of young sources, Collinder 69 and Cha I, we use specific symbols to highlight particular properties of the individual objects: a red dot for accretors and large open circles for infrared excess detection.}
\label{vrot}
\end{figure}

Activity and magnetic fields in very low-mass stars and brown dwarfs are intimately linked with angular momentum evolution, and since, as shown in Table~\ref{sources}, \citet{Joergens01} provide projected rotational velocities for eight of the members of Cha I and one upper limit, we have tried to relate this parameter with the strength of the H$\alpha$ emission and the spectral type. In Fig.~\ref{vrot} we have plotted the projected rotational velocities against the spectral type for the Cha I sample (see caption for details) along with another sample of young very low-mass stars and brown dwarfs members of Collinder 69 ($\sim$5-12 Myrs old from \citealt{Bayo11}) and a sample of M6--M7 (old disk population) field dwarfs from \citet{Mohanty03}.

We must note that given the size of the Cha I sample we cannot reach strong conclusions, but Fig.~\ref{vrot} suggests that the trend of disk-harbouring sources showing lower values and/or less dispersion in $v\sin(i)$ with respect to diskless sources (already discussed in \citealt{Bayo12}), holds for these lower-mass (and somewhat younger) objects. 

In addition, while the study from \cite{Bayo12} was dominated by upper-limits (mainly coming from FLAMES data analyzed in \citealt{Sacco08,Maxted08}) for slow rotators, this sample allows to go lower in mass and rotational velocity and we see how the gap in $v\sin(i)$ values between the young and old populations could just be caused by the censored data.

\section[]{Summary and conclussions}
\label{conclussions}

In this work we have studied in detail the properties of 14 young, late M-type, very low-mass stars and brown dwarfs belonging to the Chamaeleon I dark cloud or the TW Hydrae Association.

Based in comparison with older, field dwarfs, we have studied the goodness of a NIR VO-based index to perform simultaneously spectral classification and distinguish between old and young populations.

When combining the atmospheric parameters derived with the different techniques with isochrones and evolutionary tracks we find the previously reported feature/problem of the age/luminosity spread on both samples. In particular, in Chamaeleon I, we highlight two brown dwarfs (Cha H$\alpha$ 8 and Cha H$\alpha$ 11) that showing very similar temperatures display clearly different surface gravities. A possible explanation for this difference is that Cha H$\alpha$ 11 could be a candidate to have undergone extreme early accretion. 

In connection with accretion but at the current stage of evolution; for the two objects in our sample classified as accretors (Cha H$\alpha$ 2 and Cha H$\alpha$ 6), we find pretty distinct behaviours: one of them (the very low-mass star Cha H$\alpha$ 2) shows strong variability in H$\alpha$ emission that could be related to episodic accretion, and the other one shows consistent intense emission (the brown dwarf Cha H$\alpha$ 6) that suggest a more stable accretion connection with the disk.

Finally, the sources of our sample follow the trend proposed in \cite{Bayo12} as a down-scaled version of disk-locking \cite{Bouvier86} for the substellar domain, since the diskless sources seem to exhibit higher rotational velocities than those harbouring disks (we must note the low number statistics).

\section*{Acknowledgments}
A. Bayo wants to acknowledge J. Olofsson and I. Baraffe for fruitful and encouraging discussions during the project and financial support from the Proyecto Fondecyt de Iniciaci—n 11140572. This research has made use of the SIMBAD database and Aladin, operated at CDS, Strasbourg, France; and the NASA's Astrophysics Data System.
This publication makes use of VOSA, developed under the Spanish Virtual
Observatory project supported from the Spanish MICINN through grant
AyA2011-24052. This work was co-funded under the Spanish grant AYA2012-38897-C02-01. J.C.B. acknowledge support from CONICYT FONDO GEMINI - Programa de astronom\'ia del DRI, folio 32130012. K.P.R. was funded by the Chilean FONDECYT Postdoctoral grant 3140351.This publication also makes use of data products from the Wide-field Infrared Survey Explorer, which is a joint project of the University of California, Los Angeles, and the Jet Propulsion Laboratory/California Institute of Technology, funded by the National Aeronautics and Space Administration.

\bibliographystyle{mnras}
\bibliography{biblio}

\appendix

\section{SED and low-resolution spectral fitting for the field dwarf sample}

In this appendix we present the corresponding tables and figures from the methodology presented in Section 4 but applied to the field dwarfs sample (representative of an old M-dwarf population) described in the Introduction. For the wavelength coverage of the SEDs, all sources have infrared photometry in \cite{2MASS} (and five of them also in \citealt{DENIS}), around half of them (eight objects) have a mid-infrared counterpart in \cite{WISE}, finally five have optical counterparts in \cite{sloan}, and for most of the other objects (eight in total) we also found optical magnitudes in \cite{Casagrande08,Winters11, Monet03,Reid04,Samus03, Zacharias10,Jenkins09} or \cite{Lepine05}. Since these targets belong to an older population, for the SED fits we include the constrain for the $\log(g)$ parameter to be equal or larger than 4.5 dex. A more detailed spectral modeling for this sample (field, old objects) is out of the scope of this paper and will be presented in Rajpurohit et al. in preparation.

\begin{figure*}
\resizebox{\hsize}{!}{\includegraphics{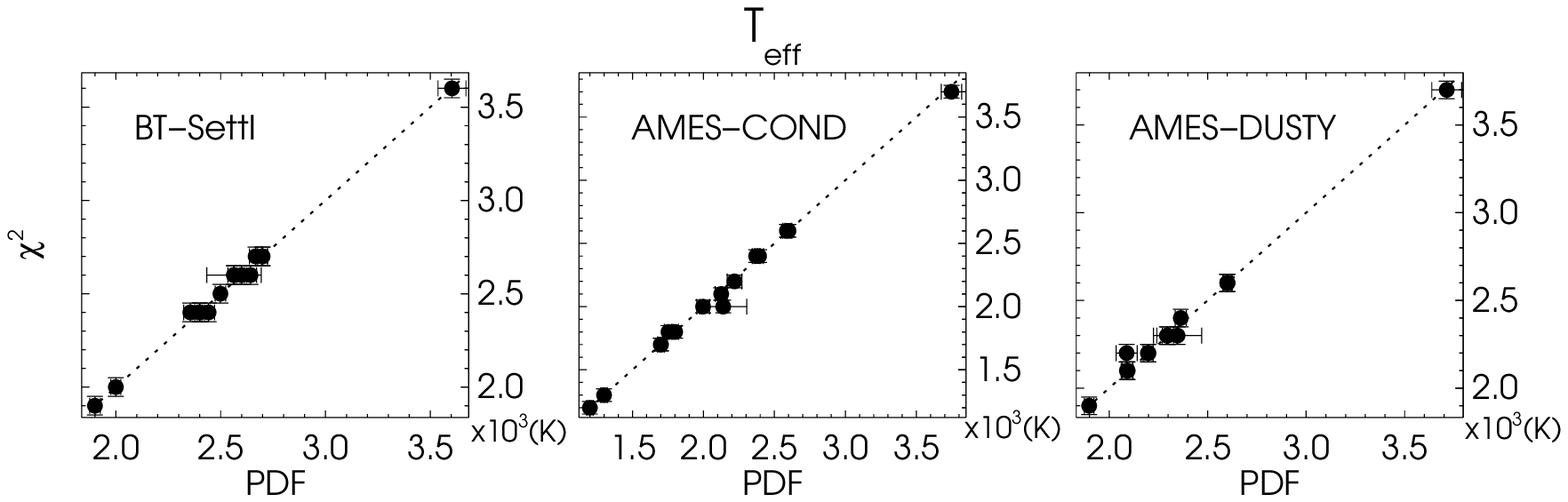}}
\caption{Comparison of the estimated T$_{\rm eff}$ via minimization of the squared differences and gaussian fit to the posterior distribution functions (see for an example Fig.~\ref{Bayesgaussfit}) considering the three dust treatments for the field dwarfs sample. Both estimations agree within the error-bars calculated as half the parameter step in the grid of models and the $\sigma$ of the gaussian fit to the PDF.}
\label{paramComparBayesChiField}
\end{figure*}

\begin{figure}
\includegraphics[width=8.8cm]{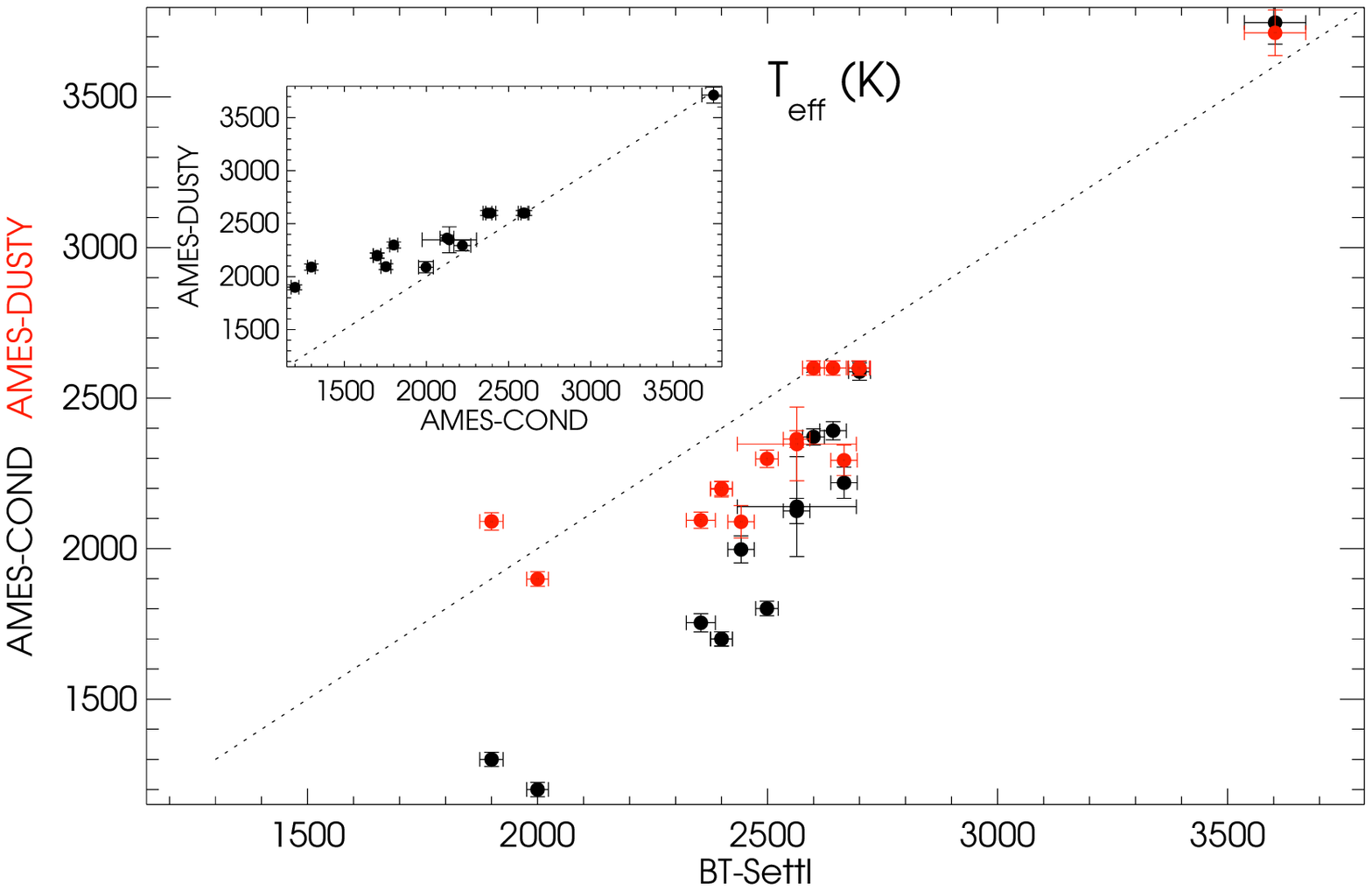}
\caption{Comparison of the estimated T$_{\rm eff}$ for the field dwarfs via gaussian fit to the posterior distribution functions (see Fig.~\ref{paramComparBayesChi} and text for details) for the three different dust treatments. For both parameters under study In the large panels we show the values obtained with the BT-Settl collection vs AMES-COND (in blak) and AMES-DUSTY (in red) and in the smaller panels we display the comparison AMES-COND vs AMES-DUSTY.}
\label{paramComparDustTreatField}
\end{figure}

\input{SED_field_fit_param.tex}
\input{best_fit_field_spec_table_compare_dust.tex}

\end{document}

%% file: sources.tex
\begin{table}
\begin{center}
\caption[Spectral types, projected rotational velocities (only for some of the ChaI objects)
and photometry for the sources analized in this paper.]{Spectral types
  and projected rotational velocities (for some of the ChaI objects) for the sources
analyzed in this work.} \label{sources}
\begin{tabular}{lccrr}
Name&RA&Dec&SpT$^{\mathrm{a}}$&$v$ sin $i$ $^{\mathrm{b}}$\\
\hline
\hline
\noalign{\smallskip}                                                  
Cha H$\alpha$ 1  & 11:07:17   &   -77:35:53 & M7.5 &7.6  $\pm$ 2.2  \\
Cha H$\alpha$ 2  & 11:07:42   &   -77:33:59 & M6.5 &12.8 $\pm$ 1.2  \\
Cha H$\alpha$ 3  & 11:07:52   &   -77:36:57 & M7   &21.0 $\pm$ 1.6  \\
Cha H$\alpha$ 4  & 11:08:19   &   -77:39:17 & M6   &18.0 $\pm$ 2.3  \\
Cha H$\alpha$ 5  & 11:08:24   &   -77:41:47 & M6   &15.4 $\pm$ 1.8  \\
Cha H$\alpha$ 6  & 11:08:40   &   -77:34:17 & M7   &13.0 $\pm$ 2.8  \\
Cha H$\alpha$ 7  & 11:07:38   &   -77:35:31 & M8   &  $\le$10       \\
Cha H$\alpha$ 8  & 11:07:46   &   -77:40:09 & M6.5 &15.5 $\pm$ 2.6  \\
Cha H$\alpha$ 9  & 11:07:19   &   -77:32:52 & M6   &--              \\
Cha H$\alpha$ 11 & 11:08:29   &   -77:39:20 & M8   &--              \\
Cha H$\alpha$ 12 & 11:06:38   &   -77:43:09 & M7   &25.7 $\pm$ 2.6  \\
\hline
SSSPMJ1102   & 11:02:10  &   -34:30:36 & M8.5$^{\mathrm{c}}$ \\
2MJ1207      & 12:07:33  &   -39:32:54 & M8$^{\mathrm{d}}$   \\
2MJ1139      & 11:39:51  &   -31:59:21 & M8$^{\mathrm{d}}$   \\
\hline
LP 803-33               & 15:48:26& -19:54:00& M5$^{\mathrm{e}}$    \\
SCR J0723-8015          & 07:24:00& -80:15:18& M6$^{\mathrm{e}}$    \\
AZ Cnc                  & 08:40:30& +18:24:09& M6$^{\mathrm{e}}$    \\
GJ 644 C                & 16:55:35& -08:23:40& M6.5$^{\mathrm{e}}$  \\
LHS 234                 & 07:40:19& -17:24:45& M6.5$^{\mathrm{e}}$  \\
SCR J0702-6102          & 07:02:50& -61:02:48& M6.5$^{\mathrm{e}}$  \\
2MASPJ125               & 12:54:37 & +25:38:50& M7.5$^{\mathrm{e}}$  \\
2MASSJ1434              & 14:34:26& +19:40:50& M8$^{\mathrm{e}}$    \\
LHS 2397 a              & 11:21:49& -13:13:08& M8$^{\mathrm{e}}$    \\
2MASSJ0858              & 08:58:18& -78:24:54& M8.5$^{\mathrm{e}}$  \\
2MASSJ1239              & 12:39:19& +20:29:52& M9$^{\mathrm{e}}$    \\
GJ 3517                 & 08:53:36& -03:29:32& M9$^{\mathrm{e}}$    \\
2MASSJ1731              & 17:31:30& +27:21:23& L0$^{\mathrm{e}}$    \\
2MASSIJ2107             & 21:07:32& -03:07:33& L0$^{\mathrm{e}}$    \\
2MASSJ2107              & 21:07:54& -45:44:06& L0$^{\mathrm{e}}$    \\
\hline
\hline
\end{tabular}
\end{center}
\begin{footnotesize}
$^{\mathrm{a}}$Spectral types from \citet{Comeron00} for the objects
of Cha I, and see further down for TWA members.\\
$^{\mathrm{b}}$Rotational velocities in km s$^{-1}$ from \citet{Joergens01} \\
$^{\mathrm{c}}$From \citet{Scholz05a} \\
$^{\mathrm{d}}$From \citet{Gizis02} \\
$^{\mathrm{e}}$As provided in Simbad \\
\end{footnotesize}
\end{table}


%% file: IRAC_Phot.tex
\begin{table*}
\begin{center}
\caption{Spitzer IRAC/MIPS photometry (when available) for the sources of Cha I observed.} \label{sources:IRAC}
\begin{tabular}{@{\extracolsep{-8pt}}lcccccc}
\hline\hline
Name&[3.6$\mu$m]&[4.5$\mu$m]&[5.8$\mu$m]&[8.0$\mu$m]&[24$\mu$m]&Class$^{\mathrm{a}}$ \\
\hline
Cha H$\alpha$ 1  & 11.53 $\pm$ 0.23 & 11.18 $\pm$ 0.23 & 10.76 $\pm$ 0.22 & 9.79  $\pm$ 0.20  & 5.94  $\pm$ 0.13 &II  \\
Cha H$\alpha$ 2  &  9.94 $\pm$ 0.20 & 9.59  $\pm$ 0.19 & 9.23  $\pm$ 0.19 & 8.58  $\pm$ 0.17  & 6.15  $\pm$ 0.13 &II  \\
Cha H$\alpha$ 3  & 10.67 $\pm$ 0.22 & 10.59 $\pm$ 0.21 & 10.48 $\pm$ 0.22 & 10.52 $\pm$ 0.22  &   ...    &III \\
Cha H$\alpha$ 4  & 10.70 $\pm$ 0.24 & 10.53 $\pm$ 0.24 & 10.37 $\pm$ 0.31 & 10.47 $\pm$ 0.29  &   ...     &III \\
Cha H$\alpha$ 5  & 10.30 $\pm$ 0.21 & 10.16 $\pm$ 0.21 & 10.10 $\pm$ 0.21 & 10.12 $\pm$ 0.21  &   ...     &III \\
Cha H$\alpha$ 6  & 10.38 $\pm$ 0.21 & 10.10 $\pm$ 0.20 & 9.83  $\pm$ 0.20 & 9.34  $\pm$ 0.19  & 6.51  $\pm$ 0.16 &II  \\
Cha H$\alpha$ 7  & 11.89 $\pm$ 0.24 & 11.70 $\pm$ 0.24 & 11.68 $\pm$ 0.25 & 11.58 $\pm$ 0.25  &   ...   &III \\
Cha H$\alpha$ 8  & 11.05 $\pm$ 0.22 & 10.92 $\pm$ 0.22 & 10.90 $\pm$ 0.23 & 10.88 $\pm$ 0.24  &   ...    &III \\
Cha H$\alpha$ 9  & 10.99 $\pm$ 0.22 & 10.55 $\pm$ 0.21 & 10.18 $\pm$ 0.21 & 9.60  $\pm$ 0.20  & 7.06  $\pm$ 0.15 &II  \\
Cha H$\alpha$ 12 & 11.34 $\pm$ 0.23 & 11.24 $\pm$ 0.23 & 10.83 $\pm$ 0.23 & 11.20 $\pm$ 0.24  &   ...   &III \\
\hline
2MJ1207                  &  11.40 $\pm$ 0.23 & 11.04 $\pm$ 0.23 & 10.59 $\pm$ 0.22 & 10.28 $\pm$ 0.21 & 8.06 $\pm$ 0.18 & III\\
2MJ1139                  &  10.94 $\pm$ 0.22 & 10.88 $\pm$ 0.22 & 10.72 $\pm$ 0.22 & 10.72 $\pm$ 0.22 & 9.73 $\pm$ 0.24 & III\\
\hline
\end{tabular}
\end{center}
\begin{footnotesize}
$^{\mathrm{a}}$ According to the IRAC [3.6]-[4.5] vs. [5.8]-[8.0] color-color diagram and the regions defined in \citet{Allen04}. In this scheme, Class III stands for diskless members and Class II are Classical TTauri stars or substellar analogs.\\
\end{footnotesize}
\end{table*}

%% file: MIR_var.tex
\begin{table}
\begin{center}
\caption{Density flux differences (in units of $erg/cm^2/s/A$) at
  $\sim$3.6 (WISE W1 and IRAC I1 channels) and $\sim$4.5 $\mu$m (WISE
  W2 and IRAC I2 channels) for sources
  with two epocs, and significance of the differences.} \label{MIRvar}
\begin{tabular}{lcccc}
\hline\hline
Name&$|W1_c-I1|$ & $ \frac{|W1-I1|}{\sqrt{eW1^2+eI1^2}} $ & $|W2_c-I2|$ & $\frac{|W2-I2|}{\sqrt{eW2^2+eI2^2}} $\\
\hline
Cha H$\alpha$ 3  & 2.83e-17 & 2.0 & 4.76e-18 & 0.8 \\
Cha H$\alpha$ 4  & 3.06e-18 & 0.1 & 7.40e-18 & 0.2 \\
Cha H$\alpha$ 5  & 4.50e-17 & 2.6 & 1.55e-19 & 0.0 \\
Cha H$\alpha$ 7  & 4.84e-18 & 0.6 & 3.25e-19 & 0.1 \\
Cha H$\alpha$ 8  & 2.36e-17 & 2.3 & 2.5e-18 & 0.5 \\
Cha H$\alpha$ 12 & 1.73e-17 & 2.4 & 9.63e-19 & 0.2 \\
\hline                                   
2MJ1207          & 1.84e-18 & 0.2 & ---     & ---- \\
2MJ1139          & 1.39e-17 & 1.4 & 1.04e-17 & 2.2 \\
\hline
\end{tabular}
\end{center}
\end{table}

%% file: SED_fit_param.tex
\begin{table*}
\footnotesize 
\begin{center}
\caption{Best fitting parameters obtained with VOSA from the SED
  fit. Step of the grid in T$_{\rm eff}$ is 100 K; 0.5 dex in $\log(g)$
  and 0.585 mag in A$_{\rm V}$} 

\label{chahachi}
%
\vspace{-0.2cm}

\begin{tabular}{@{\extracolsep{-8pt}}lcccccccccccccccc}
\hline\hline
  Object  & Model &T$_{\rm eff}^{1)}$ & $\mathcal{P}$(T$_{\rm eff}$)&$\log(g)$ & $\mathcal{P}(\log(g))$ & Md &$\dot{\chi^2}$ & F$_{\rm T}$$\pm$eF & F$_{\rm O}$/F$_{\rm T}$ & L$_{\rm bol}\pm$eL$_{\rm bol}$ & $\lambda_{\rm Max}^{2)}$ & N$_{\rm F}$/N$_{\rm T}$ & A$_{\rm V}^{1)}$ & $\mathcal{P}$(A$_{\rm V}$) & Age$^{3)}$ & Mass\\
\vspace{-0.3cm}\\
          &  & $10^2$K           &                               &         &          &  $10^{-20}$  &         &     $10^{-13}$erg/cm$^2$/s   &          & $10^{-4}$L$_{\odot}$                    &  $\mu$m          &                         &     mag             &          &Myr &10$^{-2}$M$_{\odot}$       \\    
\hline 
		& cond  &  28   & 0.29 &   3.5   & 0.61 & 1.1  &  9 & 372$\pm$48  & 0.73 & 297$\pm$94   & 2.2 & 10/20 &   2.9     & 0.53  & $<$1& $^{*)}$--- $^{6.1}_{5.2}$\\
Cha H$\alpha$1  & dust &  26   & 0.67 &   3.5   & 0.50 & 1.2  &   5 & 321$\pm$48  & 0.71 & 256$\pm$86   & 2.2 & 10/20 &   2.3     & 0.72  & $<$1&---\\
    		& bts   &  26   & 0.50 &   4.0   & 0.55 & 1.1  &  4 & 280$\pm$48  & 0.69 & 223$\pm$80   & 2.2 & 10/20 &   1.8     & 0.42  & $<$1&--- \\
\hline
\vspace{-0.3cm}\\
		& cond  &  32   & 0.32 &   4.5   & 0.50 & 3.5  &   7 & 2078$\pm$18 & 0.76 & 1658$\pm$325 & 2.2 & 10/20 & 4.7/4.1   & 0.52  &---&---\\
\vspace{-0.3cm}\\              
Cha H$\alpha$2  & dust &  30   & 0.44 &   4.0   & 0.56 & 4.0  &   6 & 1749$\pm$218 & 0.73 & 1396$\pm$276 & 2.2 & 10/20 &   4.1     & 0.63  & $<$1&---\\
\vspace{-0.3cm}\\              
    		& bts   & 32/30 & 0.37 &   4.0   & 0.57 & 3.0  &  7 & 1751$\pm$18 & 0.73 & 1397$\pm$276 & 2.2 & 10/20 &   4.1     & 0.47  & 1.8 $^{2.9}_{1.1}$ & 21 $^{25}_{18}$ \\
\vspace{-0.3cm}\\
\hline
		& cond  &  31   & 0.45 &   3.5   & 0.76 & 2.0  &  10 & 1054$\pm$20 & 0.83 & 841$\pm$174  & 7.6 & 20/20 &   2.9     & 0.70  &---&---\\
Cha H$\alpha$3  & dust &  26   & 0.50 &   4.5   & 0.67 & 2.8  &  14 & 759$\pm$20  & 0.83 & 605$\pm$129  & 7.6 & 20/20 &   1.8     & 0.55  & $<$1&---\\
    		& bts   &  27   & 0.36 &   4.5   & 0.90 & 2.2  &  10 & 676$\pm$20  & 0.80 & 539$\pm$117  & 7.6 & 20/20 &   1.2     & 0.39  & $<$1&---\\
\hline
		& cond  & 30/29 & 0.45 &   3.5   & 0.49 & 2.0  &   8 & 936$\pm$24 & 0.83 & 747$\pm$159  & 7.6 & 20/20 &   1.8     & 0.67  &---&---\\
Cha H$\alpha$4  & dust &  27   & 0.37 &   4.5   & 0.66 & 2.5  &   9 & 795$\pm$24  & 0.83 & 635$\pm$138  & 7.6 & 20/20 & 1.2/1.8   & 0.41  & $<$1&---\\
    		& bts   & 27/30 & 0.34 & 4.5/4.0 & 0.47 & 2.3  &  10 & 703$\pm$24  & 0.80 & 561$\pm$124  & 7.6 & 20/20 & 0.59/1.8  & 0.34  & $<$1&---\\
\hline
		& cond  &  30   & 0.35 &   3.5   & 0.66 & 3.0  &   8 & 1382$\pm$27 & 0.83 & 1102$\pm$228 &  12 & 20/20 &   2.9     & 0.63  &---&---\\
Cha H$\alpha$5  & dust &  30   & 0.30 & 3.5/4.5 & 0.43 & 3.0  &  10 & 1392$\pm$27 & 0.83 & 1110$\pm$230 &  12 & 20/20 &   2.9     & 0.48  & $<$1&---\\
    		& bts   &  29   & 0.32 &   4.5   & 0.50 & 2.9  &  10 & 1200$\pm$27 & 0.81 & 958$\pm$201 &  12 & 20/20 &   2.3     & 0.35  & $<$1&---\\
\hline
		& cond  &  28   & 0.34 &   3.5   & 0.63 & 3.1  &  16 & 1187$\pm$13 & 0.76 & 947$\pm$288  & 2.2 & 10/20 & 2.9/3.5   & 0.60  &$<$1&---\\
Cha H$\alpha$6  & dust &  28   & 0.54 &   3.5   & 0.64 & 3.1  &  14 & 1203$\pm$14 & 0.75 & 960$\pm$191 & 2.2 & 10/20 &   2.9     & 0.60  & $<$1&---\\
    		& bts   &  28   & 0.48 &   4.5   & 0.54 & 2.6  &  16 & 964$\pm$14 & 0.71 & 769$\pm$155  & 2.2 & 10/20 &   2.3     & 0.48  & $<$1&---\\
\hline
		& cond  &  20   & 0.18 &   4.5 & 0.37 & 2.0  &   8 & 186$\pm$7   & 0.93 & 149$\pm$34   & 12 & 20/20 &   1.8    & 0.28  &$<$1&---\\
Cha H$\alpha$7  & dust &  26   & 0.23 &   4.5  & 0.58 & 1.0  &  8 & 275$\pm$7   & 0.84 & 219$\pm$47   & 12 & 20/20 &   2.9    & 0.34  & $<$1&---\\
    		& bts   &  24   & 0.26 &  4.0  & 0.38 & 1.0 &   5 & 186$\pm$7   & 0.81 & 148$\pm$34   & 12 & 20/20 &   1.2     & 0.27  & $<$1&---\\
\hline
\vspace{-0.3cm}\\
		& cond  &  30   & 0.47 &   3.5   & 0.71 & 1.5  &  10 & 689$\pm$11  & 0.78 & 550$\pm$112  & 3.4 & 10/20 &   2.9     & 0.85  &---&---\\
\vspace{-0.3cm}\\
Cha H$\alpha$8  & dust &  30   & 0.31 &   3.5   & 0.51 & 1.5  & 11  & 695$\pm$11  & 0.77 & 555$\pm$113  & 3.4 & 10/20 &   2.9     & 0.54 &1.5 $^{2}_{1}$ & 10 $^{-}_{8.5}$\\
\vspace{-0.2cm}\\
    		& bts   & 30/29 & 0.28 &   4.0   & 0.45 & 1.5  & 10 & 705$\pm$11  & 0.76 & 562$\pm$114   & 3.4 & 10/20 & 2.9/2.3   & 0.34  & 2.0 $^{ 2.8}_{ 1.1}$ &11 $^{13}_{ 10.6}$\\
\vspace{-0.3cm}\\
\hline
\vspace{-0.3cm}\\
		& cond  & 32/30 & 0.23 & 4.5/4.0 & 0.39 & 1.6  &  10 & 1107$\pm$6  & 0.79 & 884$\pm$170  & 2.2 & 10/20 & 7.0/6.4   & 0.56  &---&---\\
\vspace{-0.3cm}\\
Cha H$\alpha$9  & dust &  30   & 0.28 &   4.0   & 0.43 & 1.9  &9 & 902$\pm$6   & 0.76 & 720$\pm$140  & 2.2 & 10/20 &   6.4     & 0.55&1 $^{1.5}_{<1}$&10 $^{-}_{8.5}$\\
\vspace{-0.2cm}\\
    		& bts   & 29/30 & 0.20 & 3.5/4.0 & 0.44 & 1.8  &  10 & 756$\pm$6   & 0.73 & 603$\pm$118  & 2.2 & 10/20 &   5.8     & 0.34  & 2.0 $^{3.0}_{1.0}$ & 10.0 $^{10.7}_{9.2}$\\
\vspace{-0.3cm}\\
\hline
		& cond  &  29   & 0.27 &   3.5   & 0.45 & 0.29 &  6 & 123$\pm$3   & 0.75 & 98$\pm$21   & 2.2 & 10/10 &   2.9     & 0.57  & 7.5 $^{9.8}_{6.2}$ & 6.7 $^{8}_{5.7}$\\
\vspace{-0.2cm}\\
Cha H$\alpha$11 & dust &  27   & 0.29 &   3.5   & 0.43 & 0.3 &5 & 106$\pm$3    & 0.73 & 85$\pm$18    & 2.2 & 10/10 &   2.3     & 0.38&$<$1 $^{2.5}_{<1}$&3 $^{3.5}_{2.7}$\\
    		& bts   & 25/26 & 0.31 & 3.5/4.0 & 0.50 & 0.30 &   4 & 69$\pm$3   & 0.68 & 55$\pm$13    & 2.2 & 10/10 & 0.59/1.2  & 0.30  & 1$^{3.0}_{<1}$&2.2$^{2.7}_{1.8}$\\
\hline
		& cond  &  25   & 0.28 &   4.5   & 0.46 & 1.6  &   8 & 469$\pm$10  & 0.86 & 295$\pm$64   &  12 & 20/20 &   1.8     & 0.50  &$<$1&---\\
Cha H$\alpha$12 & dust &  26   & 0.32 &   4.5   & 0.70 & 1.4  &   7 & 383$\pm$10  & 0.83 & 305$\pm$66   &  12 & 20/20 &   1.8     & 0.51  & $<$1&---\\
    		& bts   & 26/25 & 0.40 &   4.5   & 0.55 & 1.3  &   6 & 341$\pm$10  & 0.80 & 272$\pm$59   &  12 & 20/20 & 1.2/0.59  & 0.43  & $<$1&---\\
\hline
\vspace{-0.3cm}\\
		& cond  &  14   & 0.64 &   3.5   & 1.00 & 9.8  &   4 & 211$\pm$4   & 0.60 & 16$\pm$7     & 2.2 & 5/9   &           &       &$<$1&---\\
\vspace{-0.3cm}\\
 SSSPMJ1102     & dust &  19   & 0.56 &   4.5   & 0.95 & 2.7  & 0.3 & 228$\pm$4   & 0.56 & 18$\pm$7     & 2.2 & 5/9   &           &       & $<$1&---\\
\vspace{-0.3cm}\\
 		& bts   &  24   & 1.00 &   3.5   & 0.86 & 1.3  &  10 & 239$\pm$4   & 0.53 & 19$\pm$8     & 2.2 & 5/9   &           &       & 4.0 $^{28.3}_{1.0}$& 1.8 $^{3}_{1.2}$\\
\vspace{-0.3cm}\\
\hline
\vspace{-0.3cm}\\
		& cond  &  17   & 0.30 &   3.5   & 0.94 & 4.5  &   5 & 217$\pm$11  & 0.83 & 20$\pm$3     & 4.4 & 9/20  &           &       &$<$1&---\\
\vspace{-0.3cm}\\
 2MJ1207        & dust & 20/21 & 0.64 &   4.0   & 0.61 & 2.3  & 1.0 & 232$\pm$11  & 0.78 & 21$\pm$3     & 4.4 & 9/20  &           &       & $<$1&---\\
\vspace{-0.3cm}\\
 		& bts   &  25   & 0.86 &   3.5   & 0.70 & 1.2  &  3 & 258$\pm$11  & 0.70 & 23$\pm$4     & 4.4 & 9/20  &            &       &10.0 $^{15.7}_{2.5}$&2.3 $^{3}_{2}$\\
\vspace{-0.3cm}\\
\hline
\vspace{-0.3cm}\\
		& cond  &  18   & 0.54 &   3.5   & 0.92 & 5.2  &  20 & 304$\pm$14  & 0.87 & 24$\pm$11    &  23 & 10/10 &           &       &$<1$&---\\
\vspace{-0.3cm}\\
 2MJ1139        & dust & 20/21 & 0.49 & 4.5/3.5 & 0.34 & 3.3  &   4 & 316$\pm$14  & 0.84 & 25$\pm$11    &  23 & 10/10 &           &       & $<$1&---\\
\vspace{-0.3cm}\\
 		& bts   &  24   & 0.49 &   3.5   & 0.78 & 1.9  &  7 & 353$\pm$14  & 0.75 & 27$\pm$12    &  23 & 10/10 &           &       &1.7 $^{15.1}_{1}$&1.5 $^{2.3}_{1.2}$\\
\vspace{-0.3cm}\\
\hline
\end{tabular}
\end{center}
\vspace{-0.1cm}
\flushleft
$^{1)}$ Parameter estimation determined from $\chi^2$ minimization /
PDF maximuum (whenever both estimations do not coincide).\\
$^{2)}$ Wavelength where VOSA detects the excess (in $\mu$m).\\
$^{3)}$ Sub-superscripts in the age and mass provide the range of
possible values interpolated in the isochrones and evolutionary tracks
taking into account the error in the bolometric luminosity and half
the step of the models in the effective temperature as uncrtainty in
the latter.\\
$^{*)}$ Area in the HR diagram not covered by the isochrones and
evolutionary tracks.
\normalsize
\end{table*}

%% file: best_fit_spec_table_comparison.tex
\begin{table*}
\footnotesize
\begin{center}
\caption{Best fitting parameters obtained via optical + near-infrared, only optical and only near-infrared spectral fits. The grid of models has been linearly interpolated so that the step in T$_{\rm eff}$ is 50 K and the chosen step in A$_{\rm V}$ is 0.1 mag. No interpolation has been done in $\log(g)$ since a more precise determination has been obtained via the detailed gravity sensitive lines comparison in Section~\ref{age} displayed on Table~\ref{PysicPar1}. 
} 
\label{specfitparam}
\begin{tabular}{@{\extracolsep{-6pt}}lccccc|cccc|cccc}
&&\multicolumn{4}{|l}{optical + near-infrared}& \multicolumn{4}{|l}{optical}&\multicolumn{4}{|l}{near-infrared}\\
\hline\hline
  Object  & Model &T$_{\rm eff}$      & A$_{\rm V}$    &$\dot{\chi^2}$   &$\log(g)$ &T$_{\rm eff}$      & A$_{\rm V}$    &$\dot{\chi^2}$   &$\log(g)$ &T$_{\rm eff}$      & A$_{\rm V}$    &$\dot{\chi^2}$   &$\log(g)$ \\
          &       & K                 & mag            &                 & dex      & K                 & mag            &                 & dex      & K                 & mag            &                 & dex      \\    
\hline 
&AMES-COND&2900$_{-150}^{+100}$&3.8$_{-0.4}^{+0.8}$&2.6&3.0&3450$_{-200}^{+25}$&6.1$_{-0.9}^{+0.05}$&0.52&4.5&2800$_{-100}^{+150}$&1.7$_{-0.3}^{+0.2}$&0.26&5.0\\
\vspace{-0.25cm}\\
Cha H$\alpha$1&AMES-DUSTY&2600$_{-25}^{+150}$&2.8$_{-0.3}^{+0.8}$&1.4&3.5&3150$_{-150}^{+25}$&4.7$_{-1.1}^{+0.2}$&0.51&4.0&2650$_{-50}^{+100}$&1.5$_{-0.3}^{+0.2}$&0.22&5.0\\
\vspace{-0.25cm}\\
&BT-Settl&2500$_{-50}^{+100}$&1.1$_{-0.4}^{+0.8}$&0.83&3.0&2950$_{-150}^{+100}$&3.8$_{-1.0}^{+0.1}$&0.48&3.5&2650$_{-150}^{+100}$&1.0$_{-0.3}^{+0.3}$&0.32&5.0\\
\hline
&AMES-COND&3050$_{-100}^{+150}$&5.2$_{-0.5}^{+0.5}$&1.2&4.0&3600$_{-150}^{+25}$&7.0$_{-0.7}^{+0.4}$&0.32&5.0&3250$_{-200}^{+300}$&3.8$_{-0.6}^{+0.2}$&0.23&5.5\\
\vspace{-0.25cm}\\
Cha H$\alpha$2&AMES-DUSTY&2950$_{-150}^{+50}$&4.5$_{-0.6}^{+0.5}$&0.87&3.5&3600$_{-150}^{+25}$&7.1$_{-0.9}^{+0.3}$&0.31&5.0&3150$_{-200}^{+350}$&3.7$_{-0.4}^{+0.3}$&0.23&5.5\\
\vspace{-0.25cm}\\
&BT-Settl&2850$_{-150}^{+100}$&3.5$_{-0.7}^{+0.8}$&0.96&3.0&3600$_{-250}^{+25}$&7.2$_{-1.4}^{+0.5}$&0.40&4.5&3150$_{-250}^{+300}$&3.3$_{-0.8}^{+0.3}$&0.26&5.0\\
\hline
&AMES-COND&2950$_{-150}^{+100}$&3.6$_{-0.4}^{+0.3}$&1.8&3.5&3550$_{-200}^{+50}$&6.1$_{-0.9}^{+0.05}$&0.37&4.5&3150$_{-200}^{+200}$&1.3$_{-0.5}^{+0.5}$&0.33&3.0\\
\vspace{-0.25cm}\\
Cha H$\alpha$3&AMES-DUSTY&2750$_{-100}^{+150}$&2.8$_{-0.6}^{+0.6}$&1.1&3.5&3100$_{-100}^{+50}$&3.9$_{-0.5}^{+0.05}$&0.43&4.0&2850$_{-200}^{+200}$&1.6$_{-0.3}^{+0.3}$&0.35&5.5\\
\vspace{-0.25cm}\\
&BT-Settl&2650$_{-100}^{+100}$&1.6$_{-0.9}^{+0.6}$&0.99&3.0&2950$_{-100}^{+50}$&2.7$_{-0.7}^{+0.05}$&0.57&3.5&2750$_{-150}^{+350}$&1.1$_{-0.4}^{+0.3}$&0.41&5.0\\
\hline
&AMES-COND&3000$_{-100}^{+100}$&3.0$_{-0.6}^{+0.3}$&1.00&4.0&3200$_{-100}^{+100}$&3.9$_{-0.3}^{+0.05}$&0.52&4.5&3200$_{-150}^{+200}$&1.5$_{-0.4}^{+0.3}$&0.22&4.5\\
\vspace{-0.25cm}\\
Cha H$\alpha$4&AMES-DUSTY&2900$_{-100}^{+100}$&2.2$_{-0.4}^{+0.4}$&0.67&3.5&3000$_{-50}^{+50}$&2.7$_{-0.4}^{+0.05}$&0.53&4.0&3100$_{-200}^{+250}$&1.5$_{-0.2}^{+0.3}$&0.20&5.0\\
\vspace{-0.25cm}\\
&BT-Settl&2800$_{-100}^{+50}$&1.4$_{-0.7}^{+0.1}$&0.84&3.5&2750$_{-100}^{+100}$&1.5$_{-0.6}^{+0.05}$&0.97&3.5&3000$_{-150}^{+200}$&0.9$_{-0.4}^{+0.3}$&0.25&3.5\\
\hline
&AMES-COND&2950$_{-100}^{+150}$&4.2$_{-0.5}^{+0.5}$&1.2&4.0&3450$_{-250}^{+100}$&6.1$_{-1.3}^{+0.05}$&0.55&4.5&3100$_{-100}^{+150}$&2.7$_{-0.2}^{+0.2}$&0.15&4.5\\
\vspace{-0.25cm}\\
Cha H$\alpha$5&AMES-DUSTY&2900$_{-150}^{+100}$&3.6$_{-0.5}^{+0.4}$&0.79&3.5&3350$_{-300}^{+200}$&5.7$_{-1.8}^{+0.4}$&0.54&4.5&3000$_{-150}^{+200}$&2.7$_{-0.2}^{+0.3}$&0.13&5.0\\
\vspace{-0.25cm}\\
&BT-Settl&2800$_{-150}^{+100}$&2.7$_{-0.9}^{+0.6}$&1.0&3.0&3200$_{-150}^{+100}$&4.9$_{-1.0}^{+0.05}$&0.75&4.0&2950$_{-200}^{+150}$&2.1$_{-0.3}^{+0.3}$&0.18&3.5\\
\hline
&AMES-COND&2900$_{-150}^{+100}$&3.8$_{-0.4}^{+0.8}$&2.6&3.0&3450$_{-200}^{+100}$&6.1$_{-1.0}^{+0.05}$&0.48&4.5&2850$_{-150}^{+400}$&1.8$_{-0.9}^{+0.3}$&0.43&5.5\\
\vspace{-0.25cm}\\
Cha H$\alpha$6&AMES-DUSTY&2650$_{-50}^{+150}$&3.1$_{-0.6}^{+0.6}$&1.7&3.5&3350$_{-300}^{+25}$&5.8$_{-1.7}^{+0.3}$&0.47&4.5&2650$_{-100}^{+200}$&1.5$_{-0.3}^{+0.4}$&0.40&5.5\\
\vspace{-0.25cm}\\
&BT-Settl&2500$_{-50}^{+100}$&1.2$_{-0.6}^{+0.9}$&1.3&3.0&3100$_{-50}^{+150}$&4.8$_{-0.8}^{+0.1}$&0.63&4.0&2650$_{-150}^{+200}$&1.1$_{-0.4}^{+0.3}$&0.50&5.0\\
\hline
&AMES-COND&2000$_{-25}^{+50}$&1.5$_{-0.3}^{+0.05}$&3.3&5.5&2050$_{-50}^{+100}$&0.0$_{0.0}^{+0.7}$&2.9&5.5&2700$_{-100}^{+150}$&1.5$_{-0.3}^{+0.05}$&0.31&3.5\\
\vspace{-0.25cm}\\
Cha H$\alpha$7&AMES-DUSTY&2000$_{-25}^{+25}$&0.20$_{-0.2}^{+0.3}$&2.0&5.5&2000$_{-25}^{+25}$&0.0$_{-0.0}^{+0.4}$&1.2&5.5&2600$_{-50}^{+150}$&1.5$_{-0.4}^{+0.05}$&0.26&3.5\\
\vspace{-0.25cm}\\
&BT-Settl&2150$_{-100}^{+50}$&0.0$_{-0.0}^{+0.4}$&2.1&5.0&2000$_{-25}^{+25}$&0.0$_{-0.0}^{+0.9}$&1.5&5.0&2550$_{-150}^{+150}$&0.70$_{-0.4}^{+0.4}$&0.36&3.0\\
\hline
&AMES-COND&2900$_{-100}^{+150}$&4.1$_{-0.3}^{+0.7}$&2.1&3.5&3450$_{-100}^{+150}$&6.1$_{-0.8}^{+0.05}$&0.62&5.0&2950$_{-150}^{+200}$&2.3$_{-0.3}^{+0.2}$&0.20&4.0\\
\vspace{-0.25cm}\\
Cha H$\alpha$8&AMES-DUSTY&2750$_{-100}^{+150}$&3.4$_{-0.7}^{+0.5}$&1.1&3.5&3550$_{-300}^{+50}$&6.1$_{-1.3}^{+0.05}$&0.60&4.5&2850$_{-150}^{+150}$&2.2$_{-0.3}^{+0.2}$&0.17&4.0\\
\vspace{-0.25cm}\\
&BT-Settl&2600$_{-100}^{+100}$&1.7$_{-0.9}^{+0.8}$&0.86&3.0&3150$_{-100}^{+150}$&4.8$_{-0.9}^{+0.1}$&0.70&4.0&2800$_{-200}^{+150}$&1.6$_{-0.3}^{+0.3}$&0.23&3.5\\
\hline
&AMES-COND&3000$_{-200}^{+150}$&7.2$_{-0.6}^{+0.8}$&2.0&3.5&3050$_{-200}^{+550}$&6.2$_{-1.7}^{+2.9}$&1.8&4.0&3000$_{-150}^{+100}$&6.4$_{-0.2}^{+0.2}$&0.23&5.5\\
\vspace{-0.25cm}\\
Cha H$\alpha$9&AMES-DUSTY&2900$_{-200}^{+100}$&6.8$_{-0.9}^{+0.7}$&1.7&3.5&2800$_{-200}^{+350}$&4.3$_{-2.1}^{+2.5}$&1.5&3.5&2850$_{-150}^{+150}$&6.3$_{-0.3}^{+0.2}$&0.23&5.5\\
\vspace{-0.25cm}\\
&BT-Settl&2800$_{-200}^{+200}$&5.8$_{-1.2}^{+1.1}$&2.1&3.0&2250$_{-25}^{+650}$&0.0$_{-0.0}^{+11}$&2.2&5.0&2750$_{-100}^{+200}$&5.7$_{-0.3}^{+0.3}$&0.28&5.0\\
\hline
&AMES-COND&2800$_{-550}^{+100}$&3.7$_{-2.5}^{+0.5}$&3.6&3.0&3150$_{-250}^{+400}$&5.7$_{-1.6}^{+0.4}$&1.6&4.0&3200$_{-250}^{+200}$&0.82$_{-0.6}^{+0.4}$&0.36&4.0\\
\vspace{-0.25cm}\\
Cha H$\alpha$11&AMES-DUSTY&2050$_{-25}^{+550}$&0.0$_{-0.0}^{+3.1}$&2.2&5.5&2800$_{-150}^{+200}$&3.3$_{-1.7}^{+1.6}$&1.2&3.5&3000$_{-200}^{+300}$&0.81$_{-0.3}^{+0.4}$&0.33&5.5\\
\vspace{-0.25cm}\\
&BT-Settl&2450$_{-100}^{+50}$&0.9$_{-0.9}^{+0.6}$&1.6&3.0&2550$_{-550}^{+150}$&1.5$_{-1.5}^{+0.05}$&1.4&3.0&3000$_{-200}^{+250}$&0$_{-0.1}^{+0.6}$&0.39&3.0\\
\hline
&AMES-COND&2100$_{-25}^{+150}$&0.0$_{-0.0}^{+0.1}$&24&3.0&---&---&---&---&3000$_{-300}^{+300}$&1.2$_{-0.8}^{+0.6}$&0.70&5.0\\
\vspace{-0.25cm}\\
Cha H$\alpha$12&AMES-DUSTY&2050$_{-50}^{+50}$&0.0$_{-0.0}^{+0.1}$&18&3.5&---&---&---&---&2850$_{-250}^{+400}$&1.2$_{-0.6}^{+0.5}$&0.67&5.0\\
\vspace{-0.25cm}\\
&BT-Settl&2100$_{-25}^{+25}$&0.0$_{-0.0}^{+0.1}$&9.0&3.0&---&---&---&---&2900$_{-300}^{+300}$&0.60$_{-0.6}^{+0.5}$&0.75&3.0\\
\hline
&AMES-COND&3200$_{-50}^{+100}$&0.&1.3&3.0&---&---&---&---&3200$_{-50}^{+100}$&0.&1.3&3.0\\
\vspace{-0.25cm}\\
2MJ1139&AMES-DUSTY&2600$_{-450}^{+350}$&0.&1.4&5.0&---&---&---&---&2600$_{-450}^{+350}$&0.&1.4&5.0\\
\vspace{-0.25cm}\\
&BT-Settl&2500$_{-200}^{+450}$&0.&1.2&3.0&---&---&---&---&2500$_{-200}^{+450}$&0.&1.2&3.0\\
\hline
&AMES-COND&2800$_{-250}^{+150}$&0.&0.88&5.0&---&---&---&---&2800$_{-250}^{+150}$&0.&0.88&5.0\\
\vspace{-0.25cm}\\
2MJ1207&AMES-DUSTY&2600$_{-200}^{+150}$&0.&0.70&5.0&---&---&---&---&2600$_{-200}^{+150}$&0.&0.70&5.0\\
\vspace{-0.25cm}\\
&BT-Settl&2500$_{-100}^{+100}$&0.&0.46&3.0&---&---&---&---&2500$_{-100}^{+100}$&0.&0.46&3.0\\
\hline
&AMES-COND&3150$_{-25}^{+25}$&0.&1.1&3.0&---&---&---&---&3150$_{-25}^{+25}$&0.&1.1&3.0\\
\vspace{-0.25cm}\\
SSSPMJ1102&AMES-DUSTY&2250$_{-100}^{+350}$&0.&0.87&5.5&---&---&---&---&2250$_{-100}^{+350}$&0.&0.87&5.5\\
\vspace{-0.25cm}\\
&BT-Settl&2450$_{-100}^{+350}$&0.&0.77&3.0&---&---&---&---&2450$_{-100}^{+350}$&0.&0.77&3.0\\
\hline
\end{tabular}
\end{center}
\normalsize
\end{table*}

%% file: logg_table.tex
\begin{table}
\begin{center}
\caption{Surface gravity determination through K I doublet analysis;
  masses, radii and distances derived combining these determinatinos
  with the results from Sections~\ref{sec:SEDfit} and~\ref{specfit}
  (via Eqs.~\ref{eq:distance} and ~\ref{eq:radii}); and previously
  published estimations of radii for most of the Cha I members.}

\label{PysicPar1}
\begin{small}
\begin{tabular}{@{\extracolsep{-8pt}}lrcccccc}
Name & $\log(g)$ & Age$^{\mathrm{a}}$ & M(M$_\odot$)$^{\mathrm{a}}$ & R(R$_\odot$)$^{\mathrm{b}}$ &R(R$_\odot$)$^{\mathrm{c}}$ &R(R$_\odot$)$^{\mathrm{d}}$ &D(pc)$^{\mathrm{e}}$ \\
\hline
\hline
\noalign{\smallskip}
Cha H$\alpha$ 1  & 4.00$\pm$0.25  &   8.5 & 0.03  & 0.74 & $\le$0.46  & 0.29  &   68\\
Cha H$\alpha$ 2  & 4.00$\pm$0.25  &   5.5 & 0.14  & 1.23 & $\le$0.73  & 0.62  &   83\\
Cha H$\alpha$ 3  & 4.00$\pm$0.25  &   7.2 & 0.04  & 1.05 & $\le$0.77  & 0.33  &   50\\
Cha H$\alpha$ 4  & 4.25$\pm$0.25  &  12.0 & 0.08  & 1.07 & $\le$0.89  & 0.35  &   61\\
Cha H$\alpha$ 5  & 3.75$\pm$0.25  &   2.9 & 0.08  & 1.21 & $\le$0.83  & 0.63  &   82\\
Cha H$\alpha$ 6  & 3.75$\pm$0.25  &   1.0 & 0.03  & 1.14 & $\le$0.68  & 0.36  &   48\\
Cha H$\alpha$ 7  & 4.00$\pm$0.25  &   9.0 & 0.03  & 0.71 & $\le$0.37  & 0.26  &   66\\
Cha H$\alpha$ 8  & 4.00$\pm$0.25  &   7.0 & 0.05  & 0.87 & $\le$0.59  & 0.35  &   66\\
Cha H$\alpha$ 9  & 4.00$\pm$0.25  &   7.2 & 0.04  & 0.95 &  --        & 0.33  &   55\\
Cha H$\alpha$ 11 & 4.50$\pm$0.25  &  22.0 & 0.08  & 0.39 &  --        & 0.26  &  157\\
Cha H$\alpha$ 12 & 4.00$\pm$0.25  &   5.5 & 0.06  & 0.81 & $\le$0.66  & 0.41  &  101\\
\hline
2MJ1139          & 4.25$\pm$0.25  &  16.0 & 0.03  & 0.11 &  --        & 0.20  &   35\\
2MJ1207          & 4.00$\pm$0.25  &  10.0 & 0.02  & 0.26 &  --        & 0.25  &   52\\
SSSPMJ1102       & 3.75$\pm$0.25  &$<$1.0 & 0.01  & 0.11 &  --        & 0.22  &   44\\
\noalign{\smallskip}
\hline
\hline
\end{tabular}
\end{small}
\end{center}
\begin{footnotesize}
\vspace*{0.2cm}
$^{\mathrm{a}}$Age in Myr from isochrones of \citet{Baraffe98,Baraffe02}.\\
$^{\mathrm{b}}$Derived from the dilution factor estimated in the SED
fitting (Table 5) combined via Eq.~\ref{eq:distance} with a distance of 160 pc \citep{Wichmann98,Knude98}
for Cha I members and individualy estimated distances from \cite{Faherty09} for
the TWA sources (18, 54 and 22 for 2MJ1139, 2MJ1207 and SSSPMJ1102, respectively).\\
$^{\mathrm{c}}$Upper limits derived with $v \sin i$ values, from \citet{Joergens01}.\\
$^{\mathrm{d}}$Calculated from the mass and log g$_*$ extracted from
the evolutionary tracks and K I doublet analysis (applying Eq.~\ref{eq:radii}).\\
$^{\mathrm{e}}$Derived from the flux ratio between the model and the observation and using the radii from the previous column and
Eq.~\ref{eq:distance}\\
\end{footnotesize}
\end{table}

%% file: HaEW_Table.tex
\begin{table}
\begin{center}
\caption{Spectral types and equivalent width of H$\alpha$ from the
  literature and this work for the Cha I sample.} \label{Halpha}
\begin{tabular}{lcccc}
Name&SpT$^{\mathrm{a}}$&SpT$^{\mathrm{b}}$&EW(H$\alpha$)$^{\mathrm{c}}$&EW(H$\alpha$)$^{\mathrm{d}}$\\
\hline
\hline
Cha H$\alpha$ 1  & M 7.5    &     M 7.75   & 35   & 32 \\
Cha H$\alpha$ 2  & M 6.5    &     M 5.25   & 33   & 11 \\
Cha H$\alpha$ 3  & M 7      &     M 5.5    & 10   &  7 \\
Cha H$\alpha$ 4  & M 6      &     M 5.5    & --   &  8 \\
Cha H$\alpha$ 5  & M 6      &     M 5.5    & 6.5  &  8 \\
Cha H$\alpha$ 6  & M 7      &     M 5.75   & 48   & 35 \\
Cha H$\alpha$ 7  & M 8      &     M 7.75   & --   &  5 \\
Cha H$\alpha$ 8  & M 6.5    &     M 5.75   & --   &  9 \\
Cha H$\alpha$ 9  & M 6      &     M 5.5    & --   &  7 \\
Cha H$\alpha$ 11 & M 8      &     M 7.25   & --   & 17 \\
Cha H$\alpha$ 12 & M 7      &     M 6.5    & --   & 22 \\
\hline
2MJ1139&\multicolumn{2}{c}{M 8$^{\mathrm{e}}$} & \multicolumn{2}{c}{10.2$\pm$0.7$^{\mathrm{f}}$}  \\
2MJ1207&\multicolumn{2}{c}{M 8$^{\mathrm{e}}$} & \multicolumn{2}{c}{44.7$\pm$2.0$^{\mathrm{f}}$}  \\
\hline
\hline
\end{tabular}
\end{center}
\begin{footnotesize}
$^{\mathrm{a}}$From \citet{Comeron00}.\\
$^{\mathrm{b}}$From \citet{Luhman07b}\\
$^{\mathrm{c}}$From \citet{Natta04} \\
$^{\mathrm{d}}$From this work revising the spectra analyzed in \cite{Comeron00} \\
$^{\mathrm{e}}$From \citet{Gizis02} \\
$^{\mathrm{f}}$From \citet{Barrado04} \\
\end{footnotesize}
\end{table}

%% file: SED_field_fit_param.tex
\begin{table*}
\footnotesize 
\begin{center}
\caption{Best fitting parameters obtained with VOSA for the field
  dwarfs from the SED
  fit. Step of the grid in T$_{\rm eff}$ is 100 K; 0.5 dex in $\log(g)$
  and 0.585 mag in A$_{\rm V}$} 

\label{chahachi}
%
\vspace{-0.2cm}

\begin{tabular}{@{\extracolsep{-2pt}}lccccccccccc}
\hline\hline
  Object  & Model &T$_{\rm eff}^{1)}$ & $\mathcal{P}$(T$_{\rm eff}$)&$\log(g)$ & $\mathcal{P}(\log(g))$ & Md &$\dot{\chi^2}$ & F$_{\rm T}$$\pm$eF & F$_{\rm O}$/F$_{\rm T}$ & $\lambda_{\rm Max}^{2)}$ & N$_{\rm F}$/N$_{\rm T}$ \\
\vspace{-0.3cm}\\
          &  & K           &                               &         &          &    &         &     $10^{-13}$erg/cm$^2$/s   &          &  $\mu$m          &                         \\
\hline 
& AMES-Cond  & 2600/2600 & 0.99 & 6.0/6.0 & 0.99 & 0.22e-20  &  90 & 60$\pm$0.9   & 0.46 &  12 & 10/10 \\ 
2MASPJ125     & AMES-Dusty & 2600/2600 & 1.00 & 6.0/6.0 & 0.97 & 0.25e-20  &  20 & 60$\pm$0.9   & 0.44 &  12 & 10/10 \\ 
& BT-Settl   & 2700/2700 & 1.00 & 5.5/5.5 & 0.98 & 0.24e-20  &   8 & 70$\pm$0.9   & 0.41 &  12 & 10/10 \\ 
\hline
&AMES-Cond  & 1700/1700 & 1.00 & 4.5/4.5 & 1.00 & 1.8e-20   &  40 & 80$\pm$4     & 0.82 &  12 & 10/10 \\ 
2MASSIJ2107   & AMES-Dusty & 2200/2200 & 1.00 & 6.0/6.0 & 0.98 & 0.68e-20  &   8 & 90$\pm$4     & 0.74 &  12 & 10/10 \\ 
&BT-Settl   & 2400/2400 & 1.00 & 4.5/4.5 & 0.81 & 0.47e-20  &  30 & 90$\pm$4     & 0.73 &  12 & 10/10 \\ 
\hline
&AMES-Cond  & 2000/2100 & 0.24 & 4.5/4.5 & 0.50 & 0.24e-20  &   2 & 20$\pm$1.    & 0.72 &  12 & 8/9   \\ 
2MASSJ0858    & AMES-Dusty & 2300/2300 & 0.31 & 5.0/4.5 & 0.26 & 0.14e-20  &   2 & 20$\pm$1.    & 0.67 &  12 & 8/9   \\ 
&BT-Settl   & 2600/2600 & 0.31 & 4.5/4.5 & 0.35 & 0.096e-20 &  1. & 20$\pm$1.    & 0.61 &  12 & 8/9   \\ 
\hline
&AMES-Cond  & 1800/1800 & 0.98 & 4.5/4.5 & 0.98 & 1.1e-20   &  50 & 60$\pm$1.0   & 0.48 &  12 & 10/10 \\ 
2MASSJ1239    & AMES-Dusty & 2300/2300 & 0.56 & 6.0/6.0 & 0.56 & 0.39e-20  &  30 & 60$\pm$1.0   & 0.48 &  12 & 10/10 \\ 
&BT-Settl   & 2500/2500 & 1.00 & 4.5/4.5 & 0.78 & 0.29e-20  &  40 & 70$\pm$1.0   & 0.46 &  12 & 10/10 \\ 
\hline
&AMES-Cond  & 2100/2100 & 0.79 & 5.0/5.0 & 0.80 & 0.21e-20  &  10 & 20$\pm$0.6   & 0.46 & 4.6 & 10/10 \\ 
2MASSJ1434    & AMES-Dusty & 2400/2400 & 0.85 & 6.0/6.0 & 0.85 & 0.11e-20  &  10 & 20$\pm$0.6   & 0.46 & 4.6 & 10/10 \\ 
&BT-Settl   & 2600/2600 & 0.83 & 5.5/5.5 & 0.57 & 0.095e-20 &   9 & 20$\pm$0.6   & 0.43 & 4.6 & 10/10 \\ 
\hline
&AMES-Cond  & 1700/1700 & 1.00 & 4.5/4.5 & 1.00 & 11e-20    & 100 & 500$\pm$30   & 0.58 &  12 & 10/10 \\ 
2MASSJ1731    & AMES-Dusty & 2200/2200 & 1.00 & 6.0/6.0 & 1.00 & 4.3e-20   &  90 & 500$\pm$30   & 0.54 &  12 & 10/10 \\ 
&BT-Settl   & 2400/2400 &  1.0 & 4.5/4.5 & 0.99 & 2.9e-20   &  90 & 500$\pm$30   & 0.53 &  12 & 10/10 \\ 
\hline
&AMES-Cond  & 1200/1200 & 1.00 & 4.5/4.5 & 1.00 & 2.7e-20   & 500 & 40$\pm$0.8   & 0.64 & 4.6 & 6/6   \\ 
2MASSJ2107    & AMES-Dusty & 1900/1900 & 1.00 & 6.0/6.0 & 1.00 & 0.57e-20  &  60 & 50$\pm$0.8   & 0.53 & 4.6 & 6/6   \\ 
&BT-Settl   & 2000/2000 & 1.00 & 4.5/4.5 & 1.00 & 0.54e-20  &  10 & 50$\pm$0.8   & 0.52 & 4.6 & 6/6   \\ 
\hline
&AMES-Cond  & 1300/1300 & 1.00 & 4.5/4.5 &  1.0 & 83e-20    & 200 & 1000$\pm$10  & 0.44 & 2.2 & 5/5   \\ 
AZCnc         & AMES-Dusty & 2100/2100 & 0.99 & 6.0/6.0 & 0.99 & 13e-20    & 100 & 1000$\pm$10  & 0.43 & 2.2 & 5/5   \\ 
&BT-Settl   & 1900/1900 & 0.94 & 4.5/4.5 & 0.88 & 16e-20    &  20 & 1000$\pm$10  & 0.44 & 2.2 & 5/5   \\ 
\hline
&AMES-Cond  & 1800/1800 & 0.60 & 4.5/4.5 & 1.00 & 18e-20    & 100 & 1000$\pm$30  & 0.73 & 2.2 & 6/6   \\ 
GJ3517        & AMES-Dusty & 2100/2100 & 1.00 & 5.0/5.0 & 1.00 & 12e-20    &  10 & 1000$\pm$30  & 0.66 & 2.2 & 6/6   \\ 
&BT-Settl   & 2400/2400 & 0.62 & 4.5/4.5 & 0.96 & 6.7e-20   &  40 & 1000$\pm$30  & 0.64 & 2.2 & 6/6   \\ 
\hline
&AMES-Cond  & 2200/2200 & 0.72 & 4.5/4.5 & 0.93 & 31e-20    &   5 & 4000$\pm$100 & 0.71 & 2.2 & 7/7   \\ 
GJ644C        & AMES-Dusty & 2300/2300 & 0.77 & 4.5/4.5 & 0.43 & 27e-20    &   2 & 4000$\pm$100 & 0.67 & 2.2 & 7/7   \\ 
&BT-Settl   & 2700/2700 & 0.88 & 4.5/4.5 & 0.59 & 16e-20    &   5 & 5000$\pm$100 & 0.61 & 2.2 & 7/7   \\ 
\hline
&AMES-Cond  & 2600/2600 & 1.00 & 4.5/4.5 &  1.0 & 7.7e-20   & 700 & 2000$\pm$30  & 0.59 & 2.2 & 7/7   \\ 
LHS234        & AMES-Dusty & 2600/2600 &  1.0 & 4.5/4.0 & 1.00 & 8.7e-20   & 400 & 2000$\pm$30  & 0.55 & 2.2 & 7/7   \\ 
&BT-Settl   & 2700/2700 & 1.00 & 4.5/4.5 &  1.0 & 8.4e-20   & 300 & 3000$\pm$30  & 0.51 & 2.2 & 7/7   \\ 
\hline
&AMES-Cond  & 2000/2000 & 0.86 & 4.5/4.5 & 1.00 & 6.5e-20   &  60 & 600$\pm$20   & 0.73 & 2.2 & 8/8   \\ 
LHS2397a      & AMES-Dusty & 2200/2100 & 0.70 & 5.0/4.0 & 0.91 & 4.8e-20   &  30 & 600$\pm$20   & 0.68 & 2.2 & 8/8   \\ 
&BT-Settl   & 2400/2400 & 0.72 & 4.5/4.5 & 0.73 & 3.5e-20   &  20 & 700$\pm$20   & 0.64 & 2.2 & 8/8   \\ 
\hline
&AMES-Cond  & 3700/3700 & 0.45 & 4.5/4.5 & 1.00 & 0.059e-20 &  20 & 70$\pm$1.    & 0.50 & 4.6 & 8/8   \\ 
LP803-33      & AMES-Dusty & 3700/3700 & 0.52 & 4.5/4.5 & 0.75 & 0.060e-20 &  20 & 70$\pm$1.    & 0.50 & 4.6 & 8/8   \\ 
&BT-Settl   & 3600/3600 & 0.60 & 4.5/4.5 & 0.51 & 0.068e-20 &  40 & 70$\pm$1.    & 0.49 & 4.6 & 8/8   \\ 
\hline
&AMES-Cond  & 2400/2400 & 0.95 & 4.5/4.5 & 1.00 & 11e-20    & 400 & 2000$\pm$30  & 0.54 & 2.2 & 6/6   \\ 
SCRJ0702-6102 & AMES-Dusty & 2600/2600 & 1.00 & 5.0/5.0 & 0.59 & 7.6e-20   & 200 & 2000$\pm$30  & 0.54 & 2.2 & 6/6   \\ 
&BT-Settl   & 2600/2600 & 1.00 & 4.5/4.5 & 1.00 & 9.5e-20   & 100 & 2000$\pm$30  & 0.47 & 2.2 & 6/6   \\ 
\hline
&AMES-Cond  & 2400/2400 & 0.99 & 4.5/4.5 & 1.00 & 4.7e-20   & 300 & 900$\pm$10   & 0.53 & 2.2 & 6/6   \\ 
SCRJ0723-8015 & AMES-Dusty & 2600/2600 &  1.0 & 4.5/4.5 & 0.84 & 3.4e-20   & 200 & 900$\pm$10   & 0.53 & 2.2 & 6/6   \\ 
&BT-Settl   & 2600/2600 & 0.71 & 4.5/4.5 & 1.00 & 4.1e-20   & 100 & 1000$\pm$10  & 0.46 & 2.2 & 6/6   \\ 
\hline
\hline
\end{tabular}
\end{center}
\vspace{-0.1cm}
\flushleft
$^{1)}$ Parameter estimation determined from $\chi^2$ minimization /
PDF maximuum (whenever both estimations do not coincide).\\
$^{2)}$ Wavelength where VOSA detects the excess (in $\mu$m).\\
$^{3)}$ Sub-superscripts in the age and mass provide the range of
possible values interpolated in the isochrones and evolutionary tracks
taking into account the error in the bolometric luminosity and half
the step of the models in the effective temperature as uncrtainty in
the latter.\\
$^{*)}$ Area in the HR diagram not covered by the isochrones and
evolutionary tracks.
\normalsize
\end{table*}

%% file: best_fit_field_spec_table_compare_dust.tex
\begin{table}
\footnotesize
\begin{center}
\caption{Best fitting parameters obtained via near-infrared spectral fits. The grid of models has been linearly interpolated so that the step in T$_{\rm eff}$ is 50 K and, given teh proximity of the targets, a 0.0 mag extinction is assumed. Uncertainties correspond to 10\% degrade in $\dot{\chi^2}$} 
\label{specfitparamField}
\begin{tabular}{@{\extracolsep{-6pt}}lcccc}
\hline\hline
  Object  & Model &T$_{\rm eff}$      &$\dot{\chi^2}$   &$\log(g)$  \\
          &       & K                 &                 & dex      \\    
\hline 
&AMES-COND&3450$_{-300}^{+25}$&0.93&4.5\\
\vspace{-0.25cm}\\
2MASPJ125&AMES-DUSTY&3450$_{-450}^{+25}$&0.93&5.5\\
\vspace{-0.25cm}\\
&BT-Settl&3550$_{-300}^{+25}$&1.0&5.0\\
\hline
&AMES-COND&2450$_{-300}^{+100}$&3.5&4.5\\
\vspace{-0.25cm}\\
2MASSIJ2107&AMES-DUSTY&2000$_{-25}^{+25}$&0.87&5.5\\
\vspace{-0.25cm}\\
&BT-Settl&2050$_{-25}^{+25}$&0.97&4.5\\
\hline
&AMES-COND&2250$_{-200}^{+200}$&2.0&5.5\\
\vspace{-0.25cm}\\
2MASSJ0858&AMES-DUSTY&2500$_{-100}^{+250}$&3.1&5.5\\
\vspace{-0.25cm}\\
&BT-Settl&2500$_{-100}^{+1100}$&4.8&5.0\\
\hline
&AMES-COND&2650$_{-350}^{+25}$&3.0&5.0\\
\vspace{-0.25cm}\\
2MASSJ1239&AMES-DUSTY&2200$_{-150}^{+50}$&1.9&5.5\\
\vspace{-0.25cm}\\
&BT-Settl&2200$_{-100}^{+200}$&1.9&4.5\\
\hline
&AMES-COND&2850$_{-400}^{+200}$&2.8&5.5\\
\vspace{-0.25cm}\\
2MASSJ1434&AMES-DUSTY&2250$_{-200}^{+450}$&2.3&5.5\\
\vspace{-0.25cm}\\
&BT-Settl&2650$_{-400}^{+600}$&2.1&5.0\\
\hline
&AMES-COND&2550$_{-300}^{+25}$&2.4&4.5\\
\vspace{-0.25cm}\\
2MASSJ1731&AMES-DUSTY&2050$_{-25}^{+150}$&1.0&5.5\\
\vspace{-0.25cm}\\
&BT-Settl&2150$_{-50}^{+50}$&0.98&3.0\\
\hline
&AMES-COND&2050$_{-50}^{+25}$&5.1&3.5\\
\vspace{-0.25cm}\\
2MASSJ2107&AMES-DUSTY&2000$_{-25}^{+25}$&0.82&5.5\\
\vspace{-0.25cm}\\
&BT-Settl&2000$_{-25}^{+50}$&1.1&4.5\\
\hline
&AMES-COND&2150$_{-25}^{+25}$&2.1&3.5\\
\vspace{-0.25cm}\\
AZCnc&AMES-DUSTY&2950$_{-50}^{+25}$&0.28&5.0\\
\vspace{-0.25cm}\\
&BT-Settl&2750$_{-50}^{+25}$&0.81&3.0\\
\hline
&AMES-COND&2650$_{-150}^{+25}$&1.8&4.5\\
\vspace{-0.25cm}\\
GJ3517&AMES-DUSTY&2200$_{-150}^{+700}$&1.3&5.5\\
\vspace{-0.25cm}\\
&BT-Settl&2500$_{-250}^{+700}$&0.89&3.0\\
\hline
&AMES-COND&3150$_{-100}^{+25}$&0.62&3.5\\
\vspace{-0.25cm}\\
GJ644C&AMES-DUSTY&3000$_{-200}^{+250}$&0.54&5.5\\
\vspace{-0.25cm}\\
&BT-Settl&3100$_{-200}^{+150}$&0.44&3.5\\
\hline
&AMES-COND&3050$_{-100}^{+100}$&0.22&5.0\\
\vspace{-0.25cm}\\
LHS234&AMES-DUSTY&2900$_{-100}^{+100}$&0.18&5.0\\
\vspace{-0.25cm}\\
&BT-Settl&2900$_{-150}^{+100}$&0.30&3.5\\
\hline
&AMES-COND&2700$_{-350}^{+200}$&2.4&5.0\\
\vspace{-0.25cm}\\
LHS2397a&AMES-DUSTY&2050$_{-25}^{+150}$&0.99&5.5\\
\vspace{-0.25cm}\\
&BT-Settl&2200$_{-100}^{+50}$&1.1&4.5\\
\hline
&AMES-COND&3600$_{-150}^{+25}$&1.4&5.5\\
\vspace{-0.25cm}\\
LP803-33&AMES-DUSTY&3600$_{-150}^{+25}$&1.4&5.5\\
\vspace{-0.25cm}\\
&BT-Settl&3600$_{-150}^{+25}$&1.7&5.0\\
\hline
&AMES-COND&2600$_{-600}^{+100}$&2.5&3.0\\
\vspace{-0.25cm}\\
SCRJ0702-6102&AMES-DUSTY&2900$_{-150}^{+300}$&3.8&3.5\\
\vspace{-0.25cm}\\
&BT-Settl&2900$_{-450}^{+550}$&6.8&3.5\\
\hline
&AMES-COND&3000$_{-150}^{+200}$&0.65&4.5\\
\vspace{-0.25cm}\\
SCRJ0723-8015&AMES-DUSTY&3100$_{-150}^{+100}$&0.62&4.5\\
\vspace{-0.25cm}\\
&BT-Settl&3300$_{-450}^{+200}$&1.6&5.0\\
\hline
\end{tabular}
\end{center}
\normalsize
\end{table}